\documentclass[twocolumn,apjl]{emulateapj}

\shorttitle{DARK MATTER HALOS AND GALAXY FORMATION MODELS. II.}
\shortauthors{SEIGAR ET AL.}

\begin{document}


\title{Constraining dark matter halo profiles and galaxy formation models using spiral arm morphology. II. Dark and stellar mass concentrations for 13 nearby face-on galaxies}


\author{Marc S.\ Seigar$^{1,2}$, Benjamin L.\ Davis$^3$, Joel Berrier$^{3,4,5}$, and Daniel Kennefick$^{3,4}$}
\footnotetext[1]{Department of Physics, University of Minnesota Duluth, 1023 University Drive, MWAH 371, Duluth, MN 55812-3009}
\footnotetext[2]{Department of Physics \& Astronomy, University of Arkansas at Little Rock, 2801 S.\ University Avenue, Little Rock, AR 72204-1099}
\footnotetext[3]{Arkansas Center for Space and Planetary Sciences, 202 Field House, University of Arkansas, Fayetteville, AR 72701}
\footnotetext[4]{Department of Physics, University of Arkansas, 835 West Dickson Street, Fayetteville, AR 72701}
\footnotetext[5]{Department of Physics \& Astronomy, Rutgers University, 136 Frelinghuysen Road, Piscataway, NJ 08854}



\begin{abstract}
We investigate the use of spiral arm pitch angles as a probe of disk galaxy mass profiles. We confirm our previous result that spiral arm pitch angles (P) are well correlated with the rate of shear (S) in disk galaxy rotation curves.  We use this correlation to argue that imaging data alone can provide a powerful probe of galactic mass distributions out to large look-back times.  We then use a sample of 13 galaxies, with {\em Spitzer} 3.6-$\mu$m imaging data and observed H$\alpha$ rotation curves, to demonstrate how an inferred shear rate coupled with a bulge-disk decomposition model and a Tully-Fisher-derived velocity normalization can be used to place constraints on a galaxy's baryon fraction and dark matter halo profile.  Finally we show that there appears to be a trend (albeit a weak correlation) between spiral arm pitch angle and halo concentration.  We discuss implications for the suggested link between supermassive black hole (SMBH) mass and dark halo concentration, using pitch angle as a proxy for SMBH mass.
\end{abstract}





\section{Introduction}

The discovery of a correlation between spiral arm pitch angle and
rotation curve rate of shear (Seigar, Block \& Puerari 2004, Seigar 
et al.\ 2005, 2006) may be a fundamental property of galaxies that is
a key in understanding mass  distributions. This is because the rate
of shear is related to the slope of the outer part of the rotation 
curve, which in turn is related to the galaxy mass distribution. In
other words, shear is strongly connected to the central mass concentrations
of galaxies.

The shear, $S$, is a dimensionless quantity, and can be measured directly
from rotation curves, if one is available. It is defined as follows,
\begin{equation}
\label{shearrate}
S=\frac{A}{\omega}=\frac{1}{2}\left(1-\frac{R}{V}\frac{dV}{dR}\right),
\end{equation}
where A is the first Oort constant, $\omega$ is the angular velocity and
$V$ is the velocity at a radius $R$. The shear rate depends upon the
shape of the rotation curve. For a rotation curve that remains flat $S=0.5$,
for a falling rotation curve $S>0.5$ and for a continually rising rotation
curve $S<0.5$. As the shape of a rotation curve depends on the mass
distribution, the shear rate at any given position depends upon the mass
within that radius, or the central mass concentration. As a result, the
spiral arm pitch angle is dependent upon the central mass concentration, and 
this is consistent with the expectations of most spiral density wave models
(e.g. Bertin et al.\ 1989a, b; Bertin 1991, 1993, 1996; Bertin \& Lin 1996; 
Fuchs 1991, 2000), although density wave models predict that pitch angles
also depend on stability (i.e. the Toomre $Q$-parameter). 

The correlation between shear and spiral arm pitch angles may explain the
weak correlations found between properties like bulge-to-disk ratio ($B/D$),
spiral arm pitch angle and Hubble Type (e.g.\ de Jong 1996; Seigar \&
James 1998a, b). In these studies the correlations are weak because,
even the near-infrared $B/D$ is not a complete picture when it comes 
to determining the central mass concentration, as the dark matter
concentration is not taken into account. However, a property such as
shear is sensitive to the total baryonic and dark matter concentrations.

Given the relationship between shear and spiral arm pitch angle (Seigar
et al.\ 2006; hereafter Paper 1) it is possible to determine shear, from
imaging data alone, given the equation from Paper 1,
\begin{equation}
\label{shearpitch}
P=(64.25\pm2.87)-(73.24\pm5.53)S,
\end{equation}
where $P$ is the pitch angle in degrees and $S$ is the shear rate.
This relationship opens  up a  fundamentally new  approach for probing  mass
distributuions  in spiral galaxies.  This   approach relies on imaging
data alone without the    need for full rotation    curve information.
Specifically, the  Tully-Fisher relation (Tully \& Fisher 1977)
for spiral  galaxies coupled
with the shear rate - pitch angle relation can be  used to determine a
rotation  curve normalization  {\it and}   slope.   In this  paper  we
explicitly demonstrate how, given  a bulge-disk decomposition, a  pitch
angle determination, and a Tully-Fisher normalizaion, one can constrain
galaxy mass distributions,    dark matter halo  concentrations,   and
other galaxy  formation  parameters.  We  obtain our constraints within  the
context of the  the standard framework  of disk formation put forth by
Fall \& Efstathiou (1980) and Blumenthal et al.\ (1986).

In principle, the technique we demonstrate here and in Paper 1
can be applied generally to a large sample of galaxies and to galaxies
at high redshift, when spiral arms are detected.
This is the second in a series of papers in which we use spiral
arm pitch angles to determine the mass distribution in spiral galaxies.
In this paper we determine mass concentrations in a large sample
of disk galaxies from their spiral arm pitch angles (as this is intrinsically
related to their shear rates, i.e.\ we determine shear using equation 
\ref {shearpitch}), their disk masses
and their disk scalelengths (determined via a bulge-disk decomposition 
technique). 

This paper is arranged as follows.  Section 2 describes the data we used and
the analysis tools.  Section 3 describes how we determine the stellar mass
distribution in our sample of galaxies.  Section 4 describes the dark matter
halo density profile models we use and how we determine the halo profiles of
our sample of galaxies.  In section 5 we present a discussion of our results
and in section 6 we provide our main conclusions and present our planned
future goals.

Throughout this work we assume a Hubble
constant of $h=0.705$ in units of $100$ km s$^{-1}$ Mpc$^{-1}$ and a
cosmology with $\Omega_m = 1 - \Omega_{\Lambda} = 0.274$ (Komatsu et al.\
2009).

\section{Observations and data reduction}


\subsection{Data}

The dataset presented in this paper consists of 13 spiral galaxies observed
with {\em Spitzer IRAC} at 3.6-$\mu$m.  These galaxies were chosen as follows:
a parent sample (defined as a Southern Hemisphere ($\delta<0^{\circ}$) galaxies
with $B_T<12.9$) of 605 bright galaxies (Ho et al.\ 2011) was selected.
From this sample, we then selected all galaxies with observed H$\alpha$ 
rotation curves from Persic \& Salucci (1995) that also have {\em Spitzer}
3.6-$\mu$m imaging data available on the 
archive\footnote{http://sha.ipac.caltech.edu/applications/Spitzer/SHA/}.
This results in a sample of 13 galaxies.  We then measure spiral arm pitch 
angles, and then perform a bulge-disk decomposition, using the Spitzer 
images.

\subsection{Measurement of spiral arm pitch angles}

Spiral arm pitch angles are measured using the same technique employed
by Davis et al.\ (2012). A  two-dimensional fast-Fourier (2D-FFT) decomposition 
technique is used, which employs a program described in Schr\"oder et
al. (1994). Logarithmic spirals are assumed in the
decomposition. The resulting pitch angles are listed in Table 1.

The amplitude of each Fourier component is given by
\begin{equation}
\label{fft}
A(m,p)=\frac{\sum_{i=1}^{I}\sum_{j=1}^{J}I_{ij}(\ln{r},\theta)\exp{[-i(m\theta _p\ln{r})]}}{\sum_{i=1}^{I}\sum_{j=1}^{J}I_{ij}(\ln{r},\theta)},
\end{equation}
where $r$ and $\theta$ are polar coordinates, $I(\ln{r},\theta)$ is the 
intensity at position $(\ln{r},\theta)$, $m$ represents the number of arms
or modes, and $p$ is the variable associated with the pitch angle $P$, defined
by $\tan{P}=-(m/p)$. 
The resulting pitch angle measured using equation \ref{fft} 
is in radians, and this is later converted to degrees for ease of perception.

The images were first projected to face-on. Mean uncertainties of position
angle and inclination as a function of inclination were discussed by 
Consid\`ere \& Athanassoula (1988). For a galaxy with low inclination, there
are clearly greater uncertainties in assigning both a position angle and an\
accurate inclination. These uncertainties are discussed by Block et al.\ (1999)
and Seigar et al.\ (2005), who take a galaxy with low inclination 
($<30^{\circ}$) and one with high inclination ($>60^{\circ}$) and varied the
inclination angle used in the correction to face-on. They found that for the
galaxy with low inclination, the measured pitch angle remained the same. 
However, the measured pitch angle for the galaxy with high inclination varied 
by $\pm 10$\%. Since inclination corrections are likely to be largest for
galaxies with the highest inclinations cases where inclination is $>60^{\circ}$
are taken as the worst case scenario.
For galaxies with inclination $i>60^{\circ}$ we take into 
account this uncertainty. Our deprojection method assumes that spiral galaxy
disks are intrinsically circular in nature.

The 2D-FFT code is then deployed to work on the deprojected images.  The 
code requires a user-defined inner and outer radius.  The outer radius is 
chosen as the radius at which spiral structure (on any galactic structure)
can no longer be seen.  The inner radius is chosen as the innermost radius 
which excludes any bulge and/or bar structure.  While the outer radius is
fairly straightforward to chose, and a slightly incorrect value will not
affect the derived pitch angle, the same is not true of the inner radius.  
If the inner radius is chosen slightly incorrectly, the derived pitch 
angle can be strongly affected by (for example) an inner bar.  As a result
Davis et al.\ (2012) employed an iterative routine that chose every possible
inner radius from the very central regions of the galaxy out to the user-defined
outer radius.  The code then outputs the pitch angle as a function of
inner radius for each mode, $1 \leq m \leq 6$.  For every galaxy, a range
of inner radii always results in a pitch angle that remains approximately
constant (Davis et al.\ 2012).  This is the pitch angle that we chose, and
the variation of the pitch angle over these radii is a measure of the error.
This region is refered to as the ``stable region''.  Note that the 2D-FFT code
actually outputs a pitch angle that is either negative or positive based upon
their chirality (Davis et al.\ 2012).  For the purpose of this paper, we 
report the absolute values of the pitch angle.

\subsection{Measurement of shear}

In this paper, 13 of the galaxies have shear values that were measured
directly from their H$\alpha$ rotation curves. The shear was initially 
calculated in Paper 1 and the rotation curves were taken from Persic \& 
Salucci (1995). These 
rotation curves are of good quality with an rms error $<10$ km 
s$^{-1}$, and an error associated with folding the two sides of the galaxy
also $<10$ km s$^{-1}$. These rotation curves have been used to estimate the
shear rates in these galaxies, using the same method used by other authors
(e.g. Block et al.\ 1999; Seigar et al.\ 2005; Seigar 2005).

\begin{deluxetable*}{lccccccccccc}
\tabletypesize{\scriptsize}
\tablecolumns{10}
\tablewidth{0pc}
\tablecaption{Galaxy measurements}
\label{table}
\tablehead{
\colhead{Galaxy} & \colhead{Hubble} & \colhead{$b/a$} & \colhead{$V_{rec}$}     & \colhead{PA}        & \colhead{$P$}       & \colhead{$S$} & \colhead{$L_{disk}$}                 & \colhead{$R_d$} & \colhead{$B/D$} \\
\colhead{Name}   & \colhead{Type}   & \colhead{}      & \colhead{}              & \colhead{}          & \colhead{}          & \colhead{}    & \colhead{}                           & \colhead{}      &       \\
\colhead{}       & \colhead{}       & \colhead{}      & \colhead{(km s$^{-1}$)} & \colhead{(degrees)} & \colhead{(degrees)} & \colhead{}    & \colhead{($\times10^{10}L_{\odot}$)} & \colhead{(kpc)} & \colhead{}      \\
}
\startdata 
ESO 009-G010  & SAc    & 0.76  & 2418          & 171	   & 23.7$\pm$1.1 & 0.44$\pm$0.06 & 1.80$\pm$0.16   & 5.97$\pm$0.53 & 0.10$\pm$0.01 \\
ESO 582-G012  & SAc    & 0.60  & 2325          & 48	   & 22.6$\pm$0.6 & 0.51$\pm$0.05 & 1.65$\pm$0.15   & 6.46$\pm$0.59 & 0.14$\pm$0.01 \\
IC 4808       & SAc    & 0.42  & 5084          & 45	   & 14.1$\pm$0.4 & 0.63$\pm$0.02 & 4.71$\pm$0.45   & 7.52$\pm$0.72 & 0.06$\pm$0.01 \\
NGC 150       & SBb    & 0.49  & 1584          & 118       &  8.4$\pm$0.1 & 0.65$\pm$0.03 & 1.81$\pm$0.15   & 3.09$\pm$0.25 & 0.19$\pm$0.02 \\
NGC 578       & SABc   & 0.63  & 1628          & 110       & 18.0$\pm$0.2 & 0.62$\pm$0.06 & 1.15$\pm$0.11   & 6.75$\pm$0.65 & 0.12$\pm$0.01 \\
NGC 908       & SAc    & 0.43  & 1509          & 75	   & 12.9$\pm$0.4 & 0.59$\pm$0.04 & 3.01$\pm$0.30   & 5.71$\pm$0.57 & 0.08$\pm$0.01 \\
NGC 1292      & SAc    & 0.43  & 1366          & 7         & 29.8$\pm$1.0 & 0.49$\pm$0.04 & 0.50$\pm$0.04   & 3.72$\pm$0.30 & 0.06$\pm$0.01 \\
NGC 1300      & SBbc   & 0.66  & 1577          & 106       & 31.7$\pm$1.1 & 0.50$\pm$0.03 & 2.64$\pm$0.25   & 9.34$\pm$0.88 & 0.16$\pm$0.02 \\
NGC 1353      & SAbc   & 0.41  & 1525          & 138	   & 36.6$\pm$1.0 & 0.34$\pm$0.05 & 2.03$\pm$0.20   & 4.66$\pm$0.46 & 0.20$\pm$0.02 \\
NGC 1365      & SBb    & 0.55  & 1636          & 32	   & 35.4$\pm$1.7 & 0.53$\pm$0.03 & 6.38$\pm$0.62   & 9.42$\pm$0.92 & 0.14$\pm$0.01 \\
NGC 1964      & SABb   & 0.38  & 1659          & 32        & 13.8$\pm$0.3 & 0.60$\pm$0.02 & 1.69$\pm$0.17   & 6.45$\pm$0.65 & 0.20$\pm$0.02 \\
NGC 3223      & SAbc   & 0.61  & 2891          & 135       & 10.7$\pm$2.0 & 0.70$\pm$0.02 & 6.72$\pm$0.68   & 9.40$\pm$0.95 & 0.14$\pm$0.01 \\
NGC 3318      & SABb   & 0.54  & 2775          & 78        & 36.9$\pm$6.5 & 0.52$\pm$0.03 & 2.40$\pm$0.22   & 6.03$\pm$0.55 & 0.10$\pm$0.01 \\
\enddata
\tablecomments{Column (1) is galaxy name; Column (2) is galaxy Hubble type taken from de Vaucouleurs et al.\ (1991; hereafter RC3); Column (3) is axis ratio taken from RC3; Column (4) is the recessional velocity taken from the NASA Extragalactic Database (NED); Column 5 is the Position Angle taken from RC3; Column (6) is the measured pitch angle in degrees; Column (7) is the measured shear; Column (8) is the disk luminosity determined from the bulge-disk decomposition; Column (9) is the disk scalelength determined from the bulge-disk decomposition; and Column (10) is the bulge-to-disk light ratio determined from the bulge-disk decomposition.
}
\end{deluxetable*}

\begin{figure}
\label{shearpitchfig}
\plotone{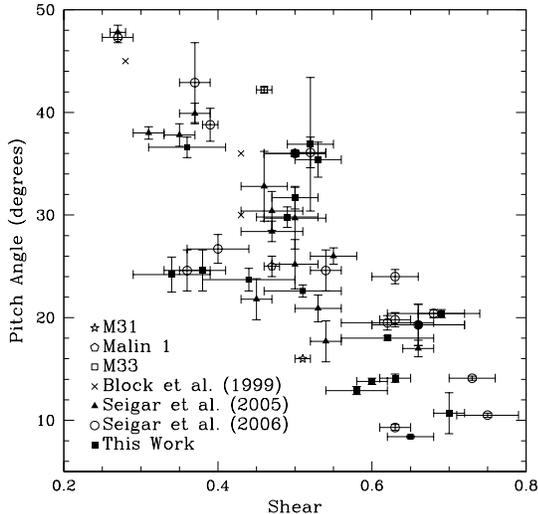}
\caption{Spiral arm pitch angle vs.\ rotation curve shear rate, showing a
correlation. The filled triangles represent galaxies with data measured by 
Seigar et al.\ (2005), the open circles are galaxies from Seigar et al.\ 
(2006), the crosses are galaxies from Block et al.\ (1999), the open square 
represents data for M33 (Seigar 2011), the open pentagon represents data for
Malin 1 (Seigar 2008), the star represents data for M31 (Seigar, Barth, \& 
Bullock 2008), and the filled squares represent the data from the present 
sample.}
\end{figure}

Using equation \ref{shearrate}, we have calculated the shear rates for these
galaxies, over the same radial ranges (i.e. the stable region) for which the 
pitch angles were measured.  The dominant sources of error on the shear rate
are the rms error in the rotation curve and the error associated
with folding the two sides of the galaxy. This is typically $<10$\%. In
order to calculate the shear rate, the mean value of $dV/dR$ measured in
km s$^{-1}$ arcsec$^{-1}$ is calculated by fitting a line of constant
gradient to the outer part  of the rotation curve (i.e.\ past the radius
of turnover and any bar or bulge that may exist in the galaxy) over the 
``stable region'', i.e., the region where the pitch angle remains approximately
constant.  Our measured shears are listed in Table 1. 

With values for the pitch angle and shear for each galaxy, we then decided
to revisit the correlation between pitch angle and shear (Seigar et al.\
2005, 2006).  Figure 1 shows a plot of pitch angle in degrees as a function
of shear as determined above.  A strong correlation still exists, although the
scatter has increased a little from that presented in Seigar et al.\ (2006).

\section{Determination of the stellar mass distribution}

For each of our galaxies we produce surface brightness profiles
using the IRAF {\bf Ellipse} routine, which fits ellipses to
isophotes in an image, using an iterative method described
by Jedrzejewski (1987). From the surface brightness profile
we then determine the disk and bulge B-band luminosity
using  an exponential disk and a S\'ersic-law bulge. We utilize 
a 1-D bulge-disk decompostion routine, which performs Levenberg-Marquardt
least-squares minimization.
Explicitly we fit a S\'ersic law profile for each bulge via
\begin{equation}
\mu(R)=\mu_{e}\exp{\left\{-b_{n}\left[\left(\frac{R}{R_{e}}\right)^{1/n}-1\right]\right\}},
\label{bulge}
\end{equation}
where $R_e$ is the effective radius, containing 50\% of the total light of the
bulge and $\mu_e$ is the surface brightness at $R_e$.
The factor $b_{n}$ is a function of the 
shape parameter, $n$, such that $\Gamma(2n)=2\gamma(2n,b_{n})$, where $\Gamma$
is the gamma function and $\gamma$ is the incomplete gamma function
(see Graham \& Driver 2005). As given
by Capaccioli (1989), $b_n$ can be well approximated by $1.9992n-0.3271$ 
for $1<n<10$. In the case where $n=1$, the S\'ersic model is equivalent to an 
exponential, and when $n=4$ it is equivalent to the $R^{1/4}$ model (de 
Vaucouleurs 1948, 1957).
We fit the disk component using
\begin{equation}
\mu(R)=\mu_0 \exp{(-R/R_d)},
\label{disk}
\end{equation}
where $\mu_0$ is the central surface brightness and $R_d$ 
is the scalelength of the disk.

\begin{figure*}
\label{b2d}
\includegraphics[width=4.1cm]{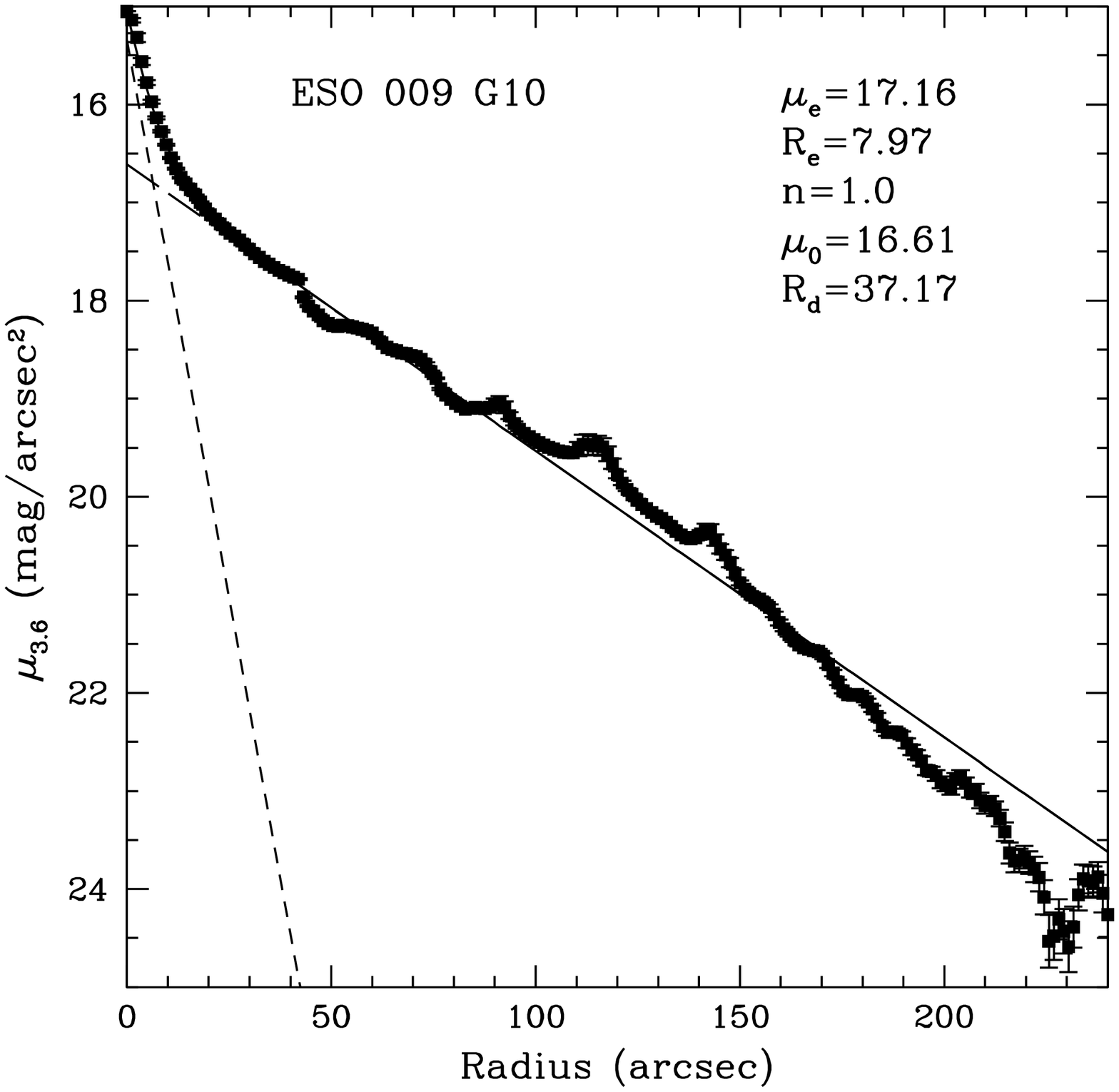}
\includegraphics[width=4.1cm]{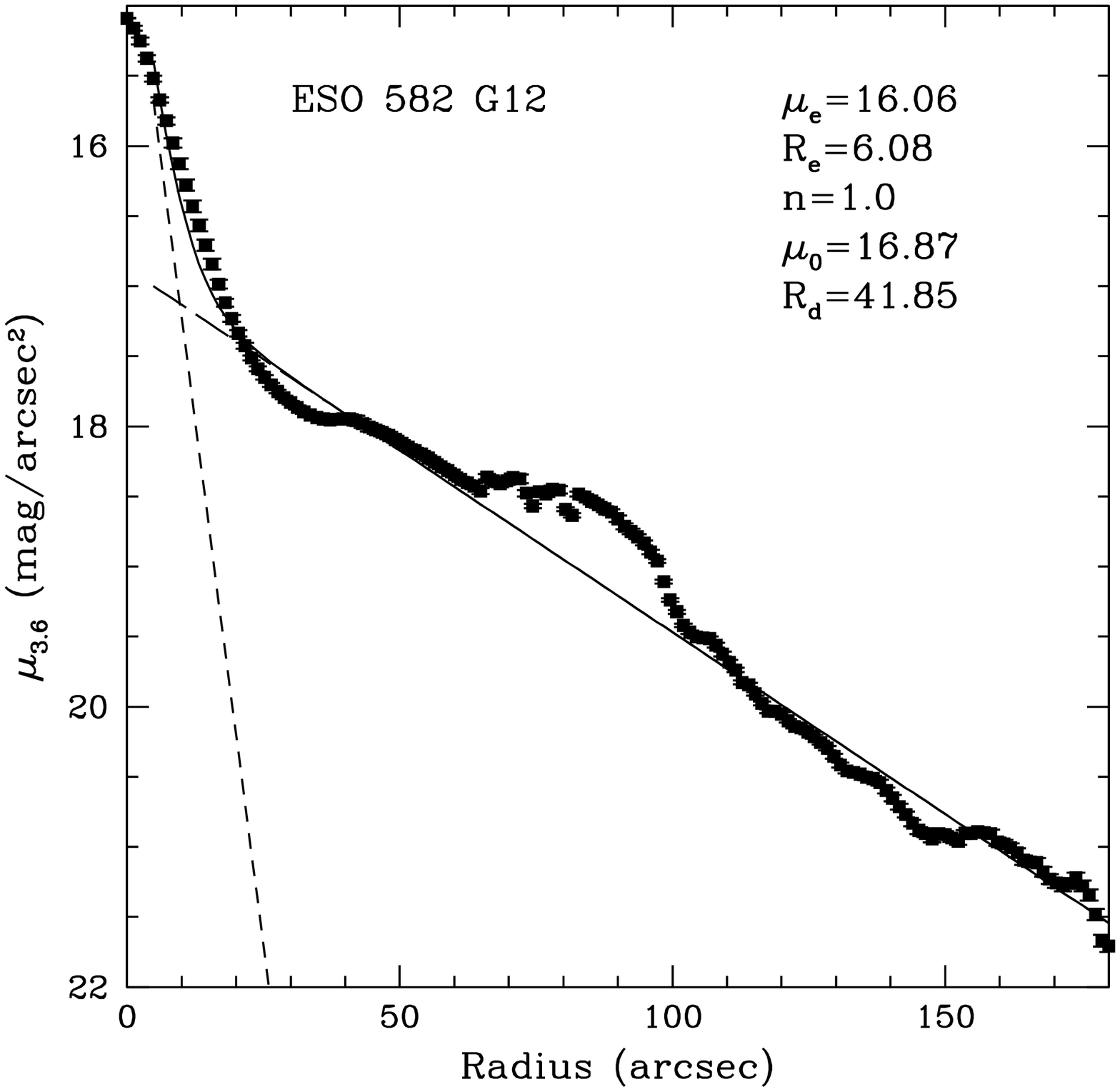}
\includegraphics[width=4.1cm]{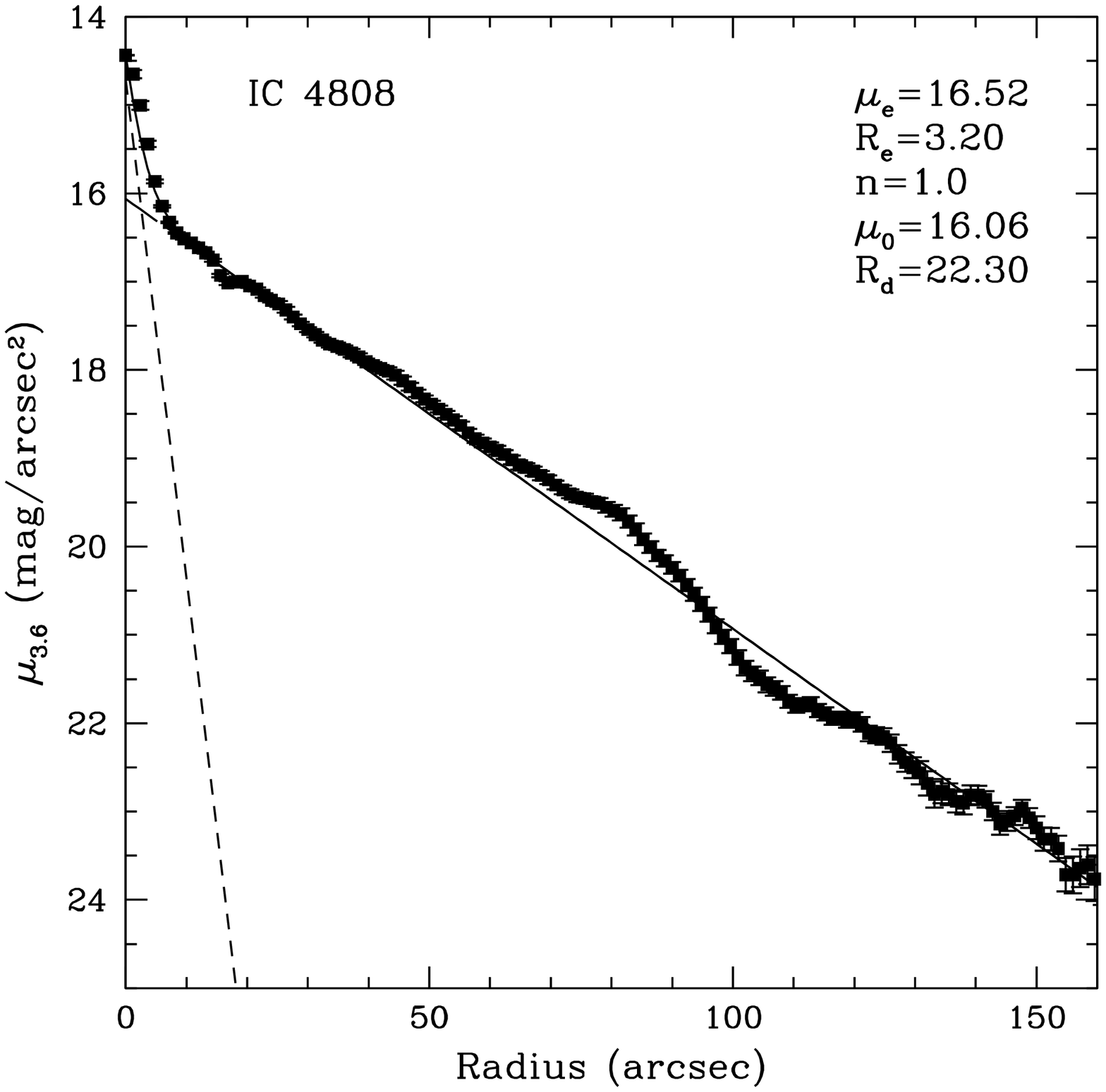}
\includegraphics[width=4.1cm]{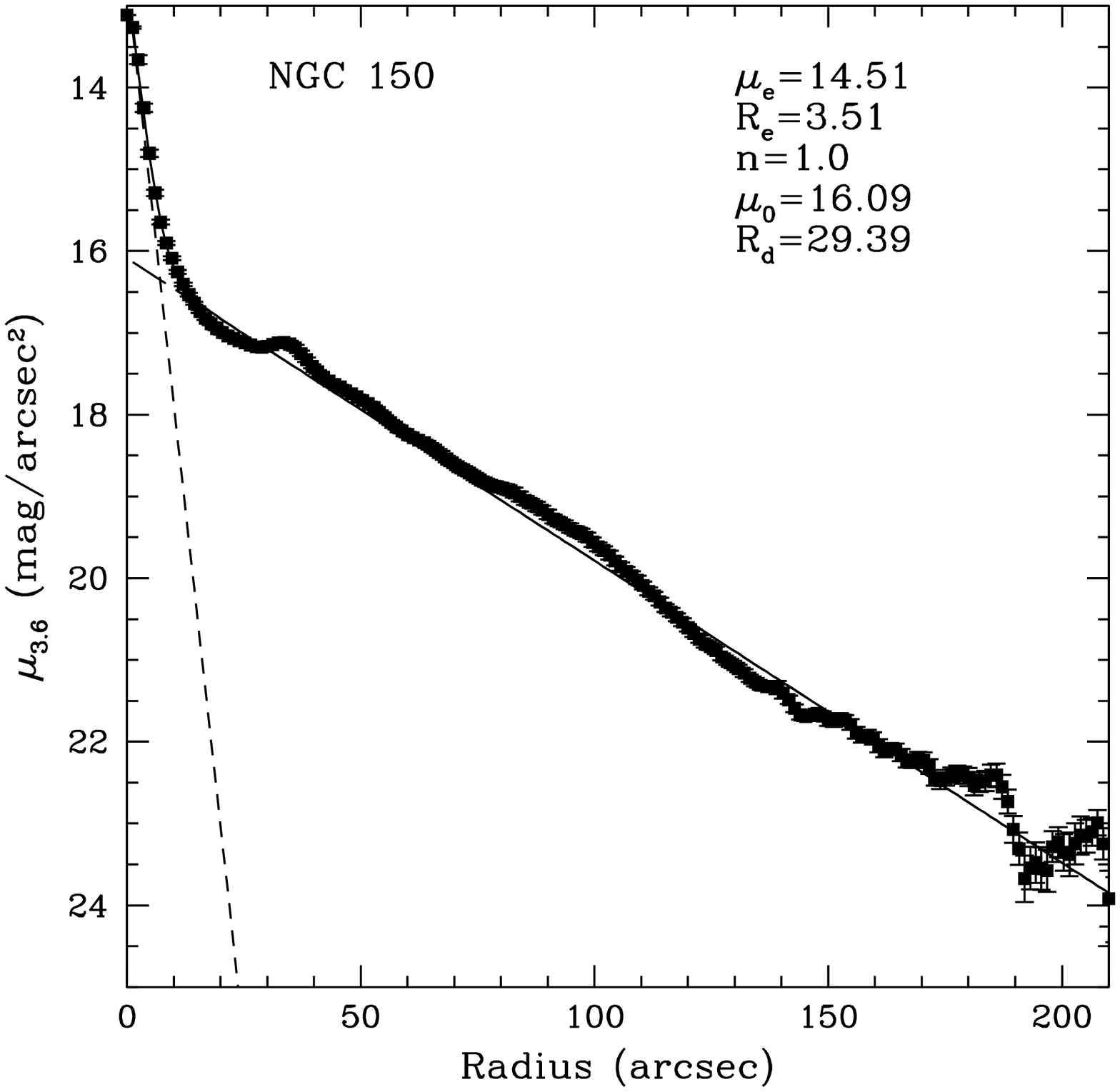}\\
\includegraphics[width=4.1cm]{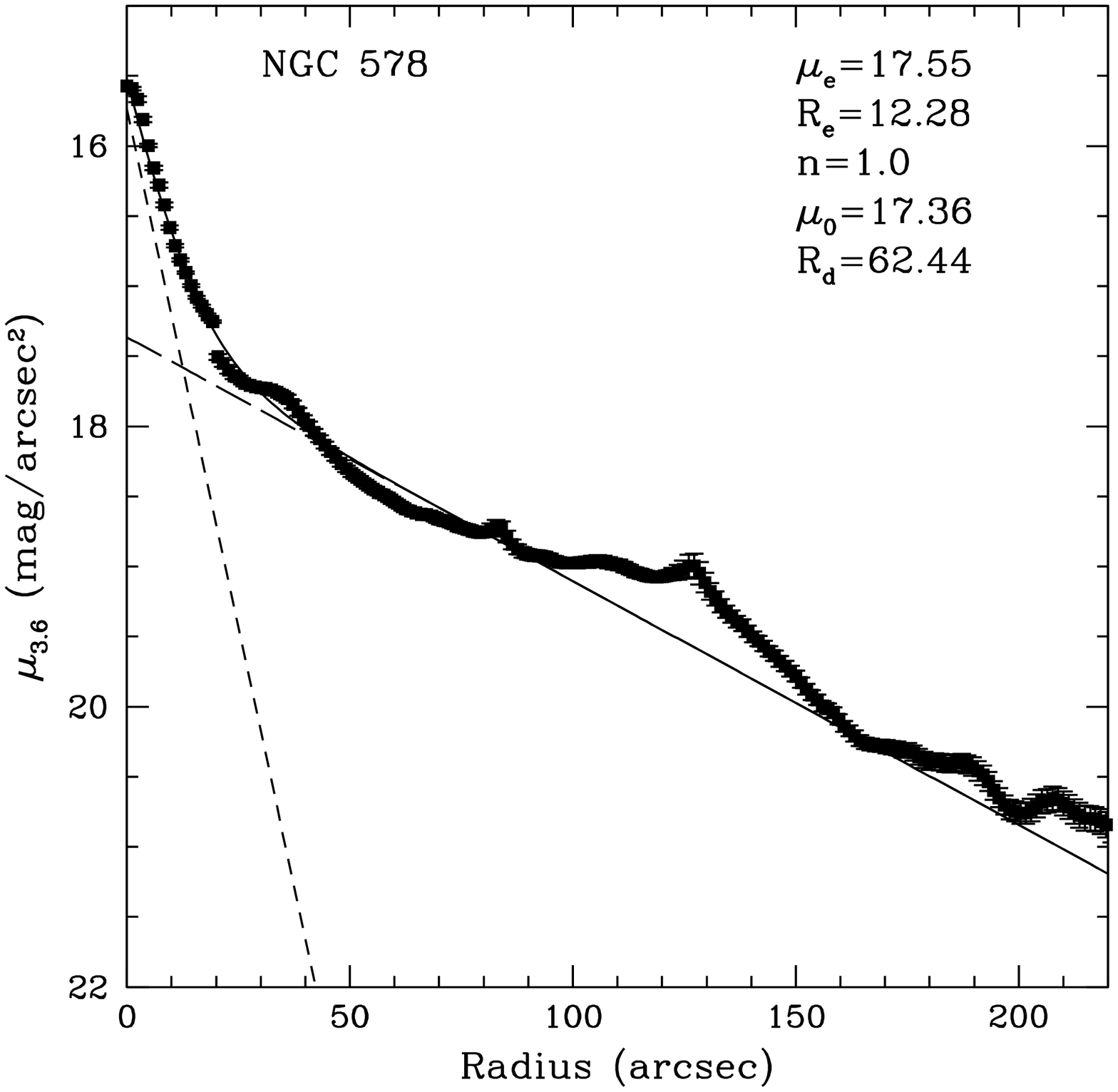}
\includegraphics[width=4.1cm]{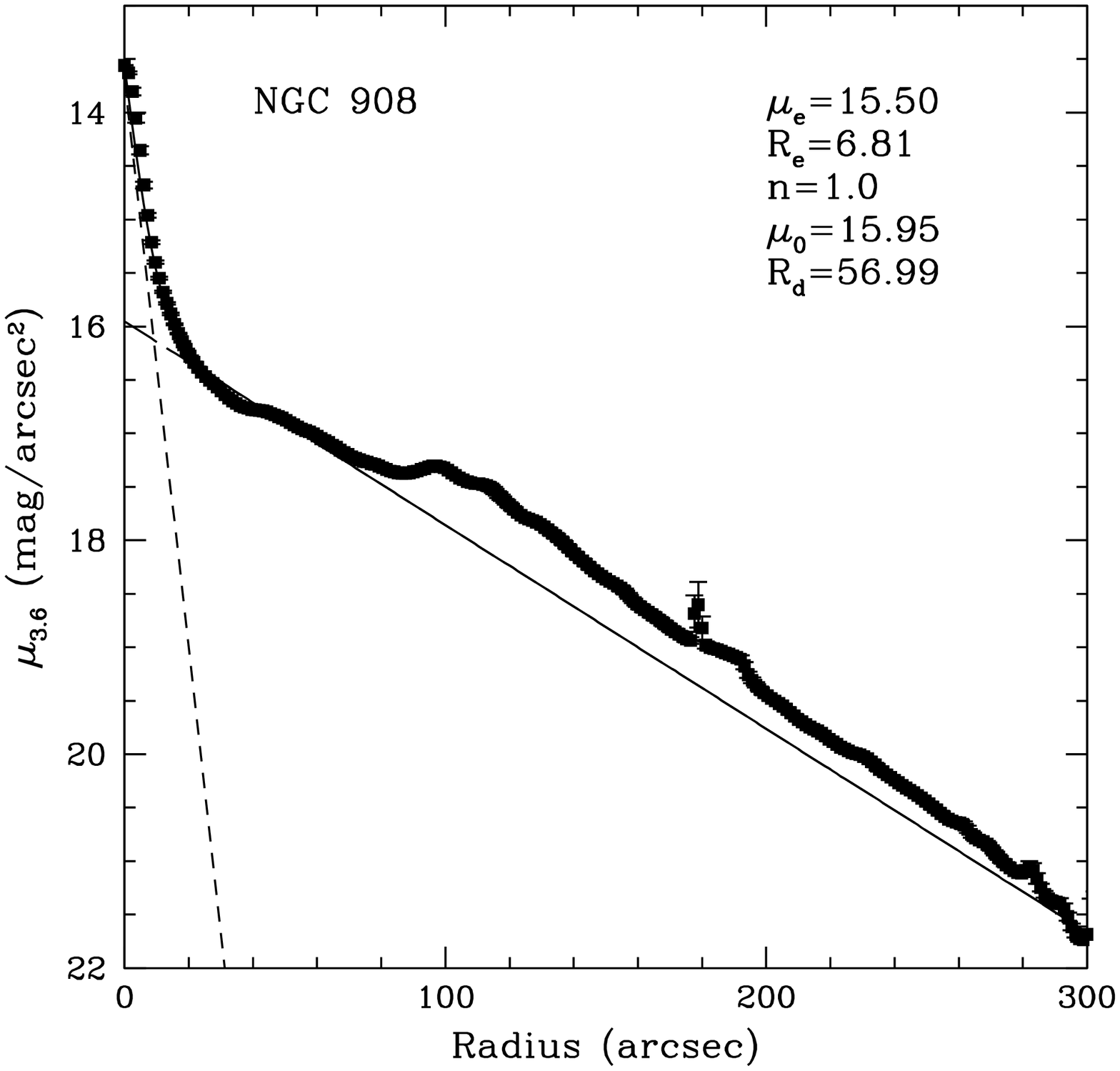}
\includegraphics[width=4.1cm]{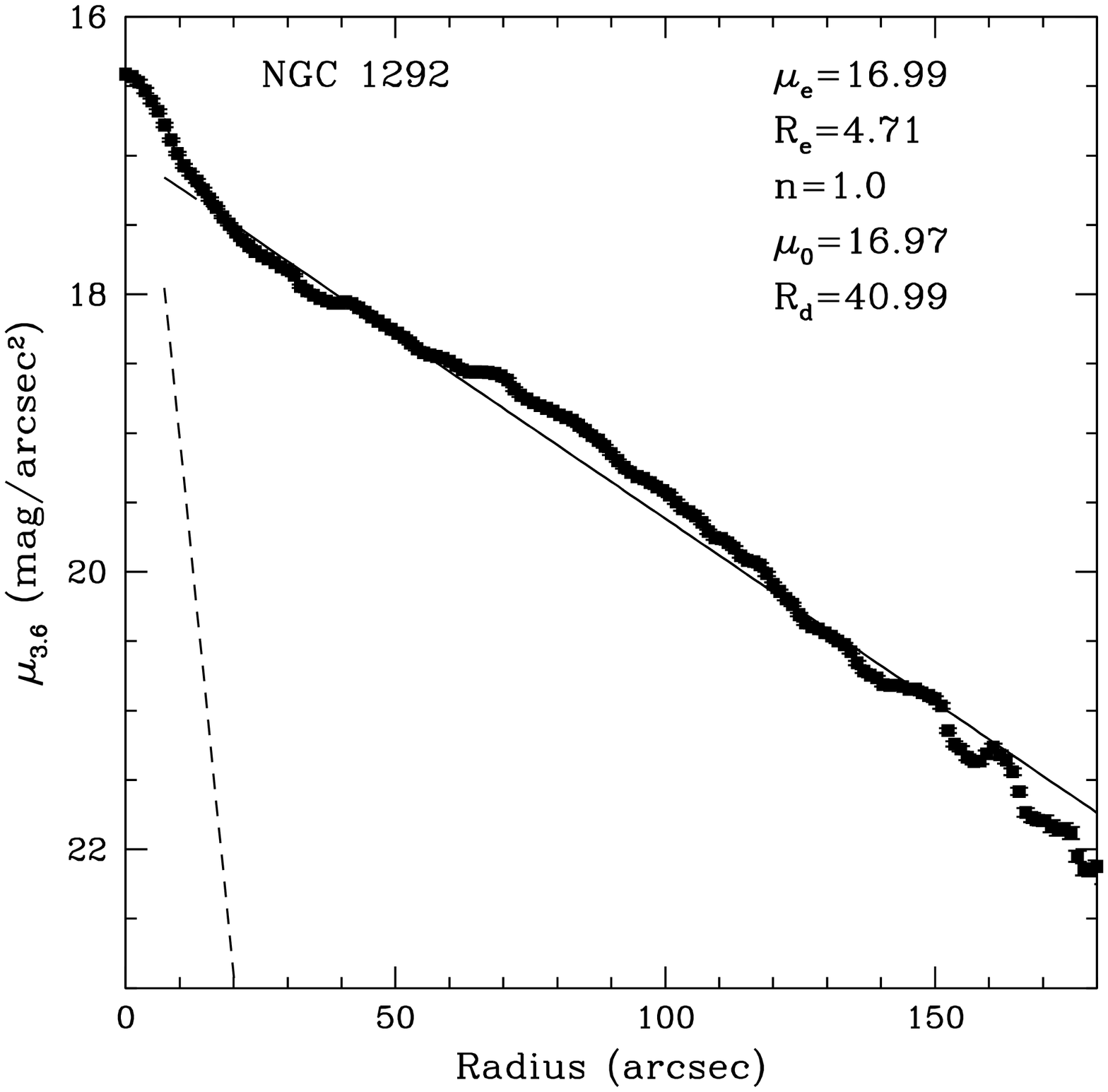}
\includegraphics[width=4.1cm]{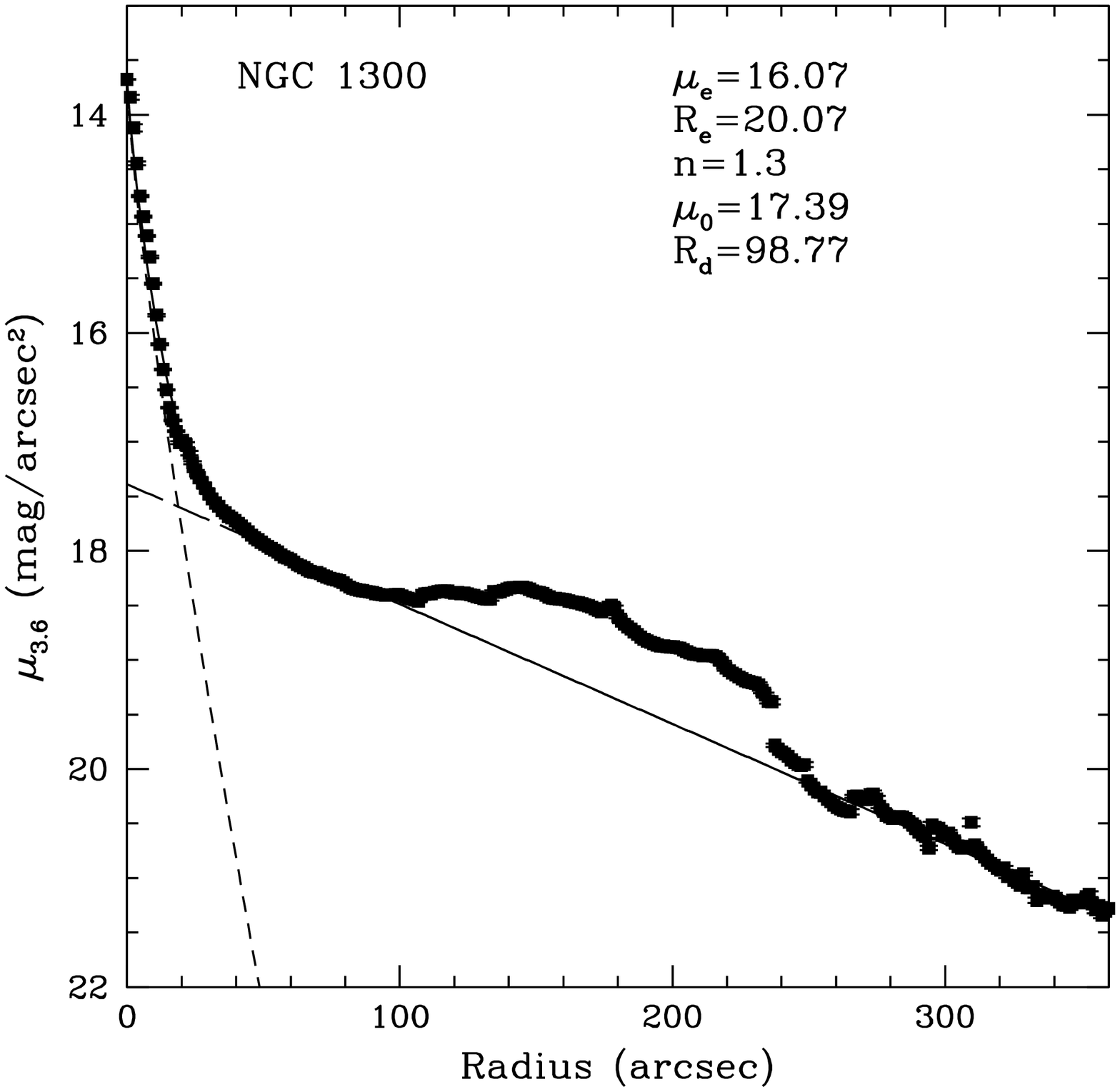}\\
\includegraphics[width=4.1cm]{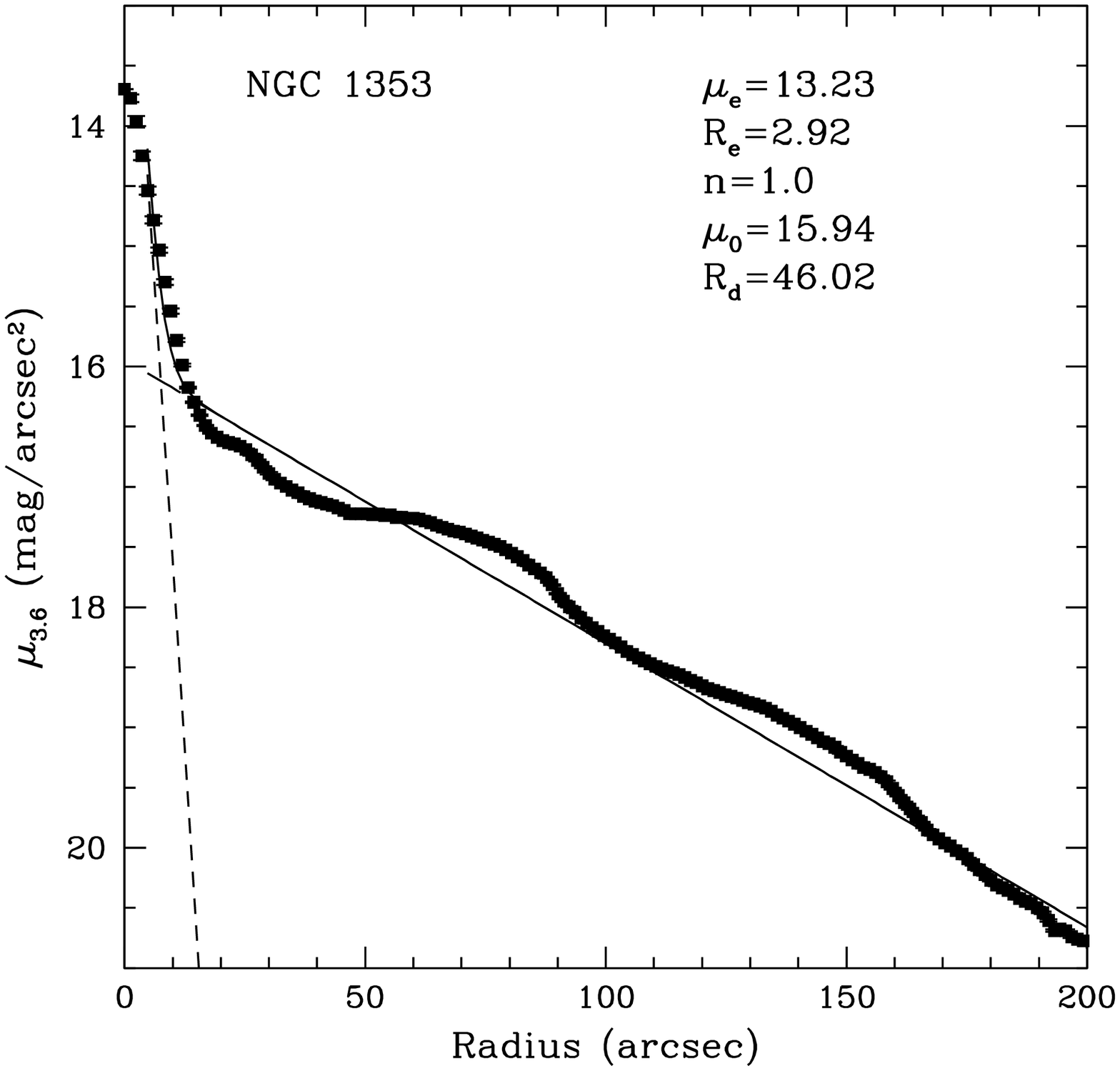}
\includegraphics[width=4.1cm]{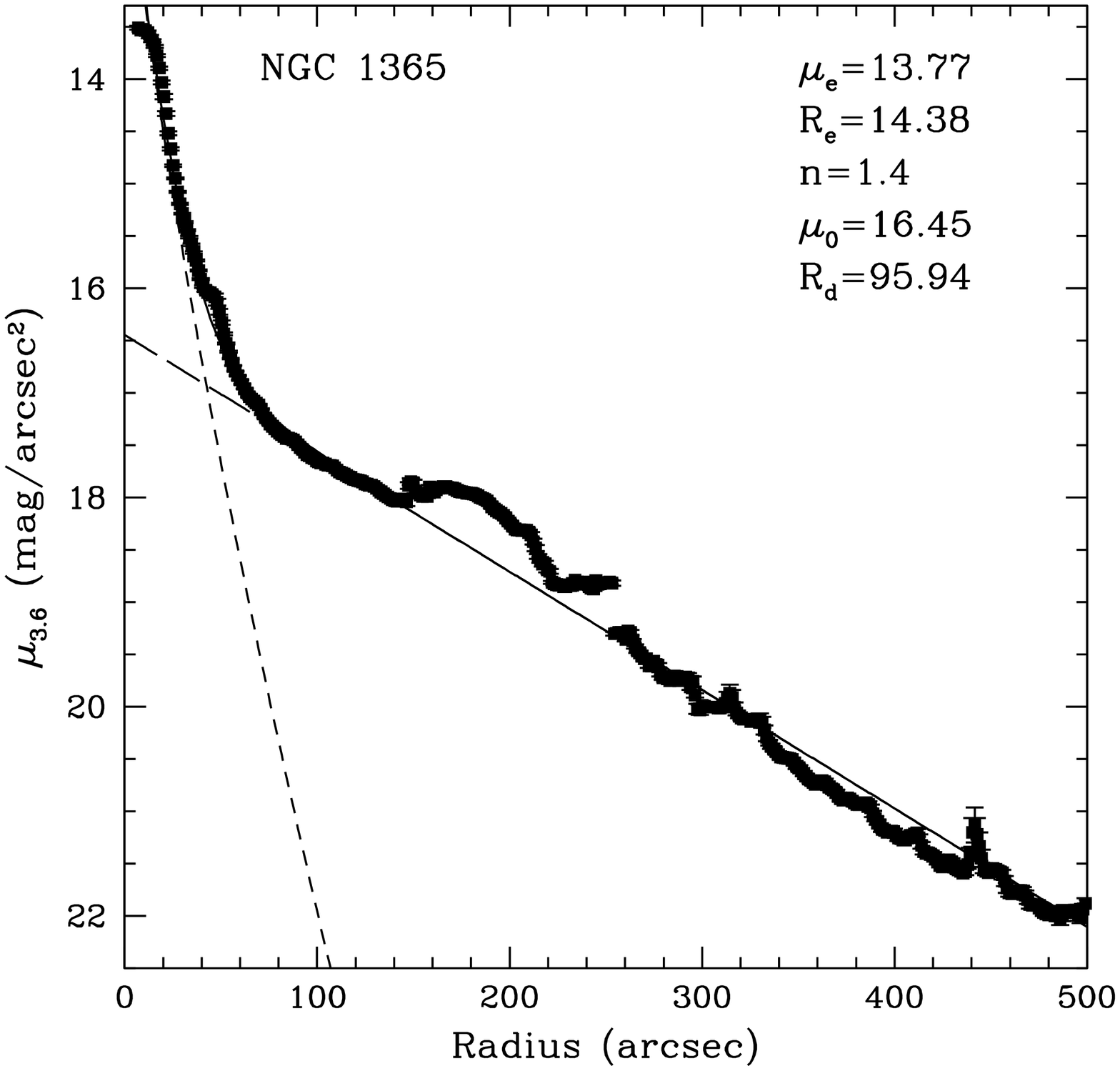}
\includegraphics[width=4.1cm]{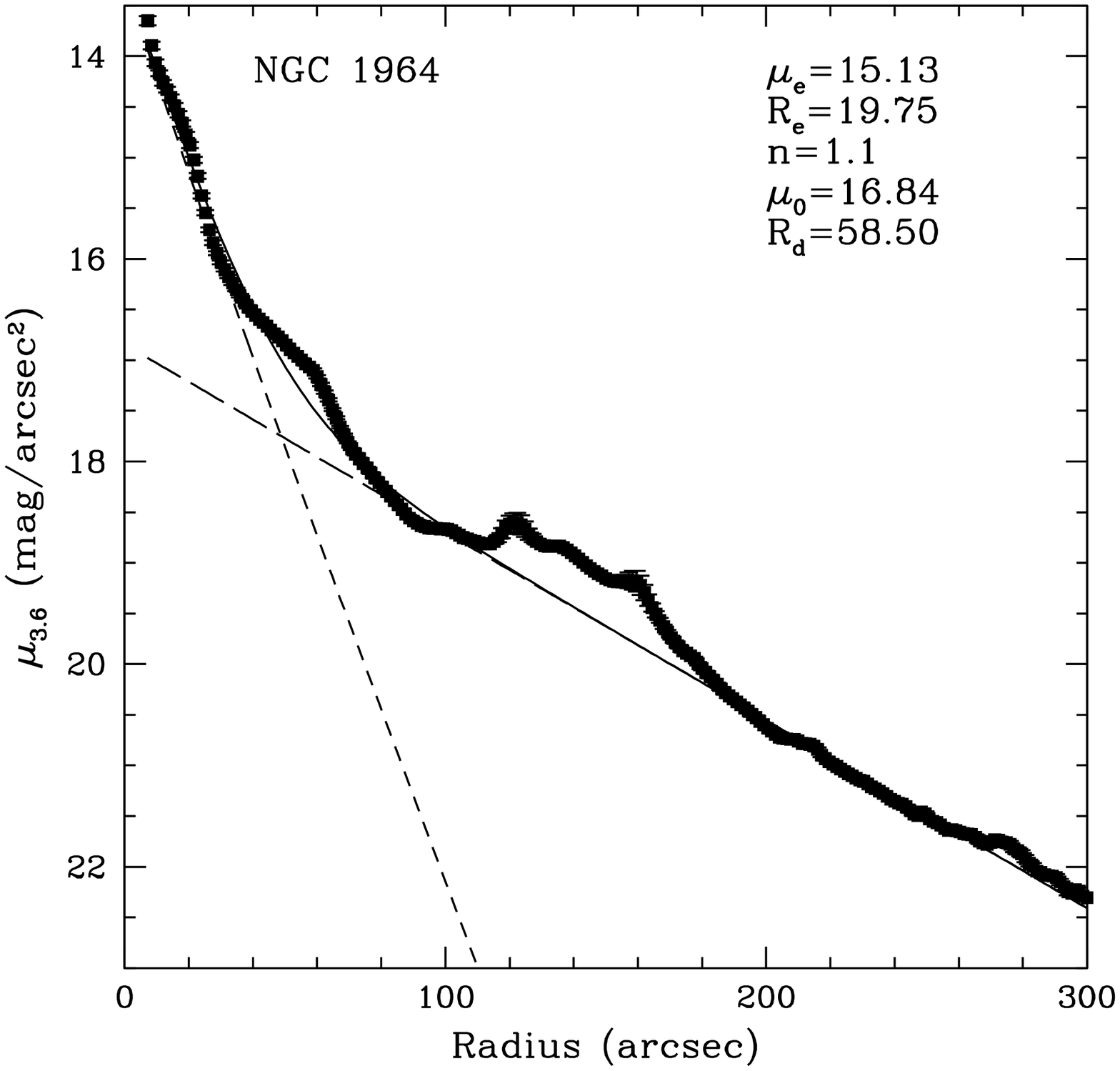}
\includegraphics[width=4.1cm]{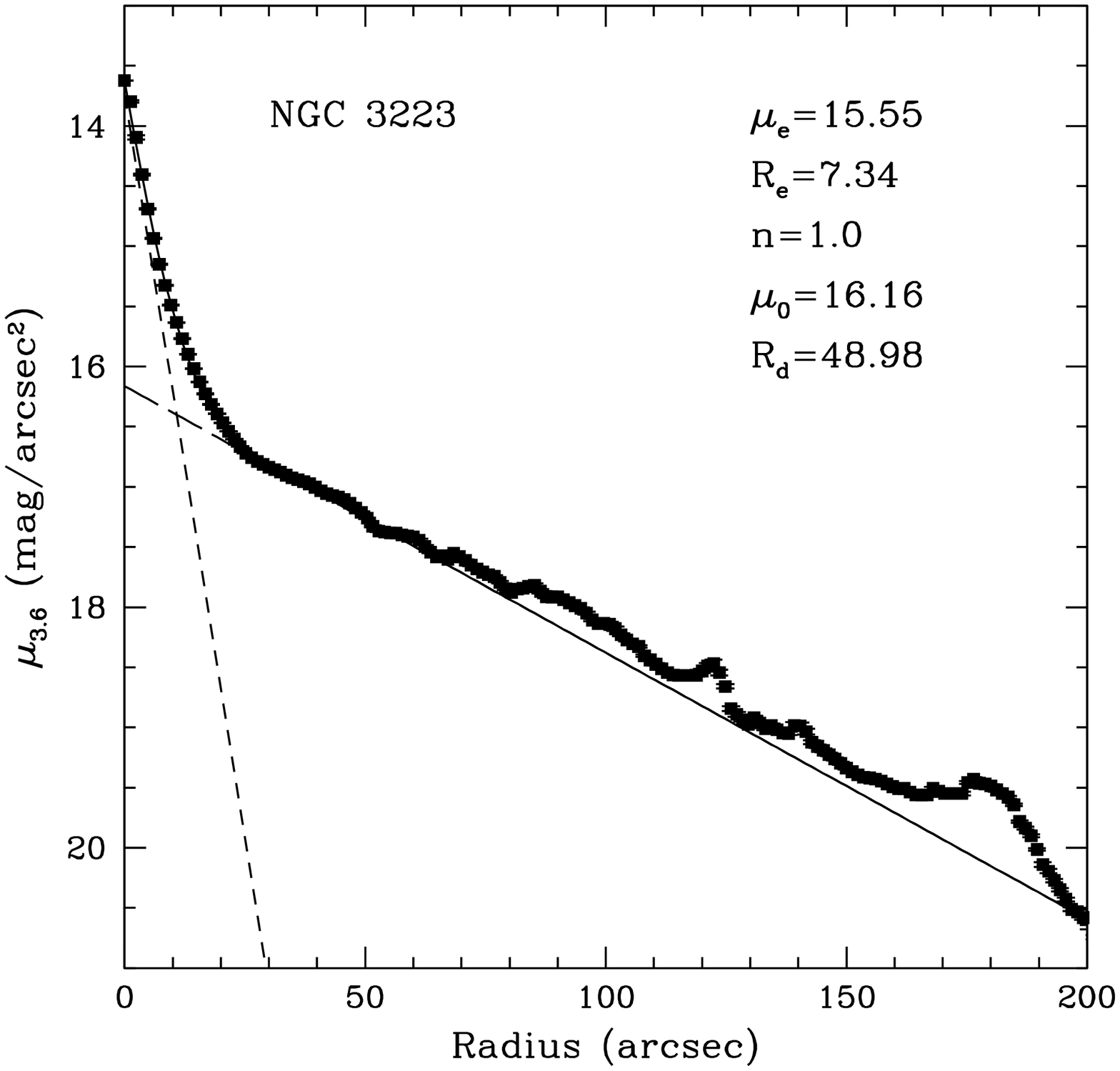}\\
\includegraphics[width=4.1cm]{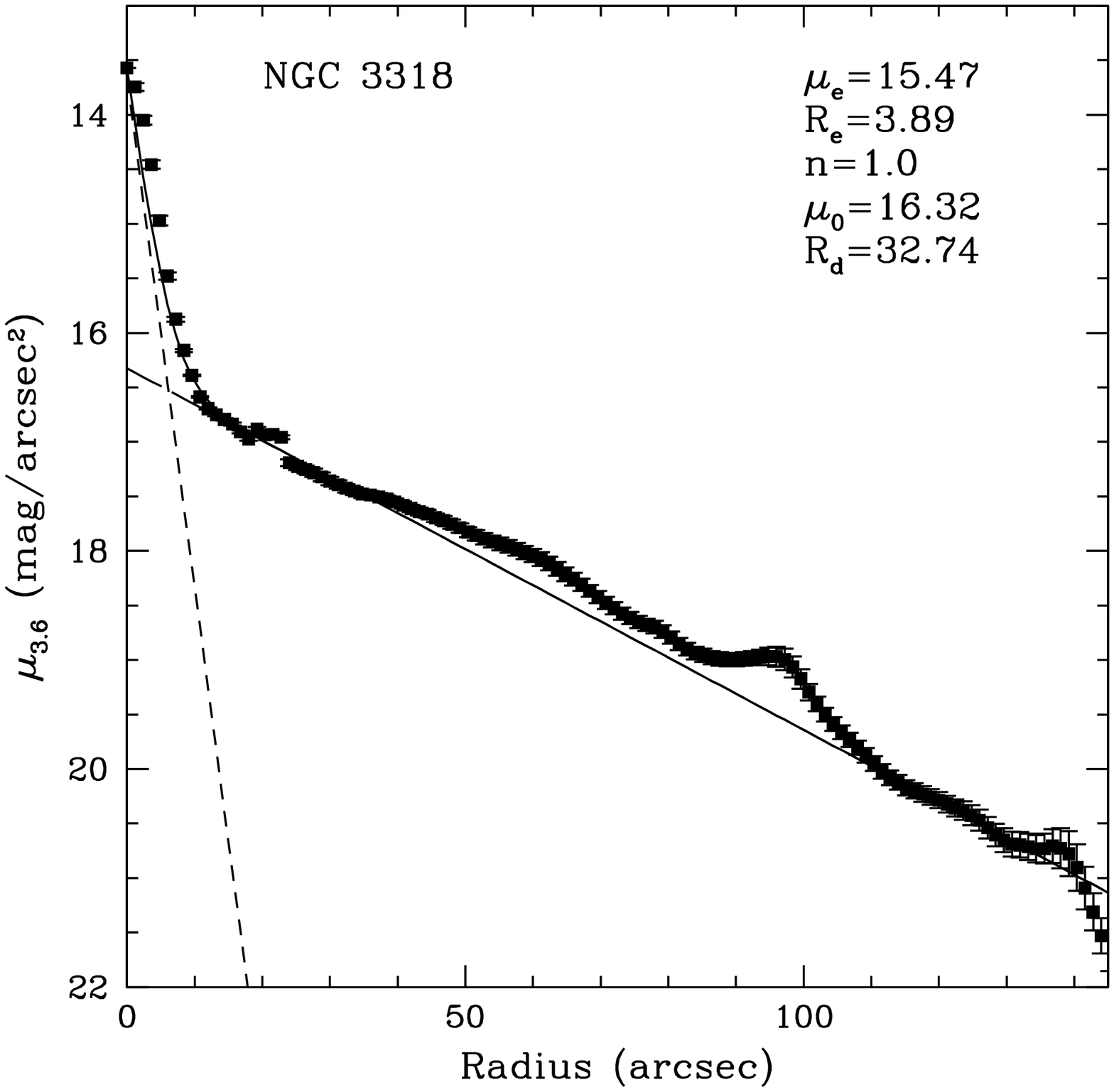}
\caption{Surface brightness profiles for 13 galaxies.  The short-dashed line is the S\'ersic fit to the bulge, the long-dashed line is the exponential disk fit, and the solid line is the total bulge+disk fit.}
\end{figure*}

From the disk luminosity determined from the bulge-disk decompostion, 
we assign masses  to the disk  and bulge components  using a  range of
stellar K$_s$-band mass-to-light ratios from Bell et al.\
(2003) and convert this to a 3.6-$\mu$m mass-to-light ratio using the values
given for a typical disk galaxy in Seigar et al.\ (2008a).
 Specifically, in our rotation curve models we allow
mass-to-light  ratios of $(M/L) =   0.7,    0.8, 0.9, 1.0$  and  $1.1$ 
(measured in 3.6-$\mu$m-band solar units), and use our
photometrically-derived disk  and bulge  light  profiles 
$L_{B} = L_{disk} + L_{bulge}$
 to determine the stellar mass
contribution to each rotation curve: 
$M_{*} = (M/L) L_{B}$.

The results of our one-dimensional bulge-disk decompositions are given in
Figure 2.

\section{Determination of the dark matter halo density distribution}

We now explore a range of allowed dark matter halo masses
and density profiles by adopting two extreme models for disk galaxy formation. 
In the first we assume that the dark matter halos
surrounding these galaxies do not undergo adiabatic contraction (AC) as 
a disk galaxy forms.  As such, the dark matter halo density profile follows
the following equation,
\begin{equation}
\rho(r)=\frac{\rho_s}{(r/r_s)(1+r/r_s)^2}
\end{equation}
where $r_s$ is a characteristic ``inner'' radius and $\rho_s$ is the density
at that radius.  This is the same profile shape of Navarro et al.\ (1997;
hereafter NFW).  It is a two-parameter function, and can be completely
specified by chosing two independent parameters, such as the virial mass
$M_{\rm vir}$ and halo concentration $c_{\rm vir}=R_{\rm vir}/r_s$ (Bullock
et al.\ 2001b).

\begin{figure*}
\label{mod1}
\includegraphics[width=5.4cm]{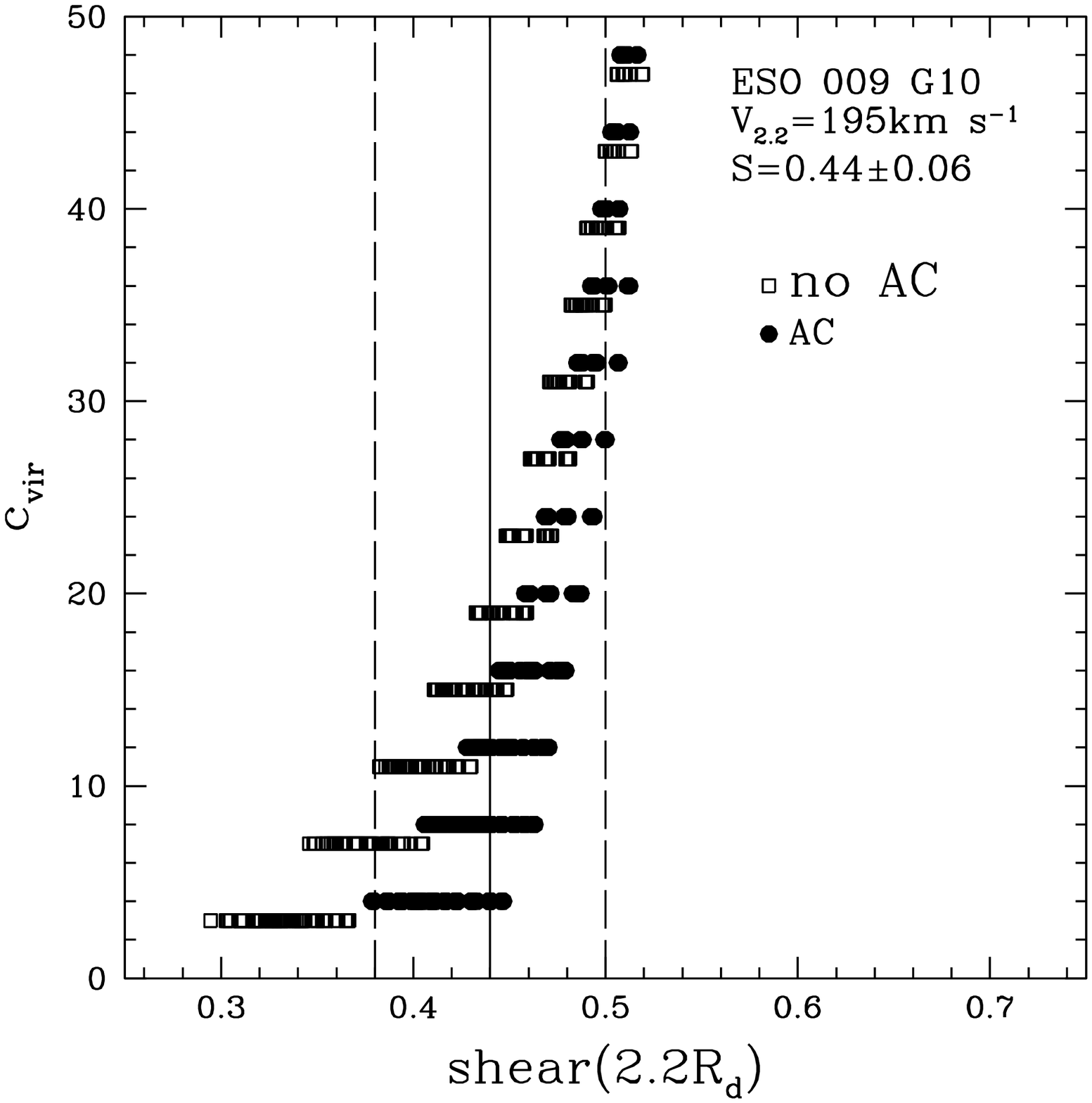}
\includegraphics[width=5.4cm]{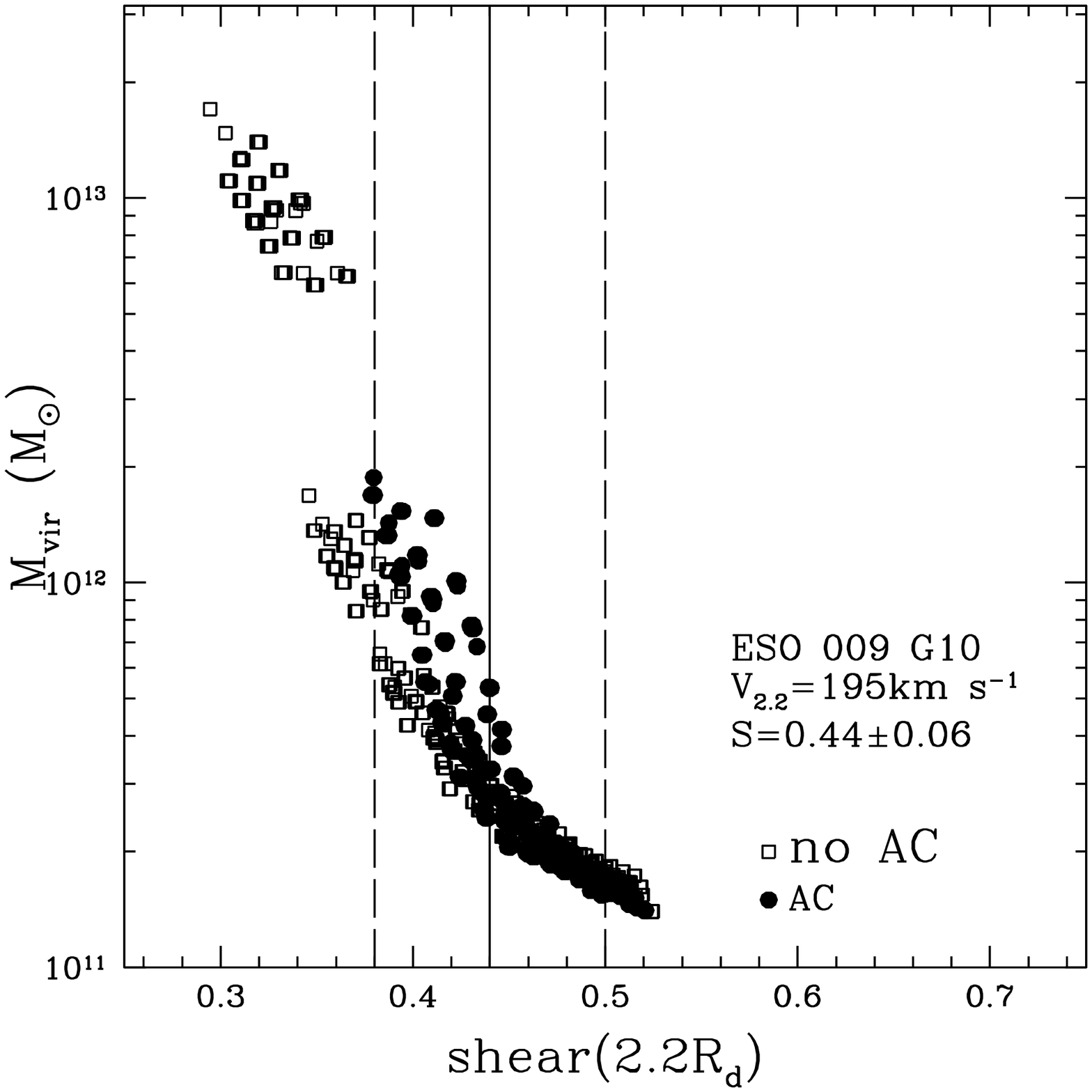}
\includegraphics[width=5.4cm]{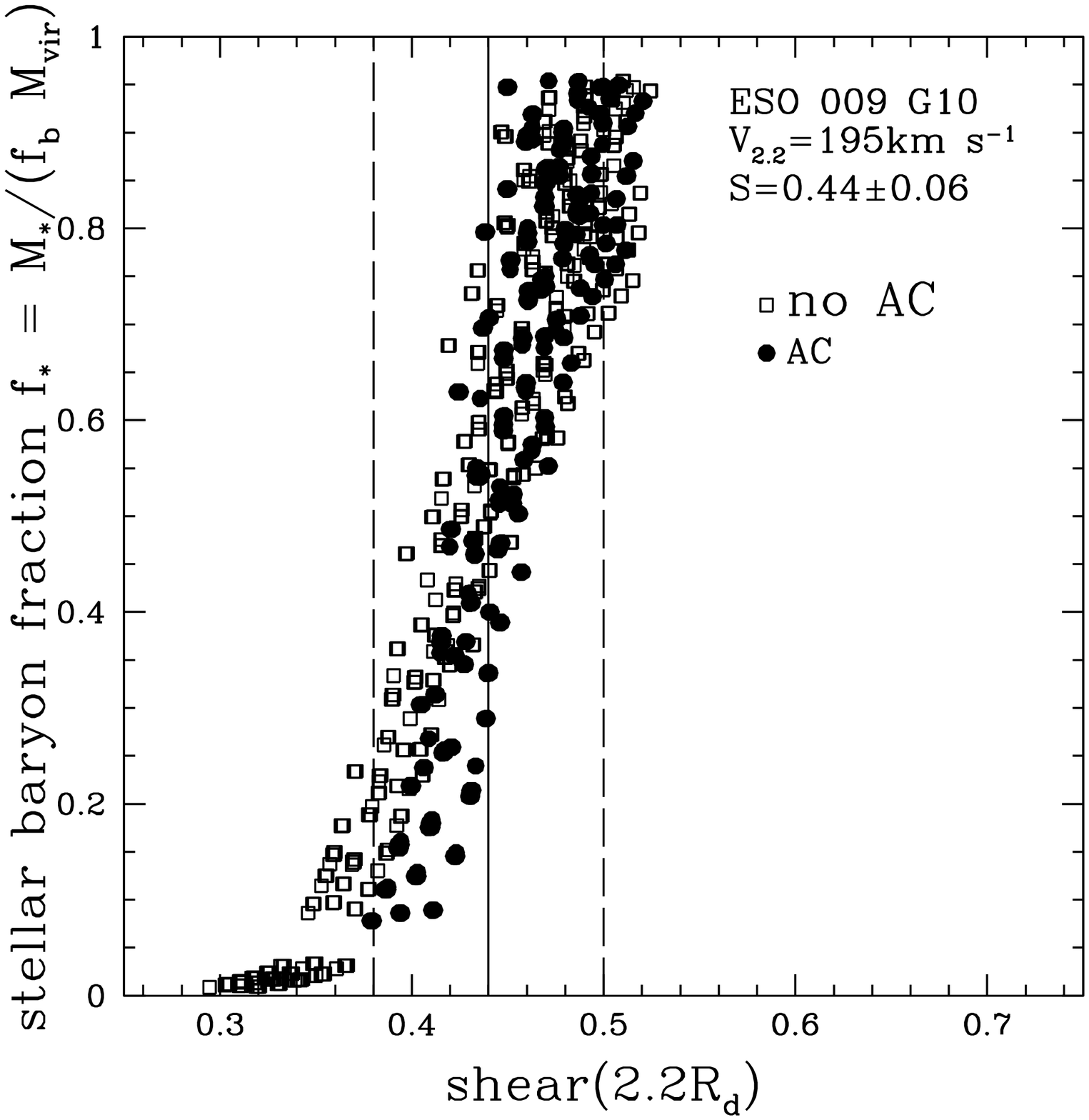}\\
\includegraphics[width=5.4cm]{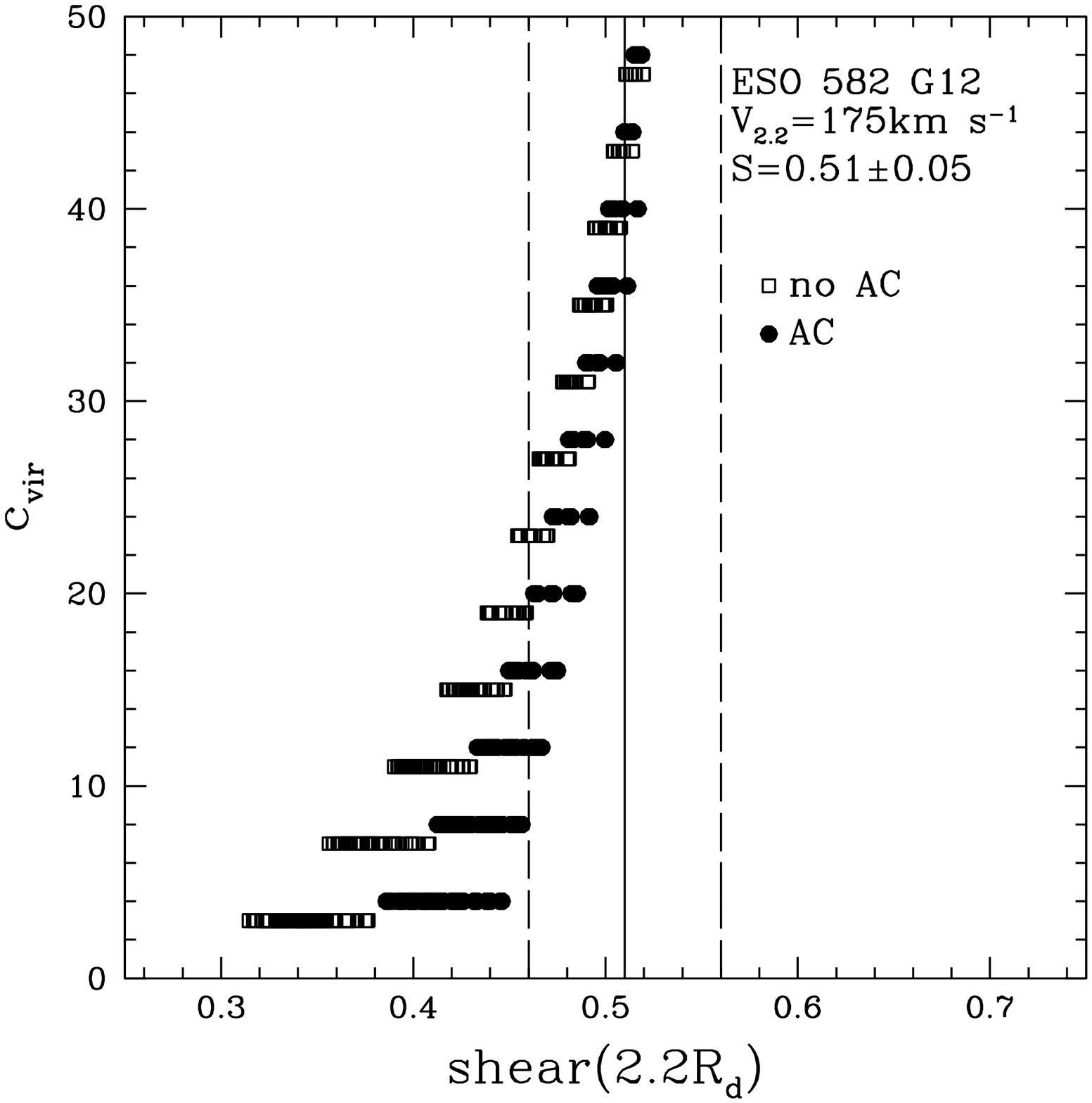}
\includegraphics[width=5.4cm]{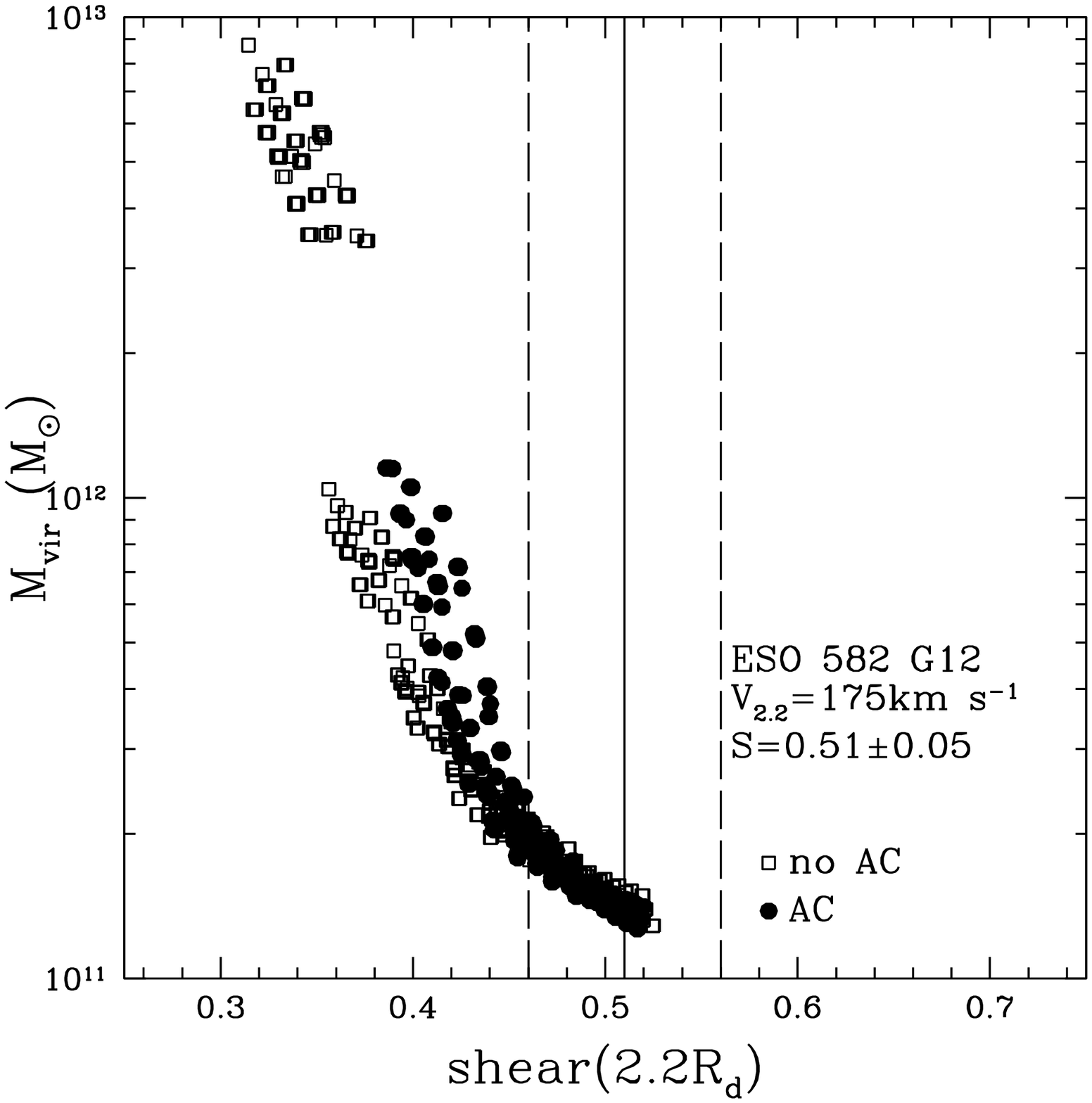}
\includegraphics[width=5.4cm]{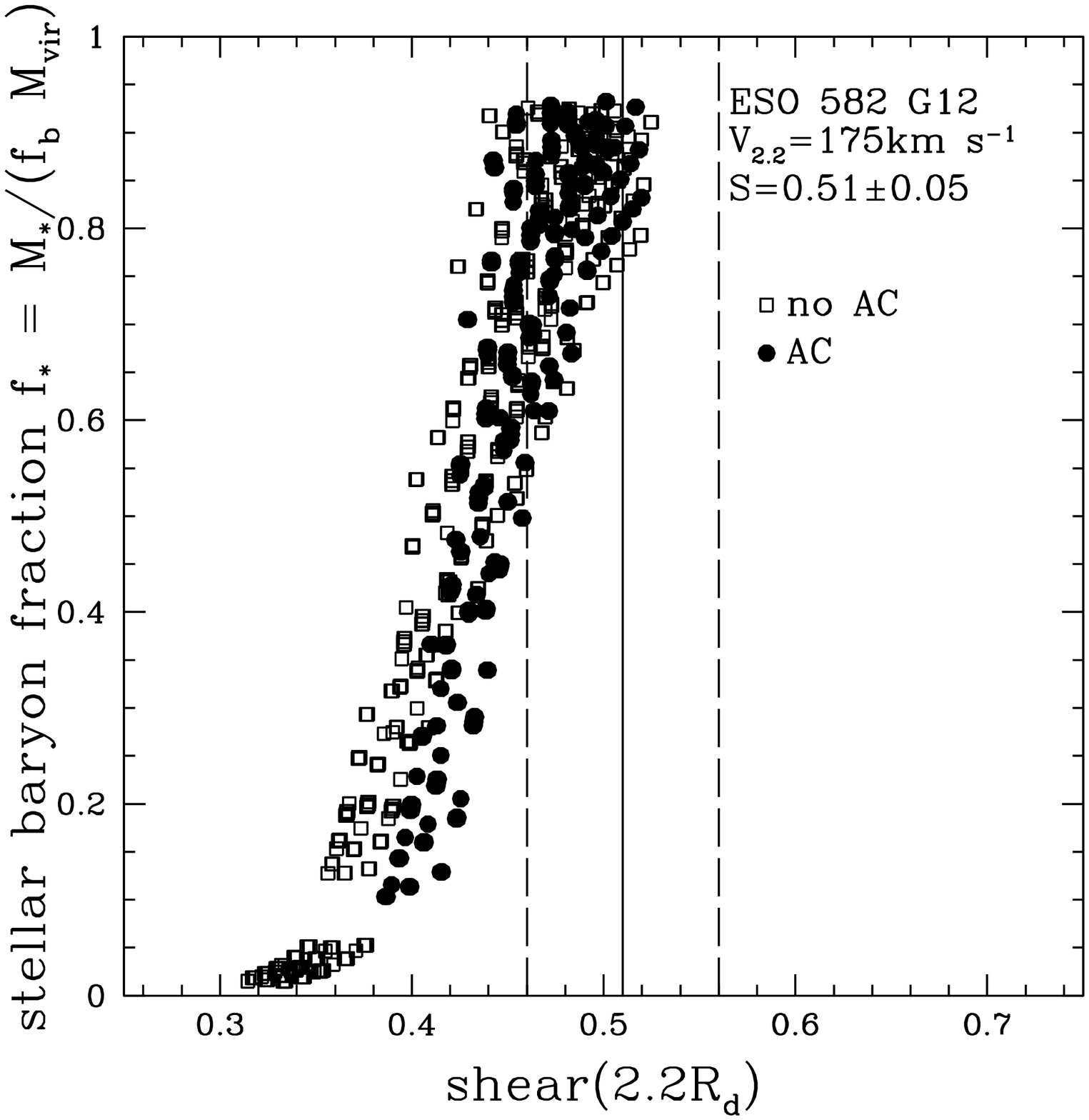}\\
\includegraphics[width=5.4cm]{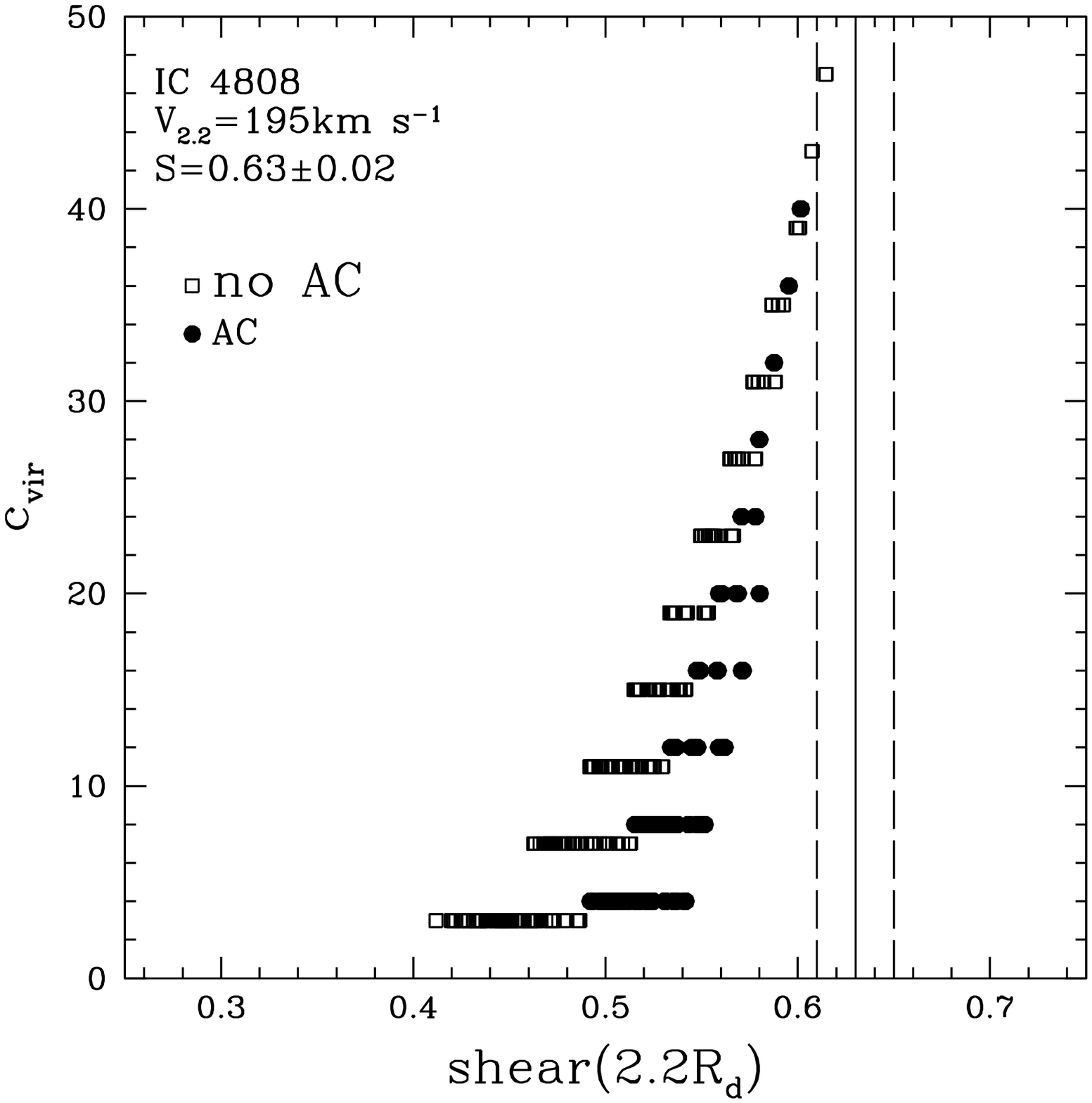}
\includegraphics[width=5.4cm]{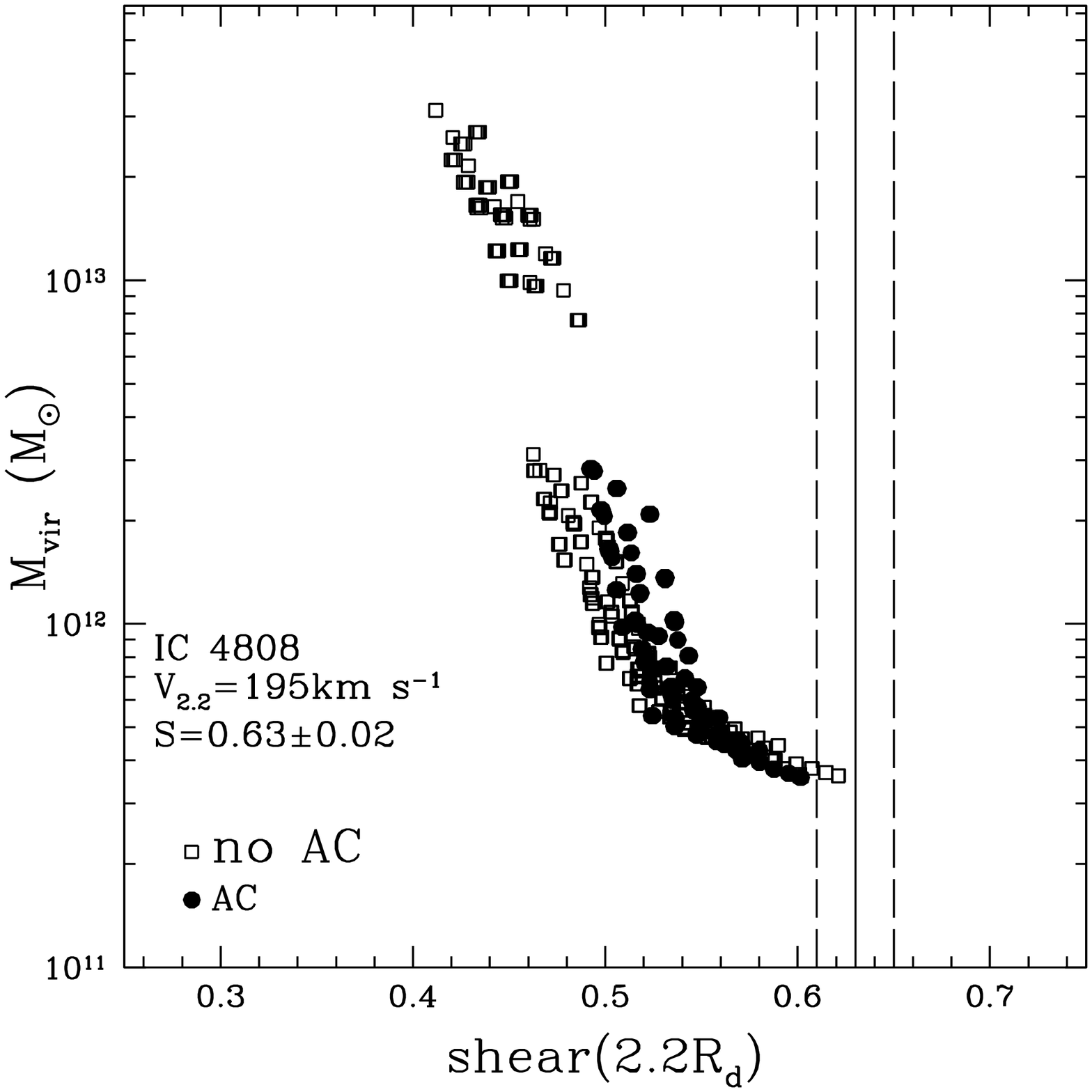}
\includegraphics[width=5.4cm]{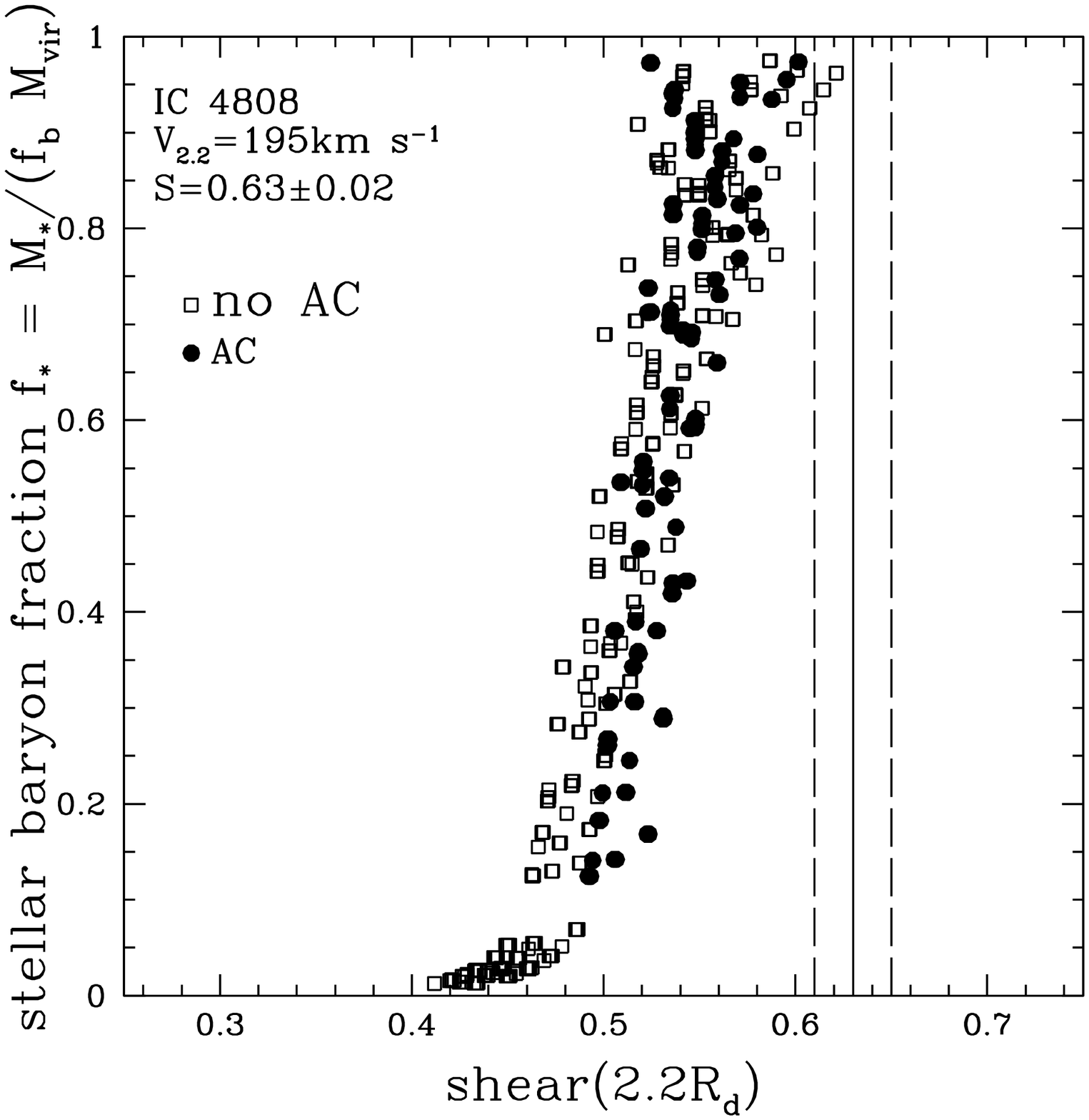}\\
\includegraphics[width=5.4cm]{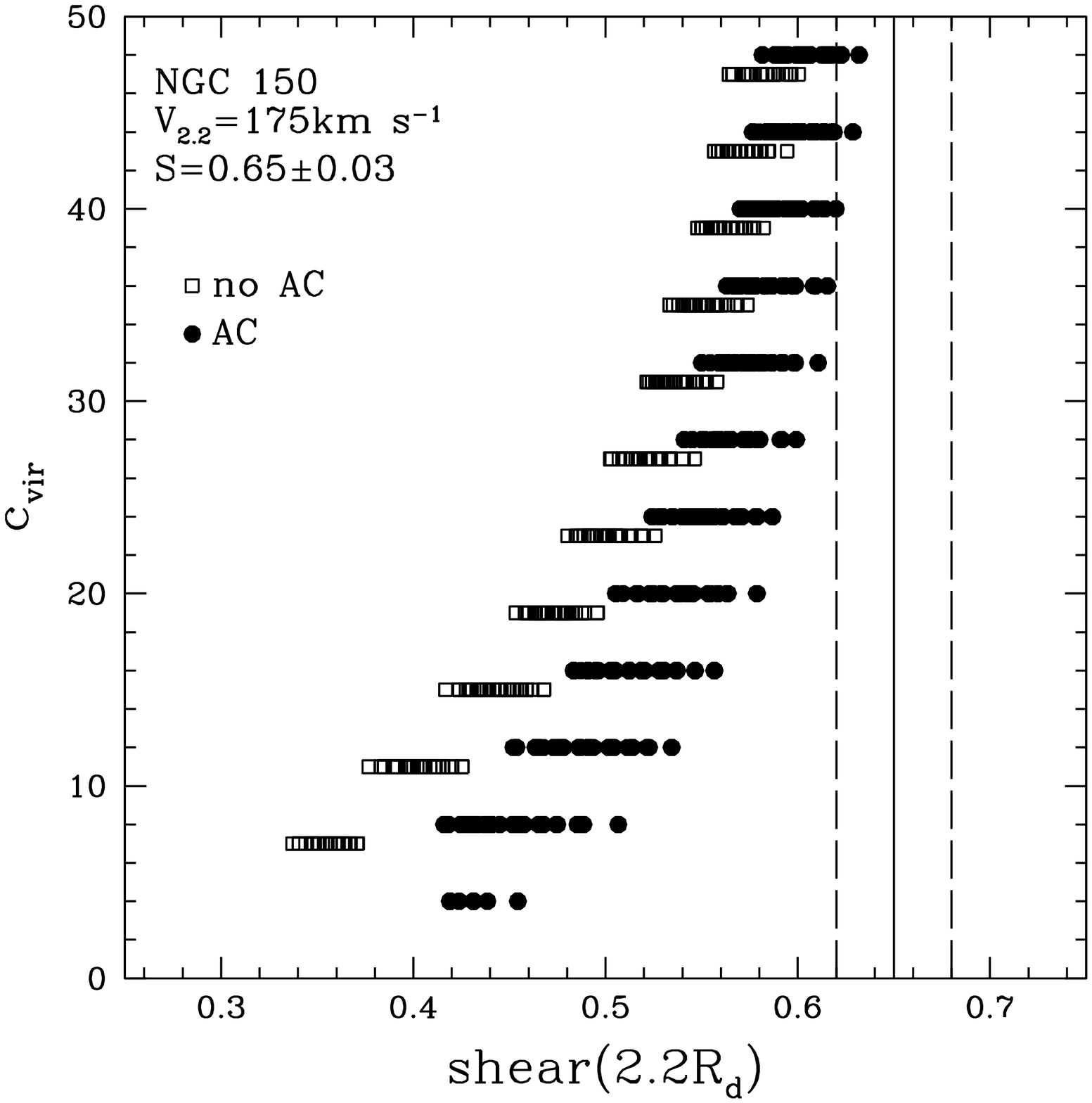}
\includegraphics[width=5.4cm]{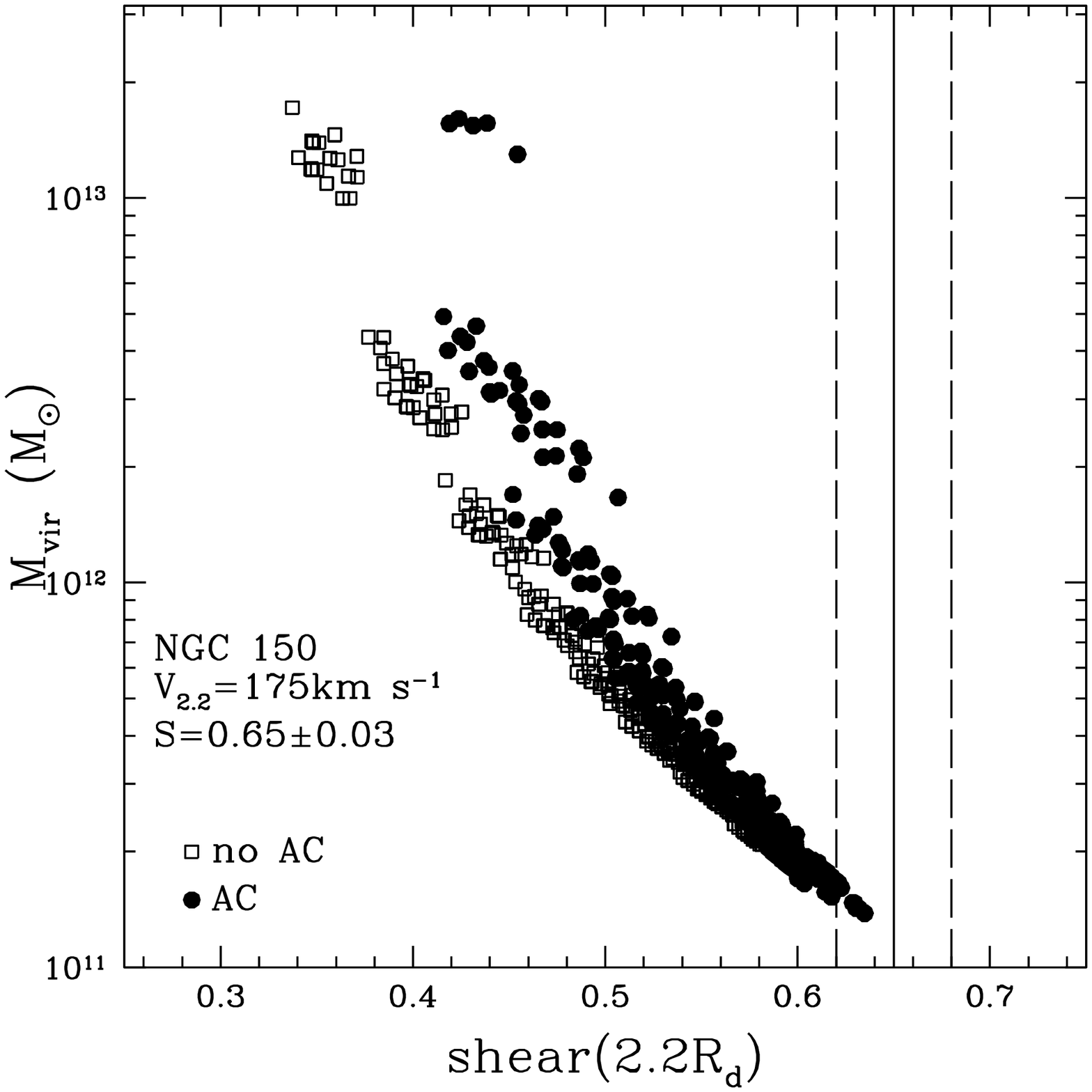}
\includegraphics[width=5.4cm]{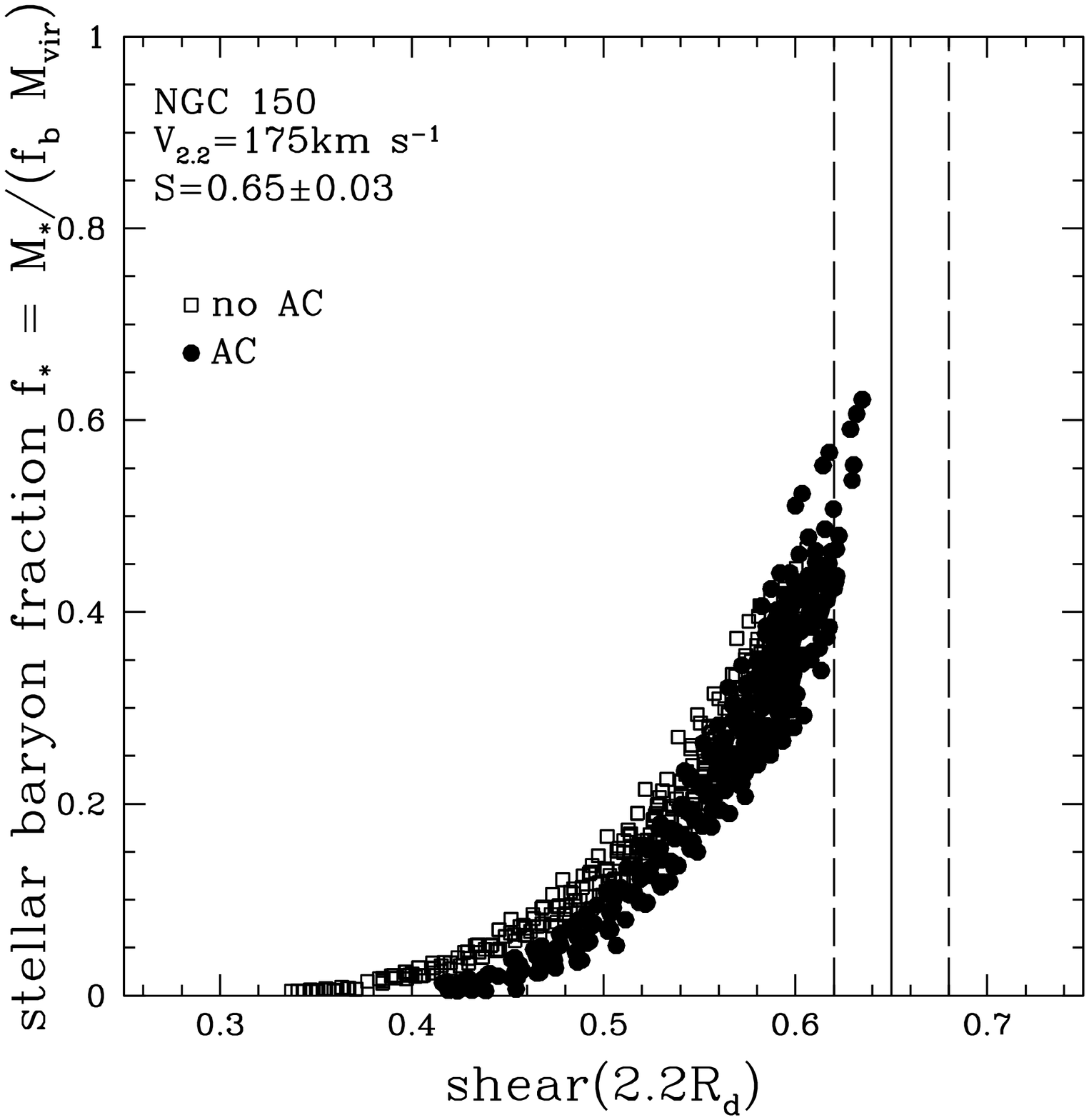}
\end{figure*}

\begin{figure*}
\includegraphics[width=5.4cm]{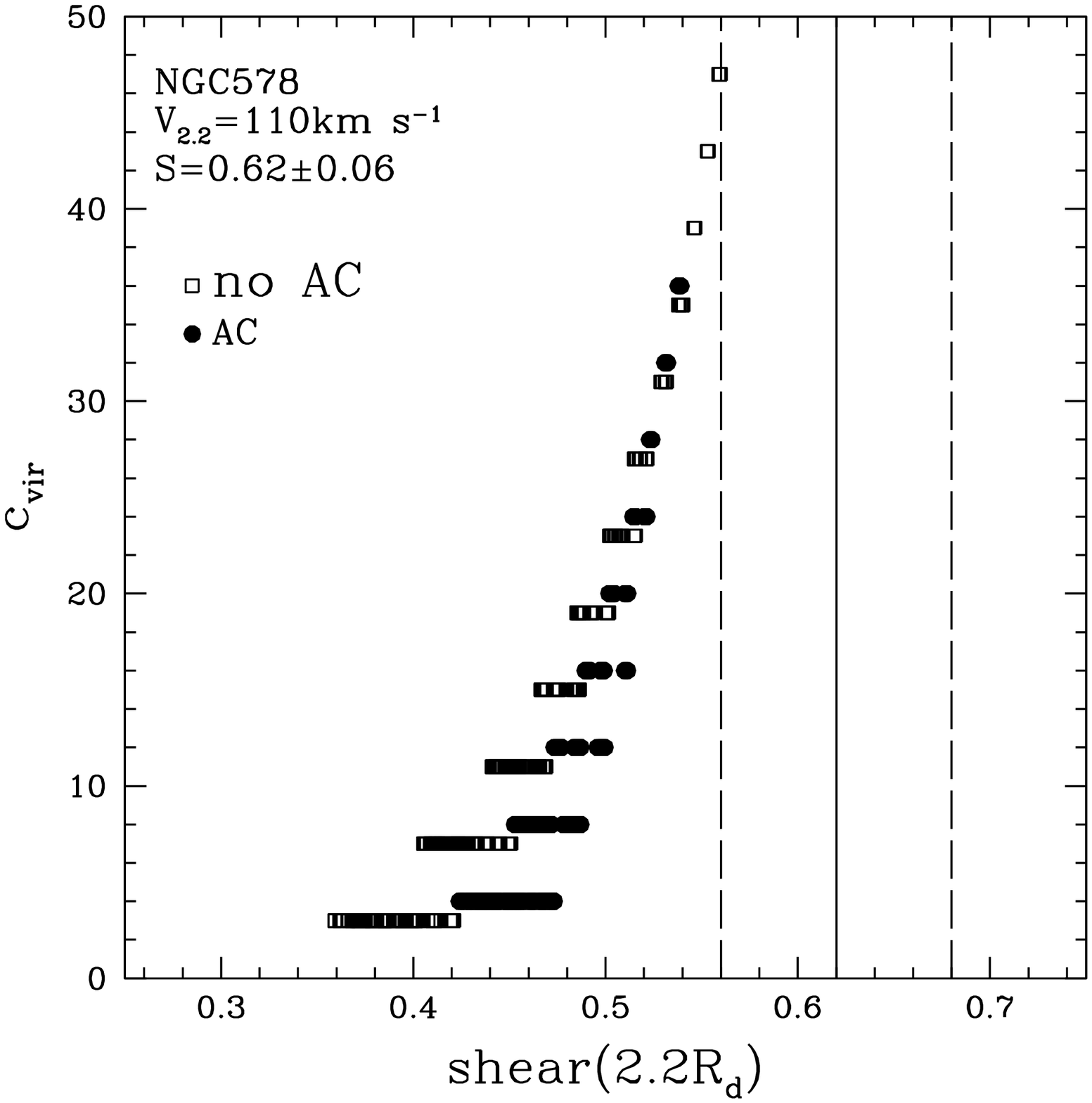}
\includegraphics[width=5.4cm]{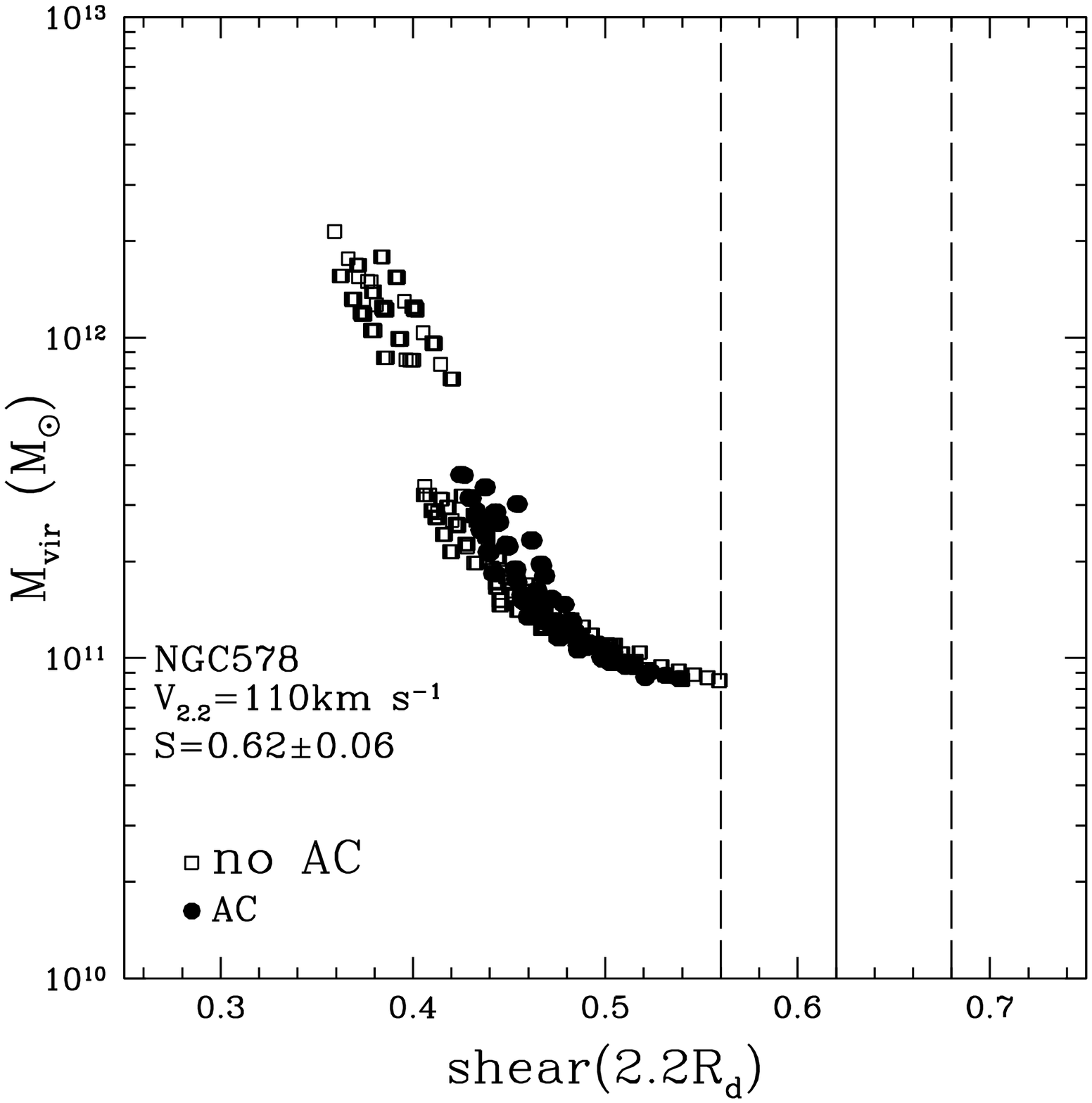}
\includegraphics[width=5.4cm]{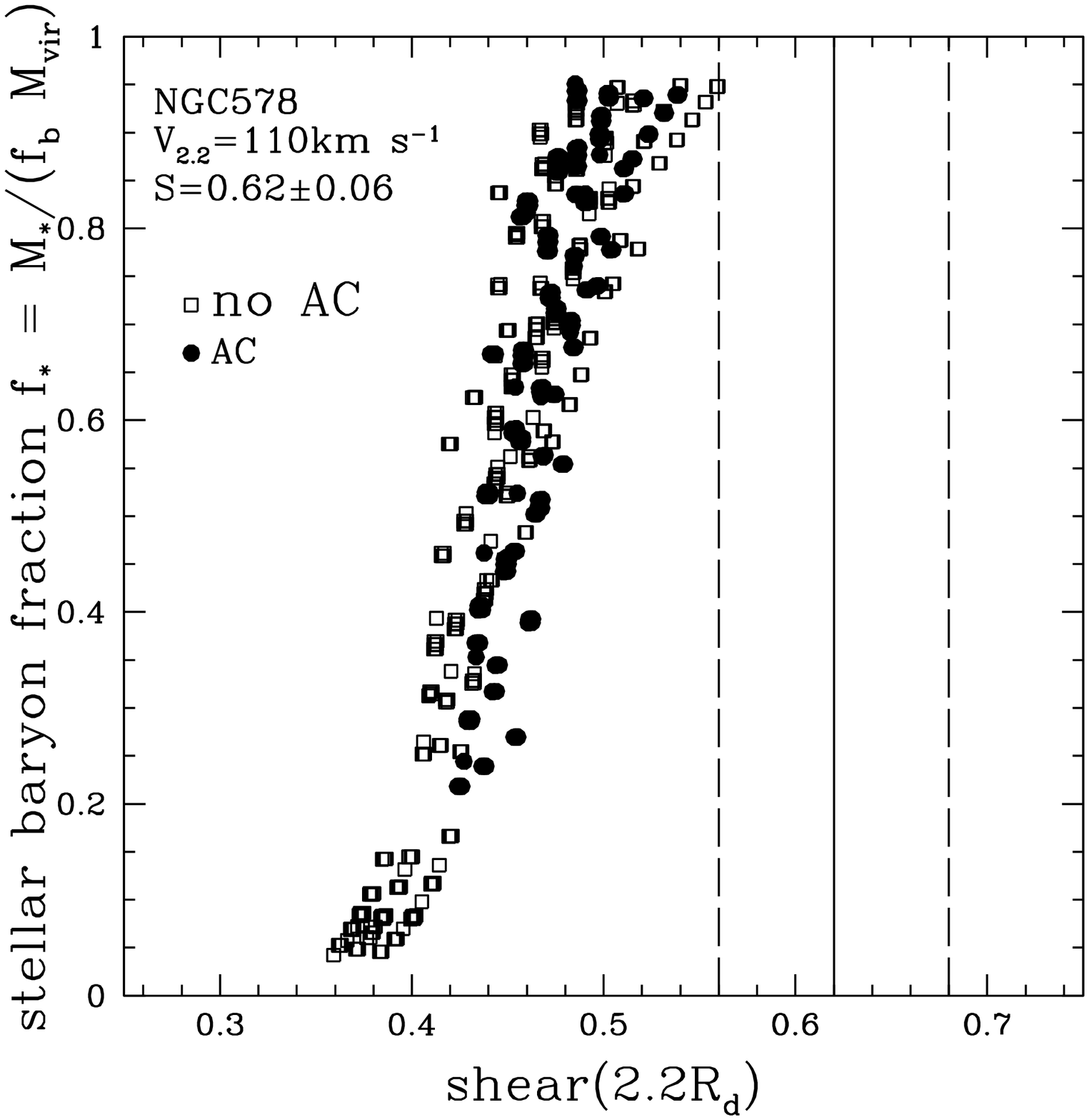}\\
\includegraphics[width=5.4cm]{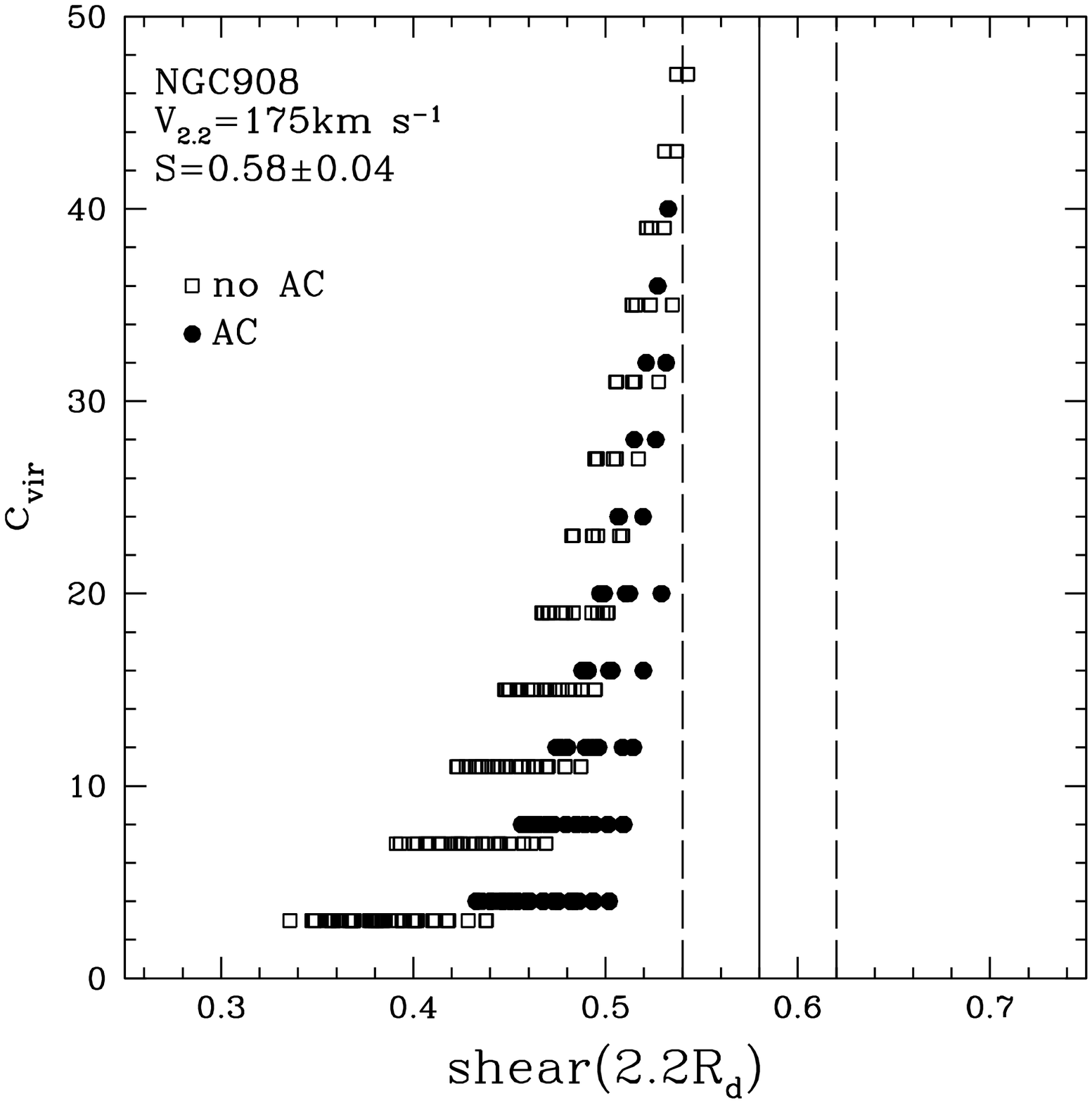}
\includegraphics[width=5.4cm]{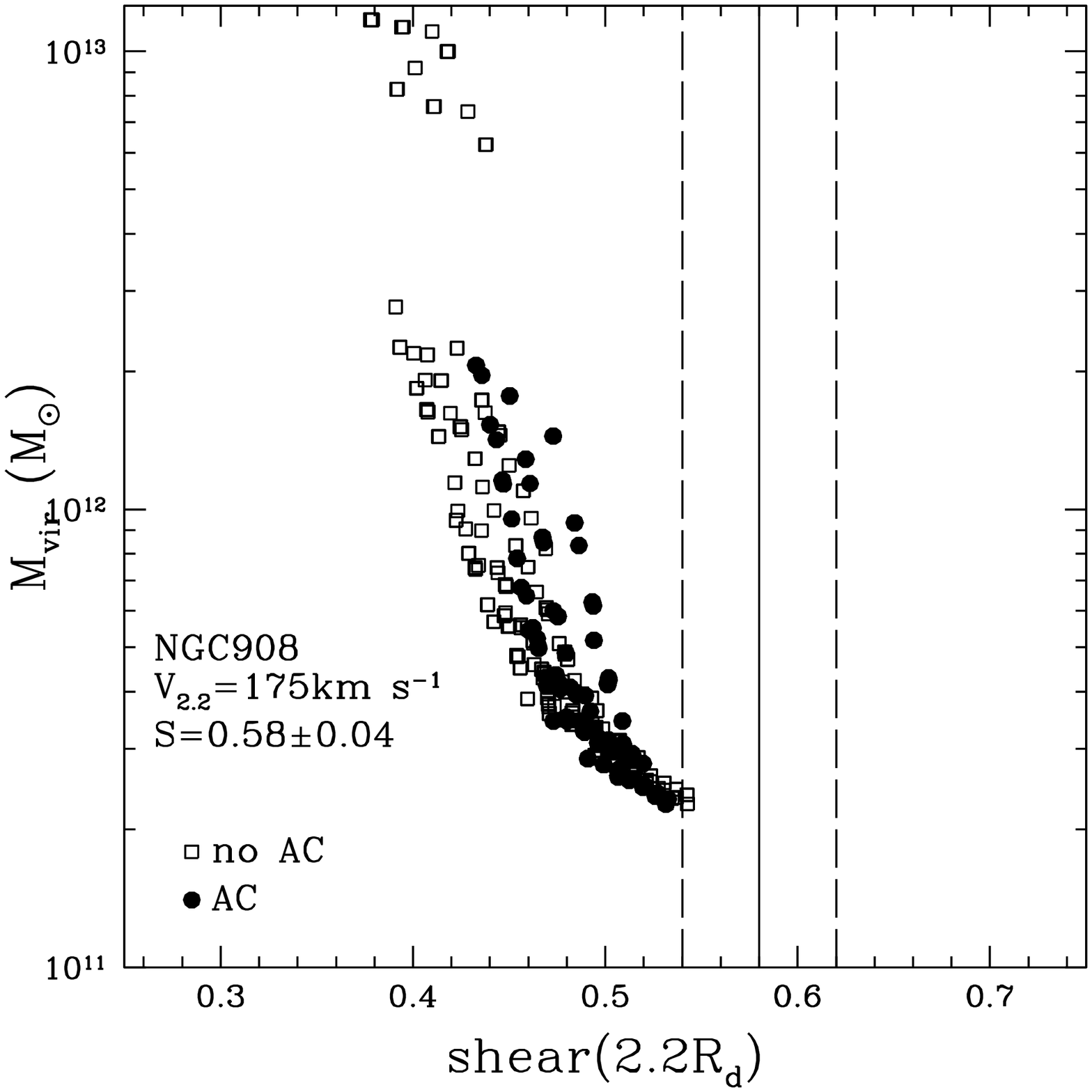}
\includegraphics[width=5.4cm]{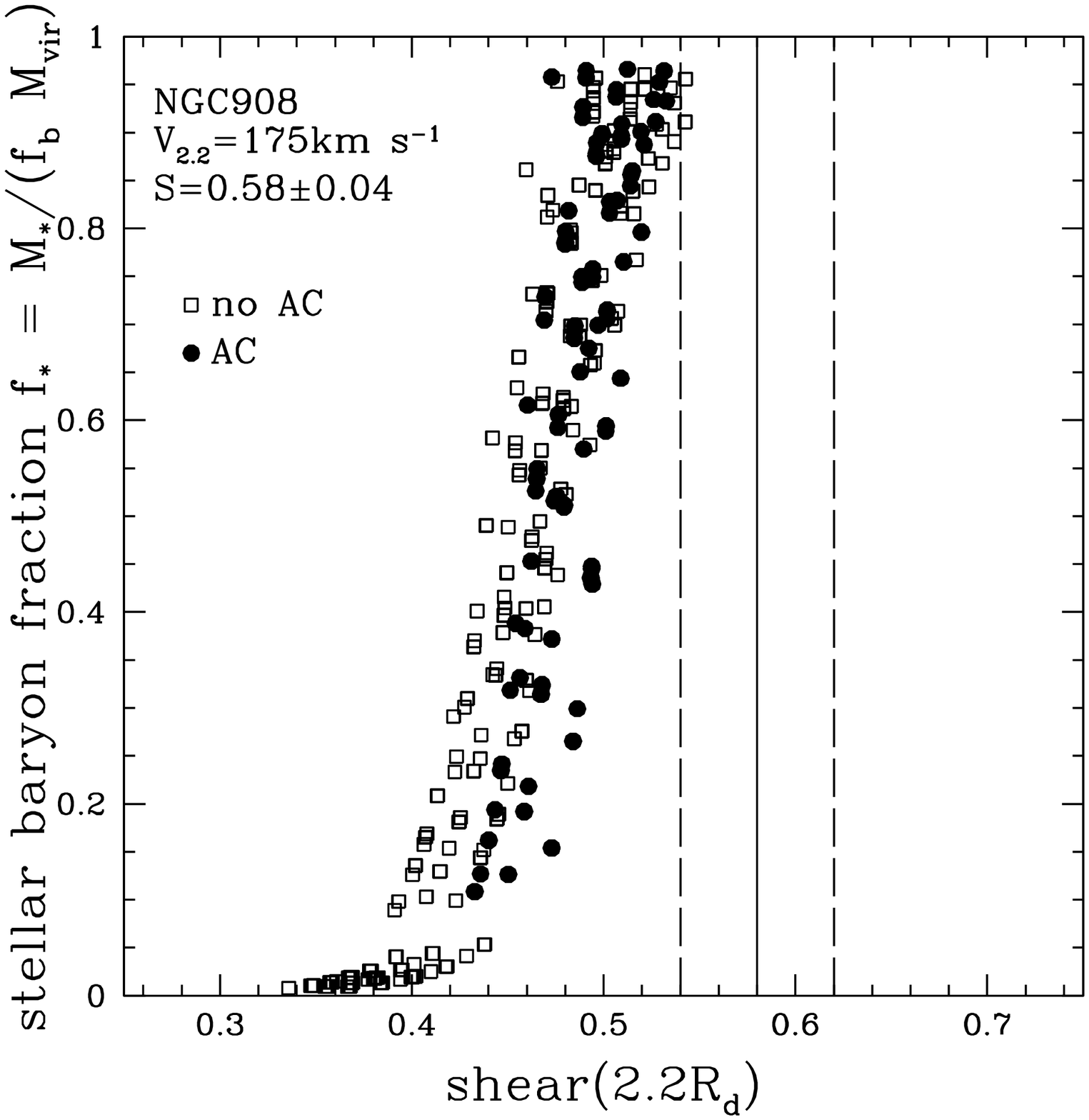}\\
\includegraphics[width=5.4cm]{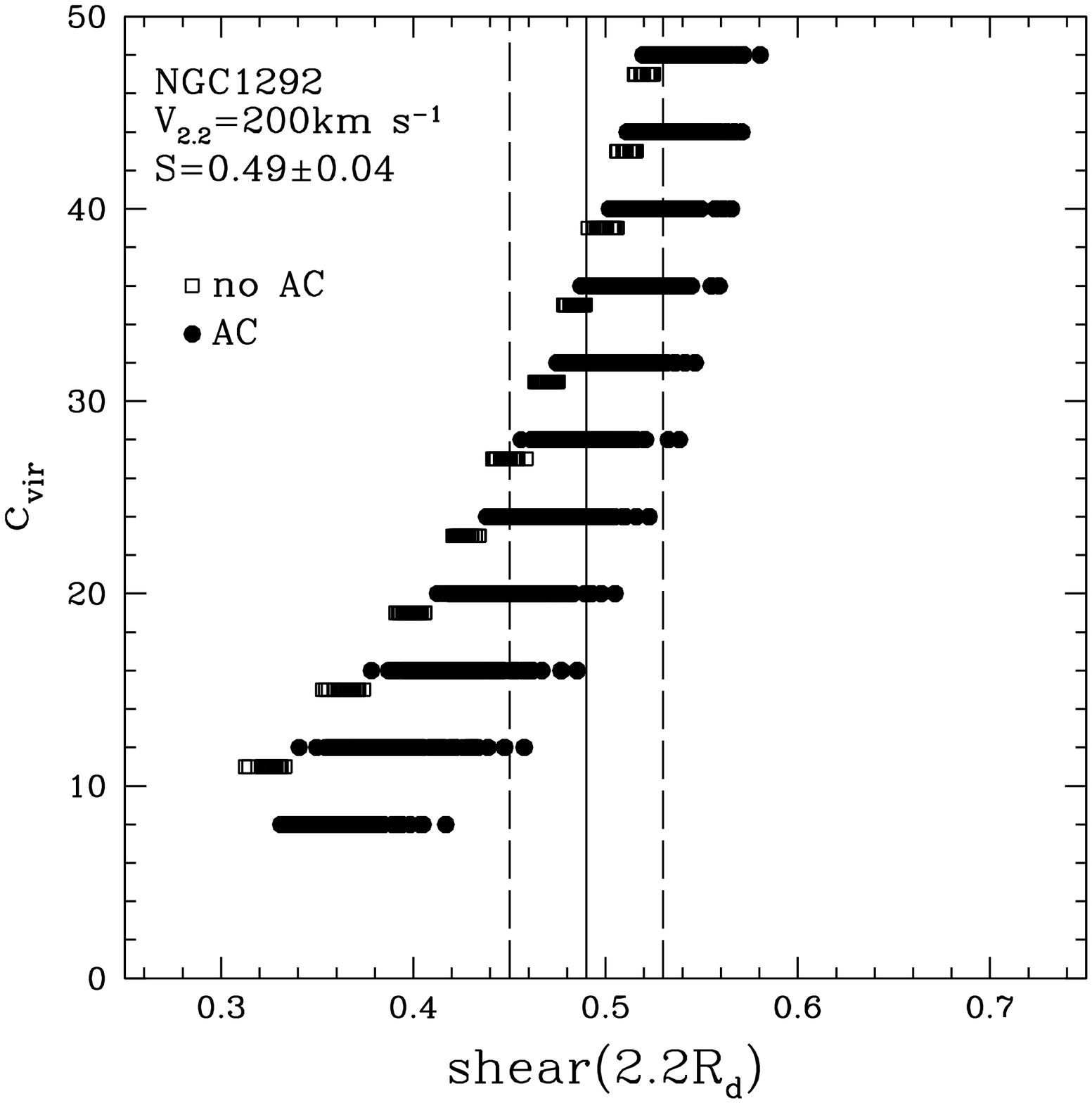}
\includegraphics[width=5.4cm]{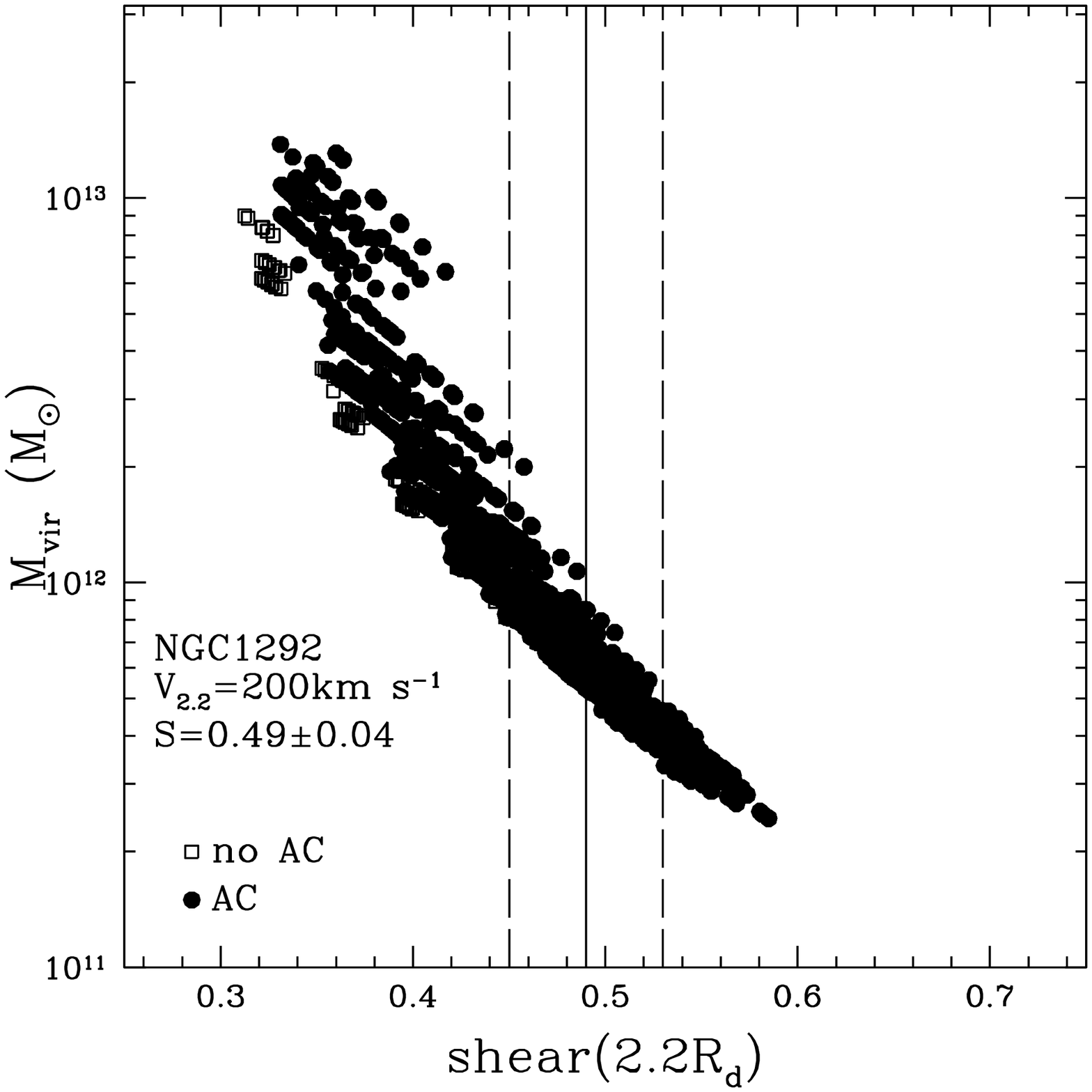}
\includegraphics[width=5.4cm]{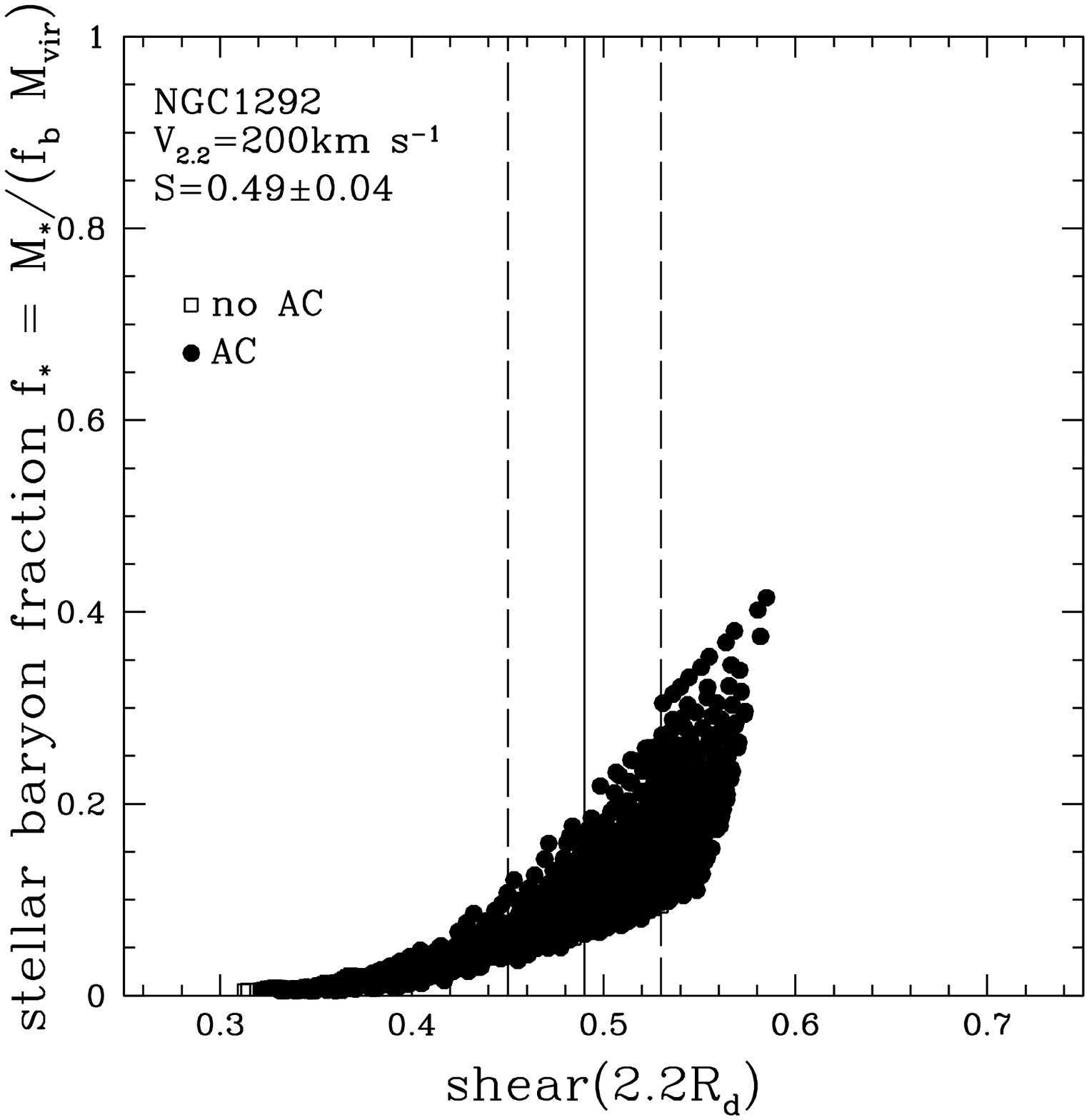}\\
\includegraphics[width=5.4cm]{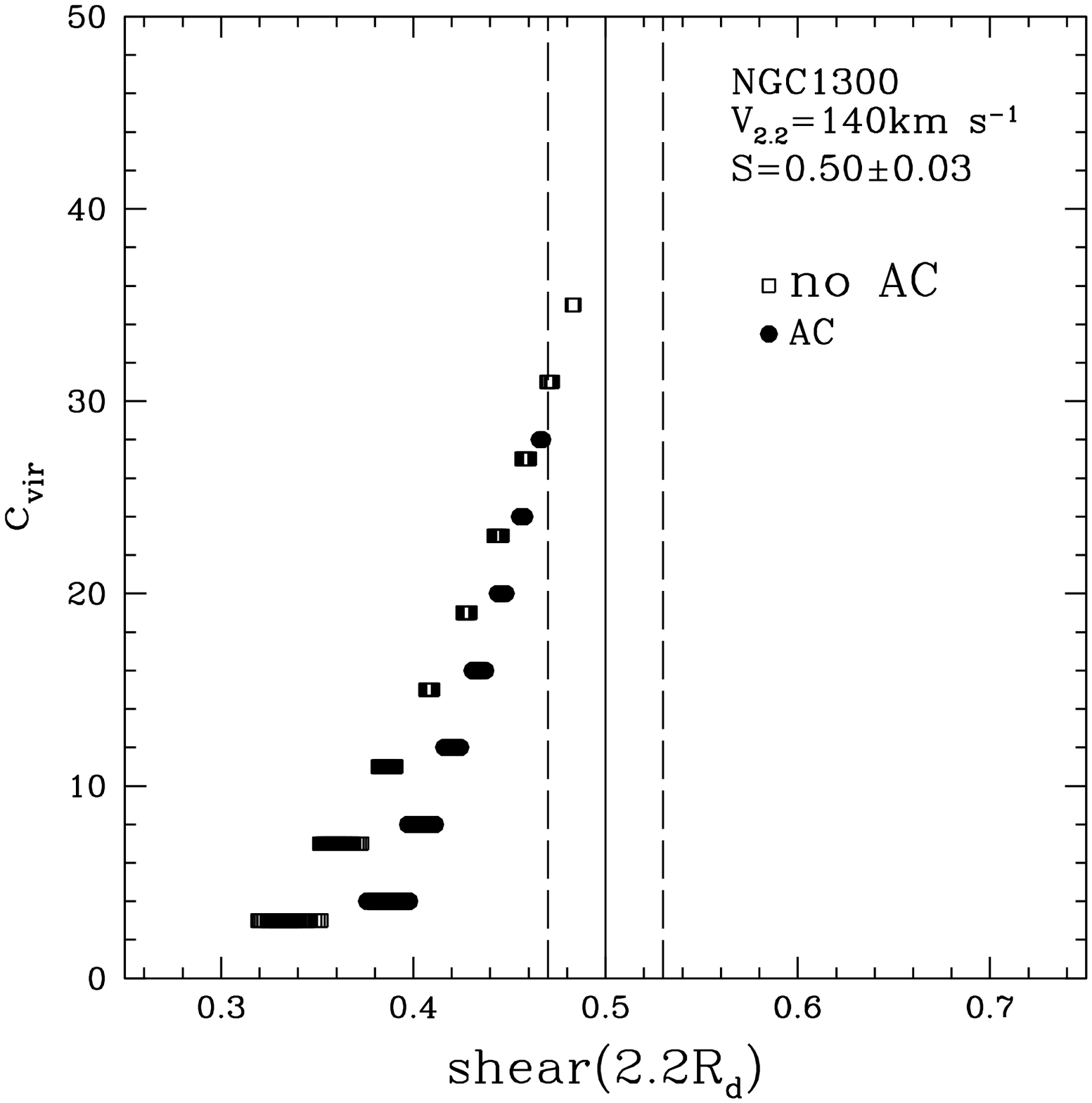}
\includegraphics[width=5.4cm]{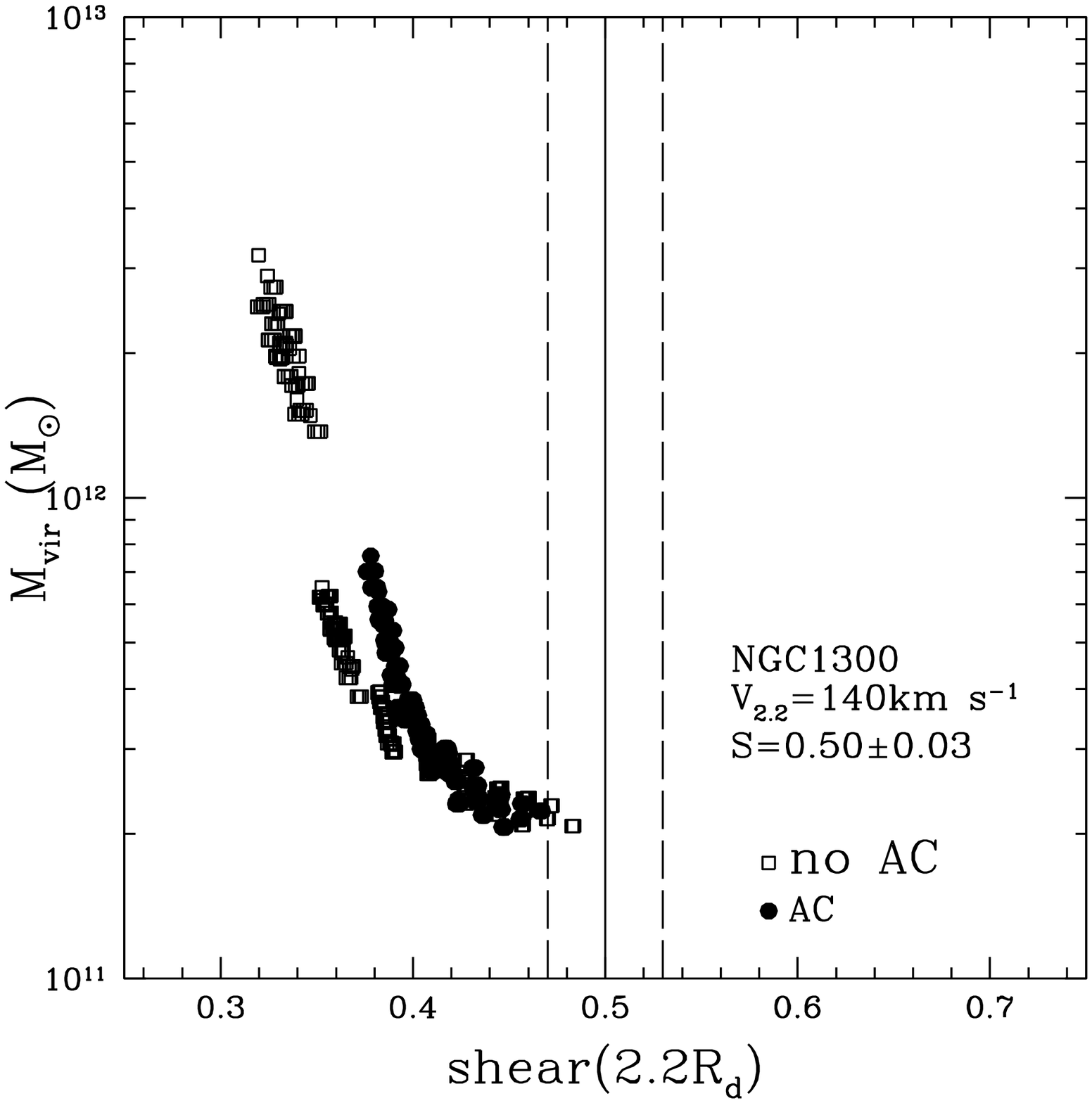}
\includegraphics[width=5.4cm]{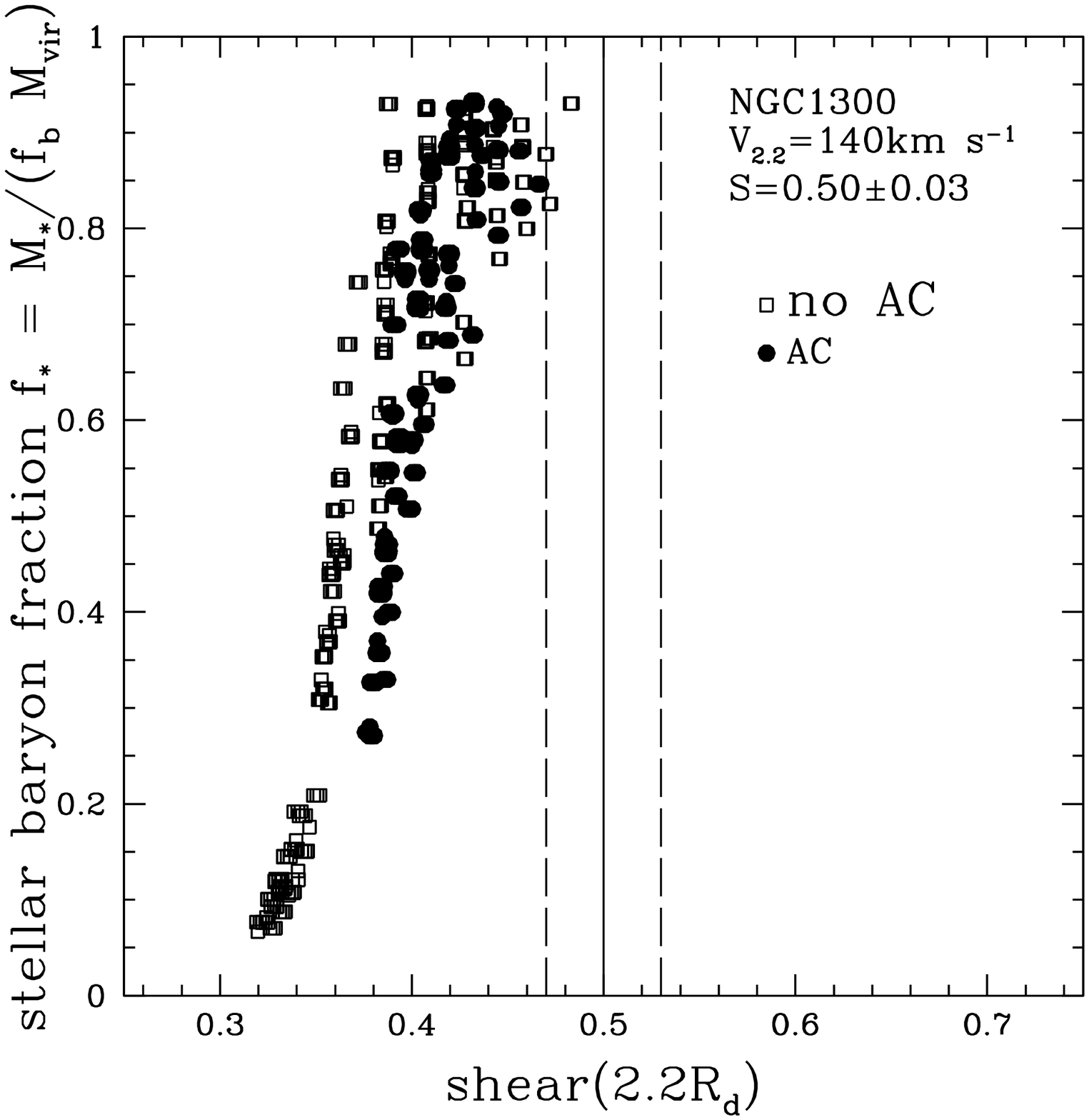}
\end{figure*}

\begin{figure*}
\includegraphics[width=5.4cm]{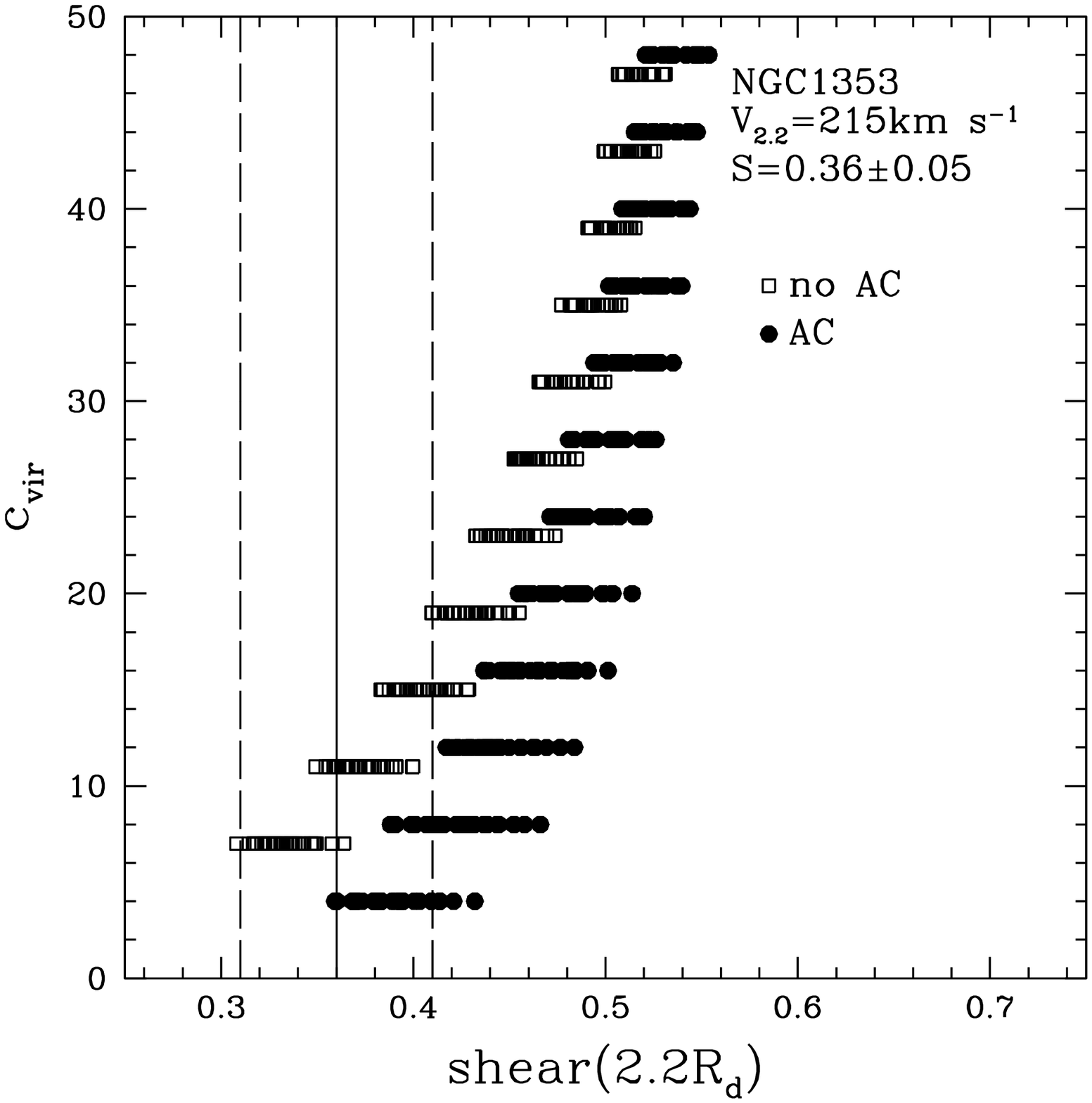}
\includegraphics[width=5.4cm]{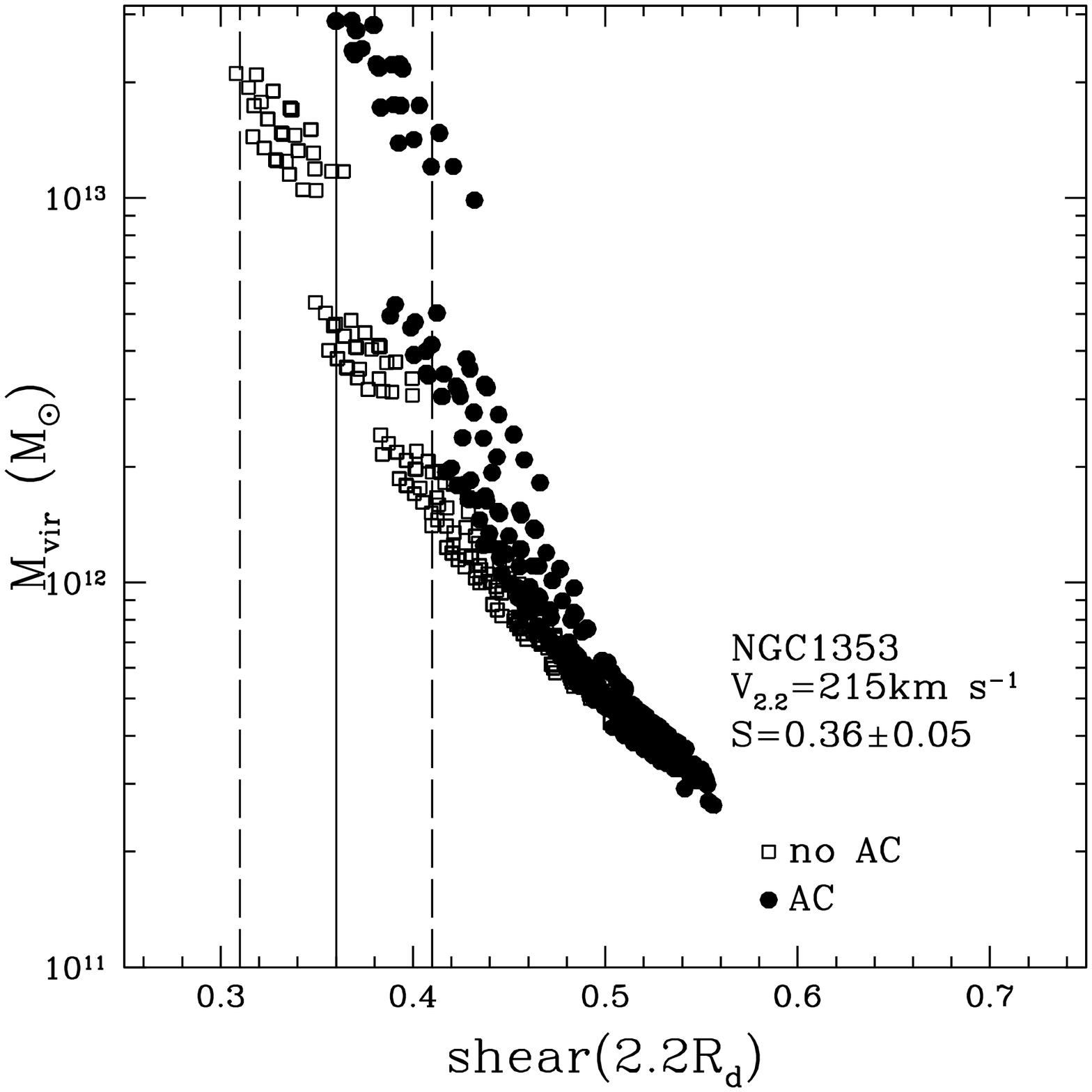}
\includegraphics[width=5.4cm]{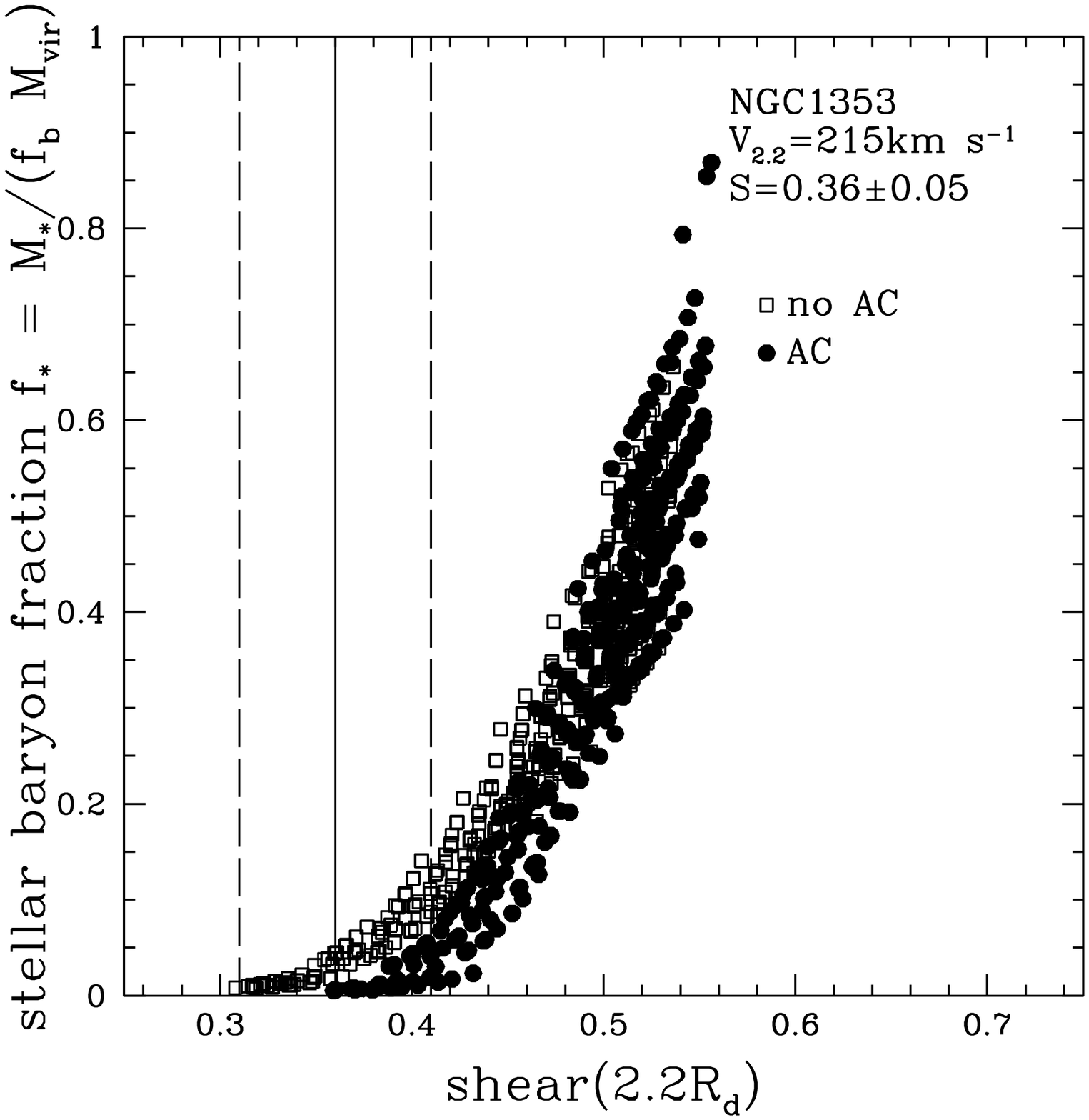}\\
\includegraphics[width=5.4cm]{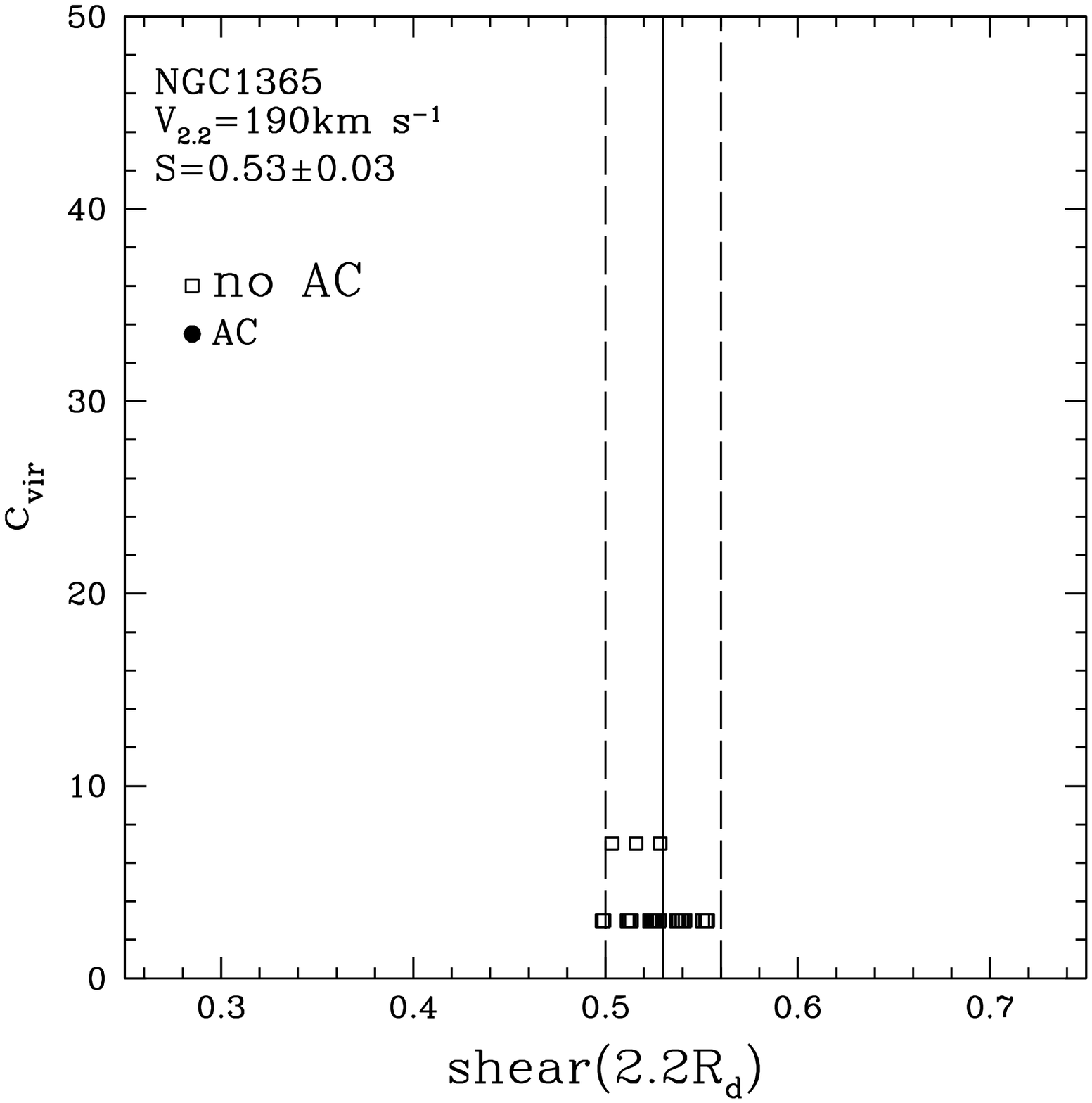}
\includegraphics[width=5.4cm]{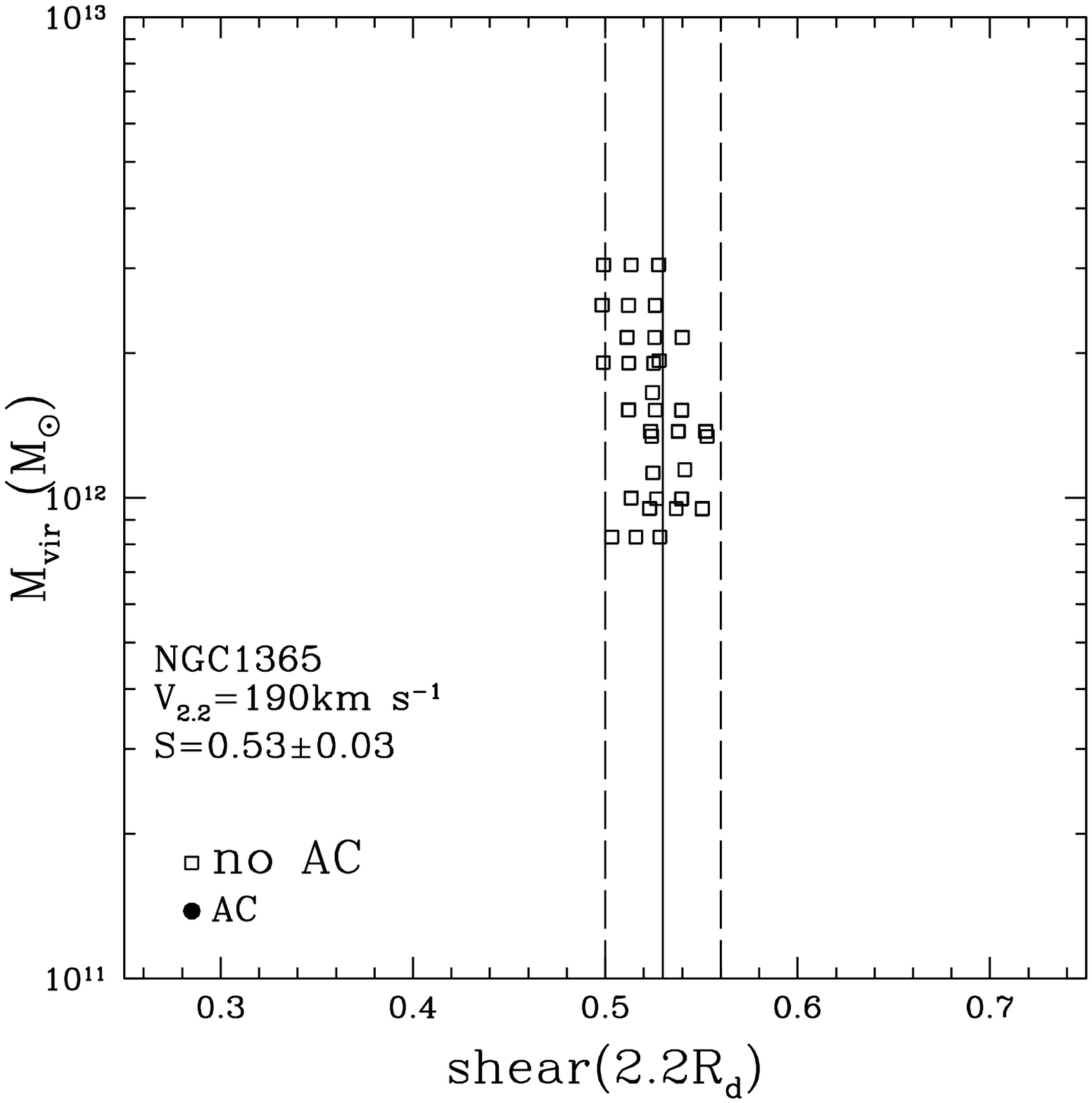}
\includegraphics[width=5.4cm]{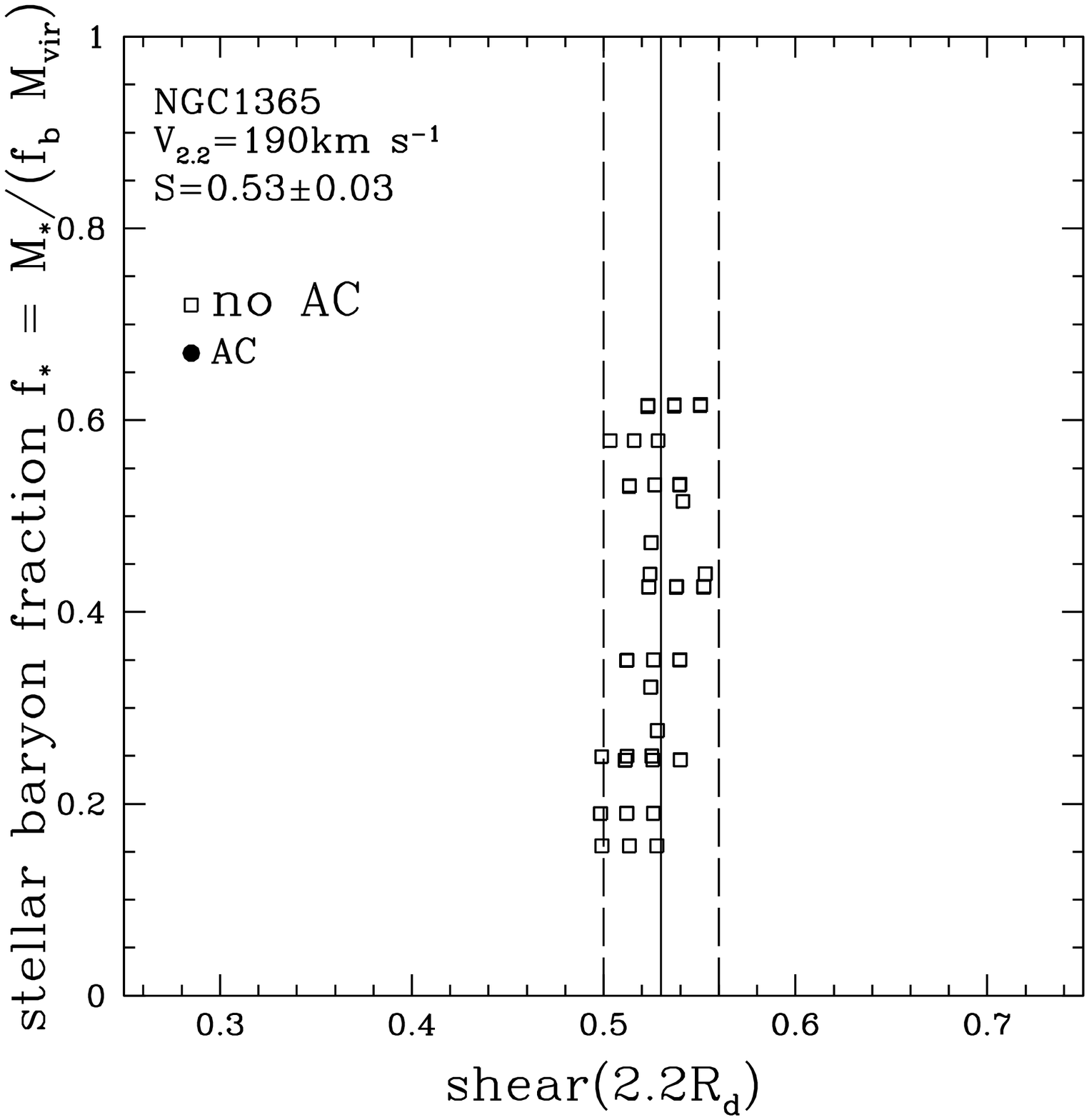}\\
\includegraphics[width=5.4cm]{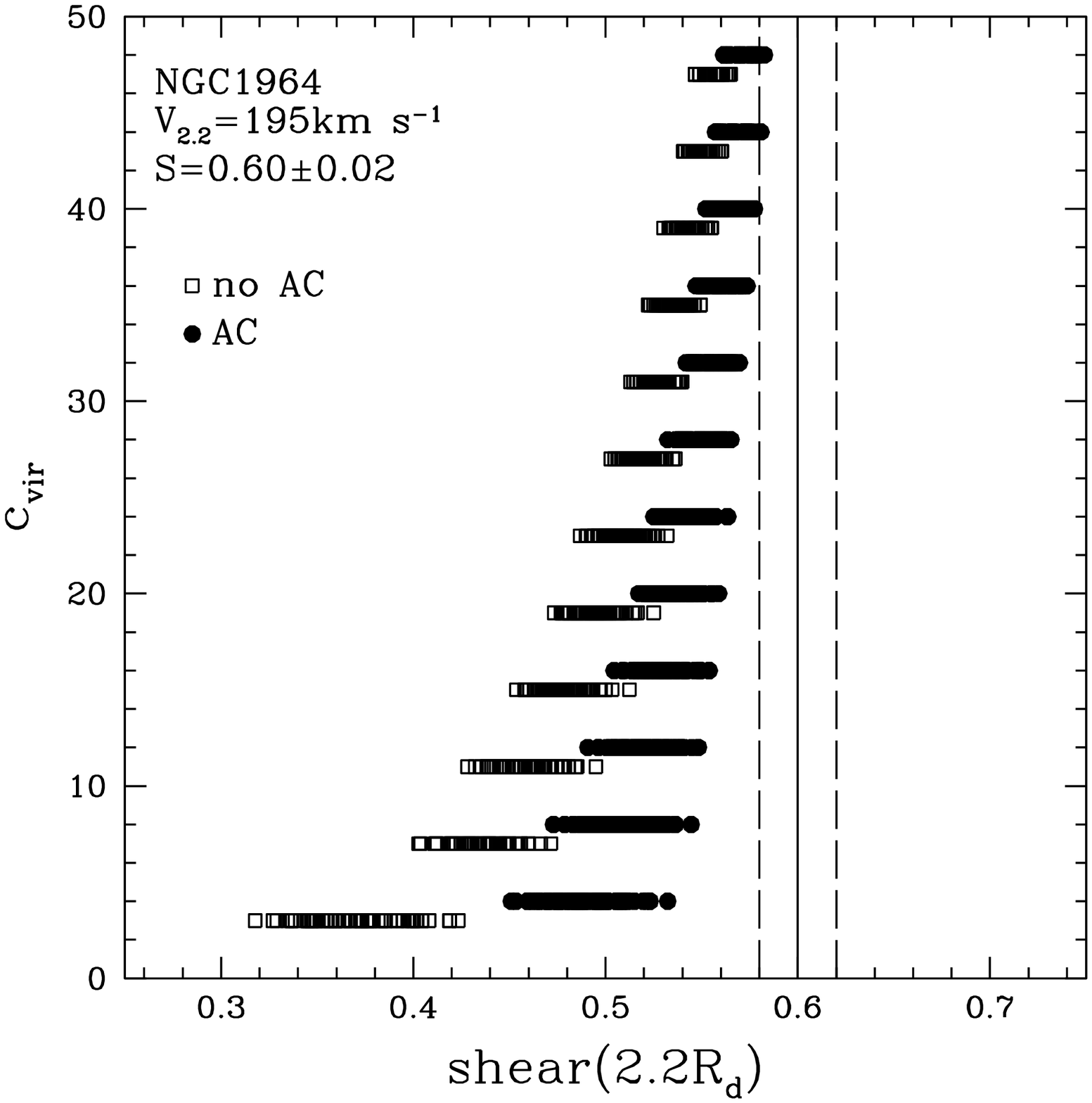}
\includegraphics[width=5.4cm]{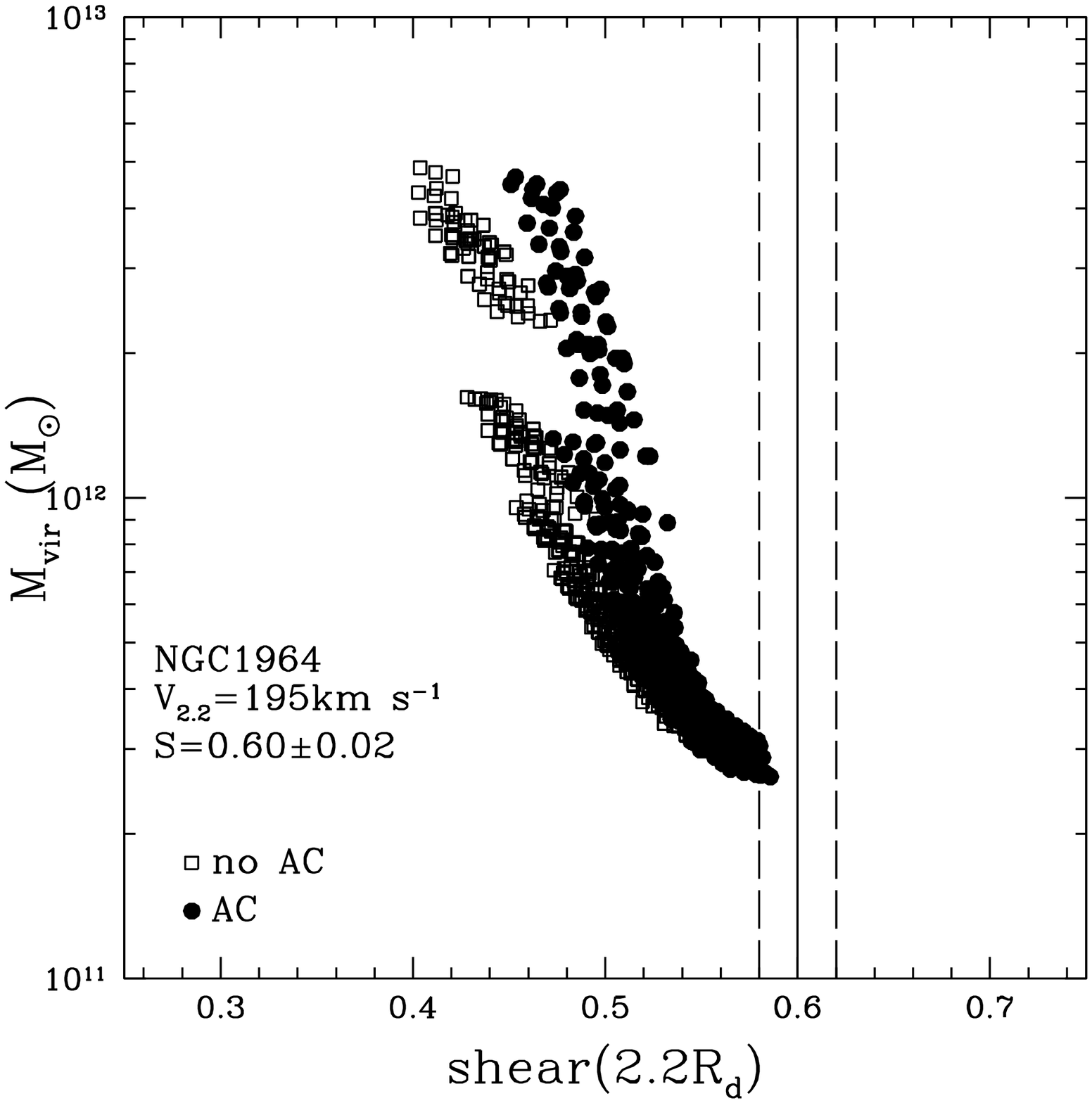}
\includegraphics[width=5.4cm]{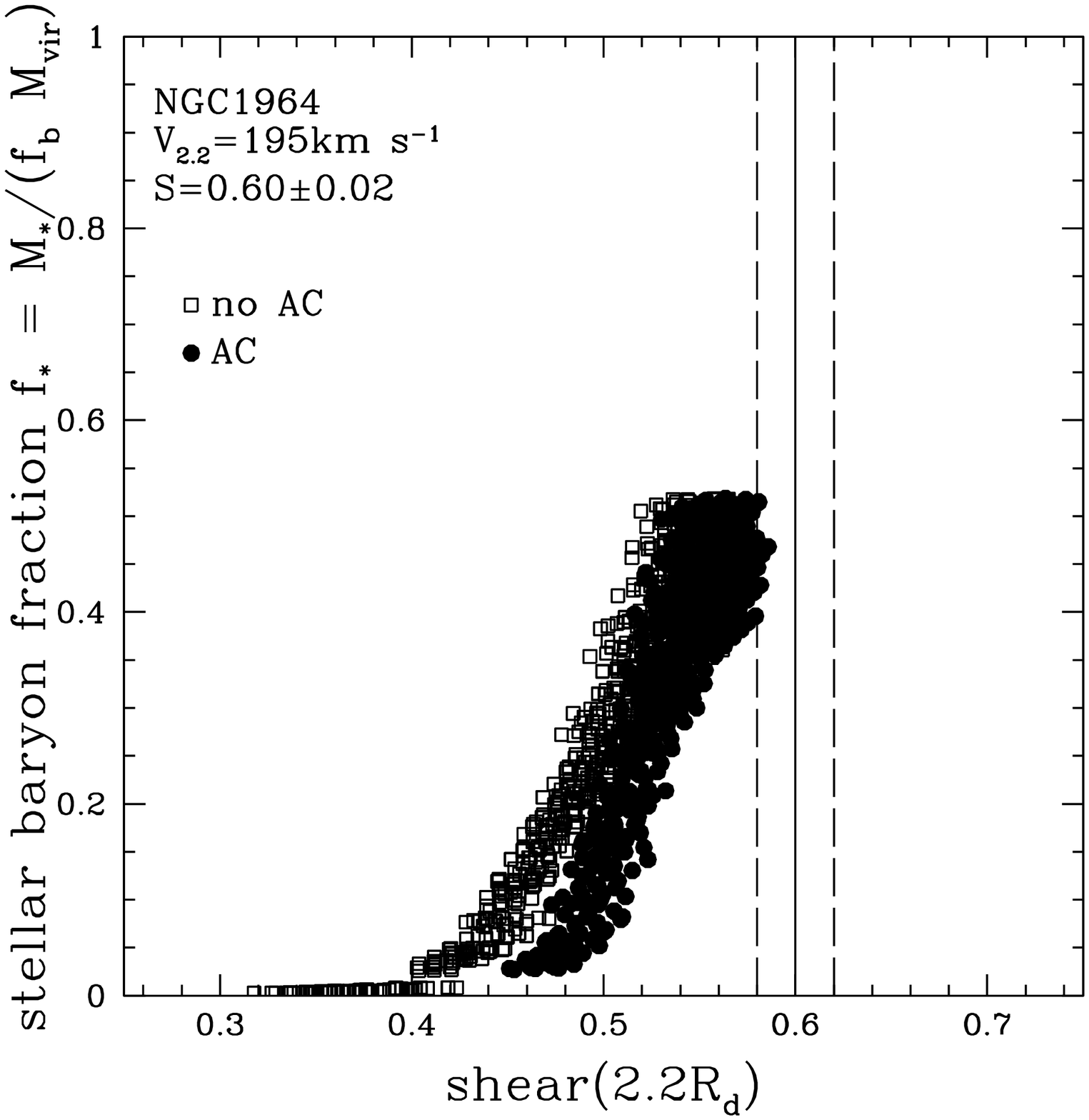}\\
\includegraphics[width=5.4cm]{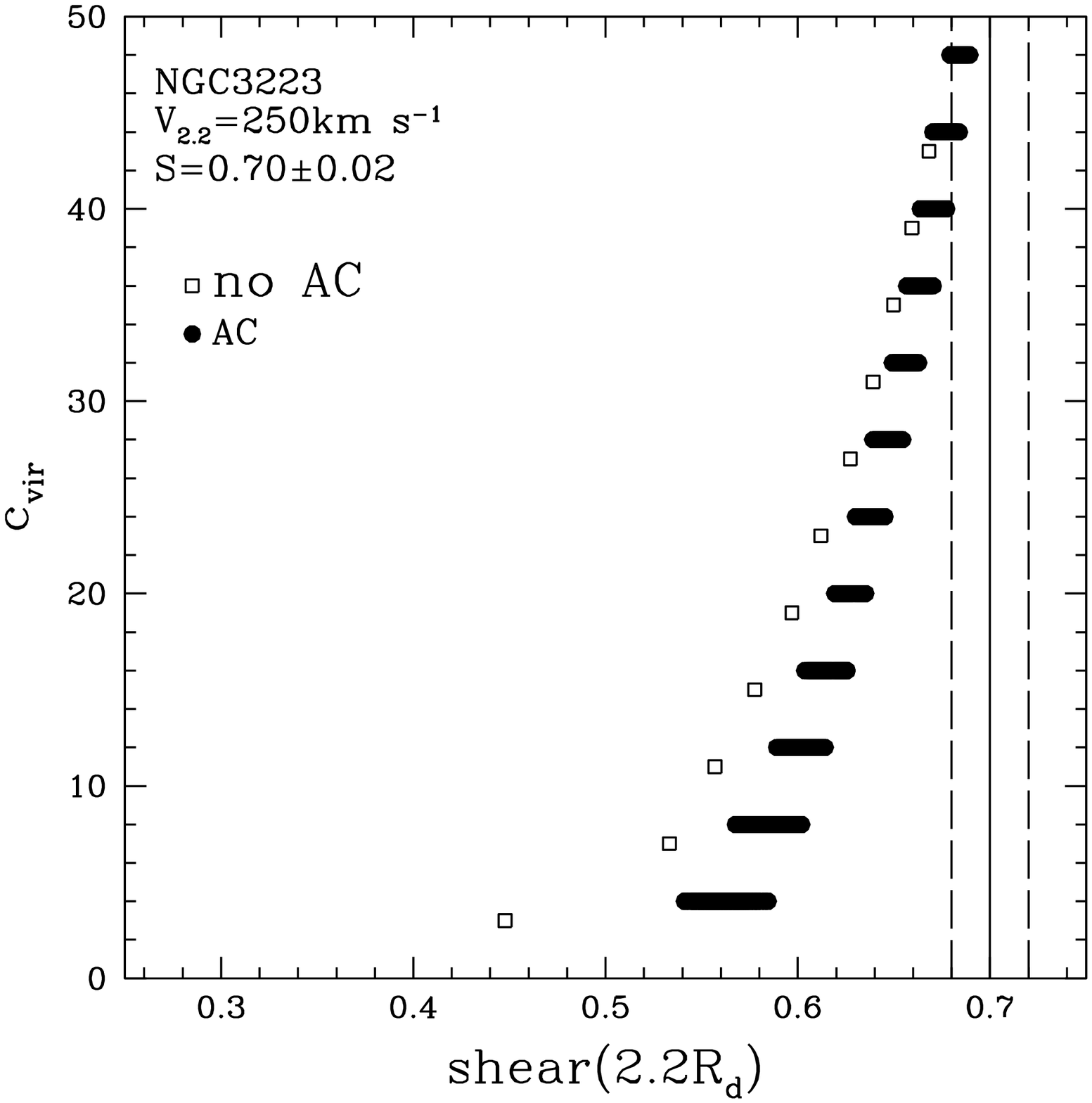}
\includegraphics[width=5.4cm]{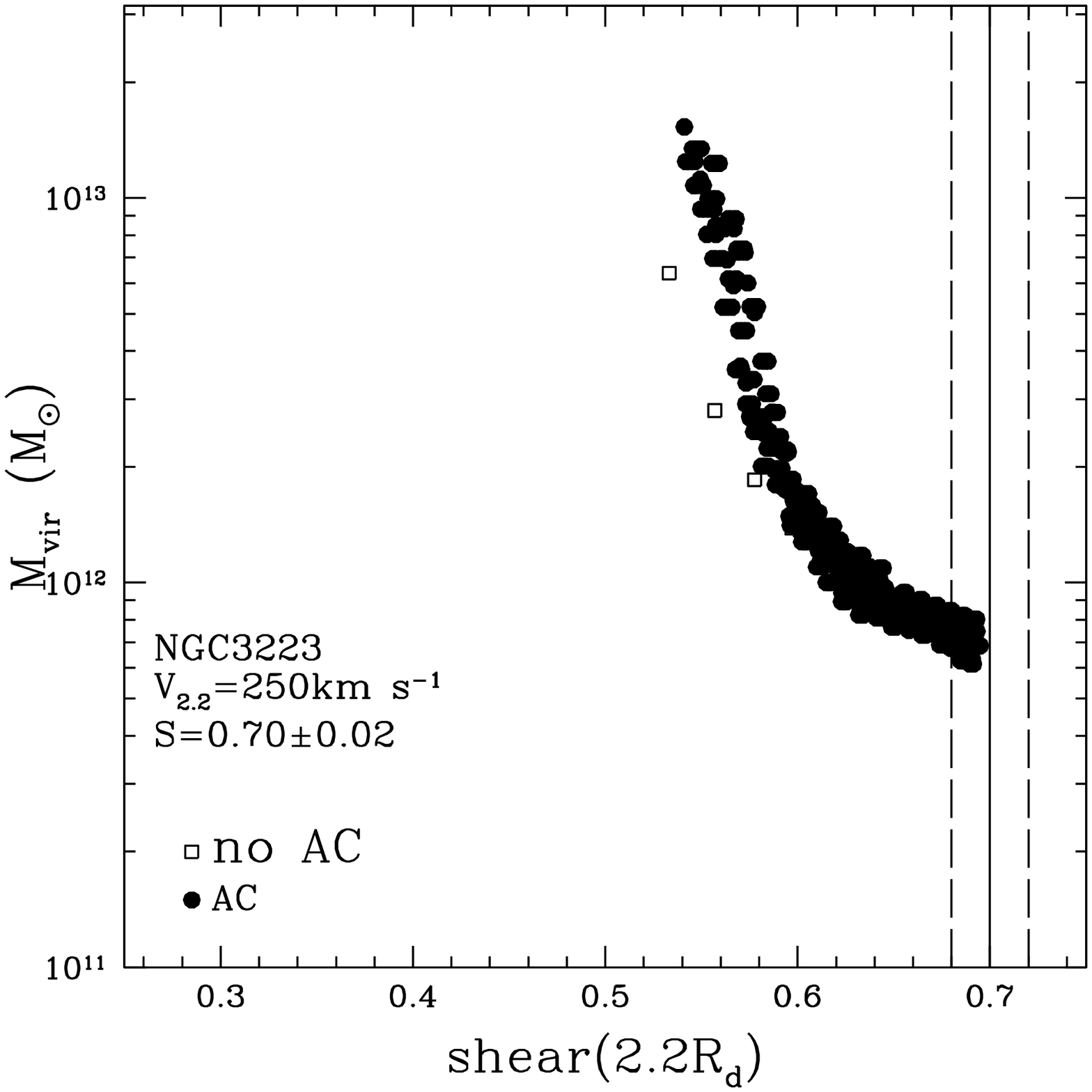}
\includegraphics[width=5.4cm]{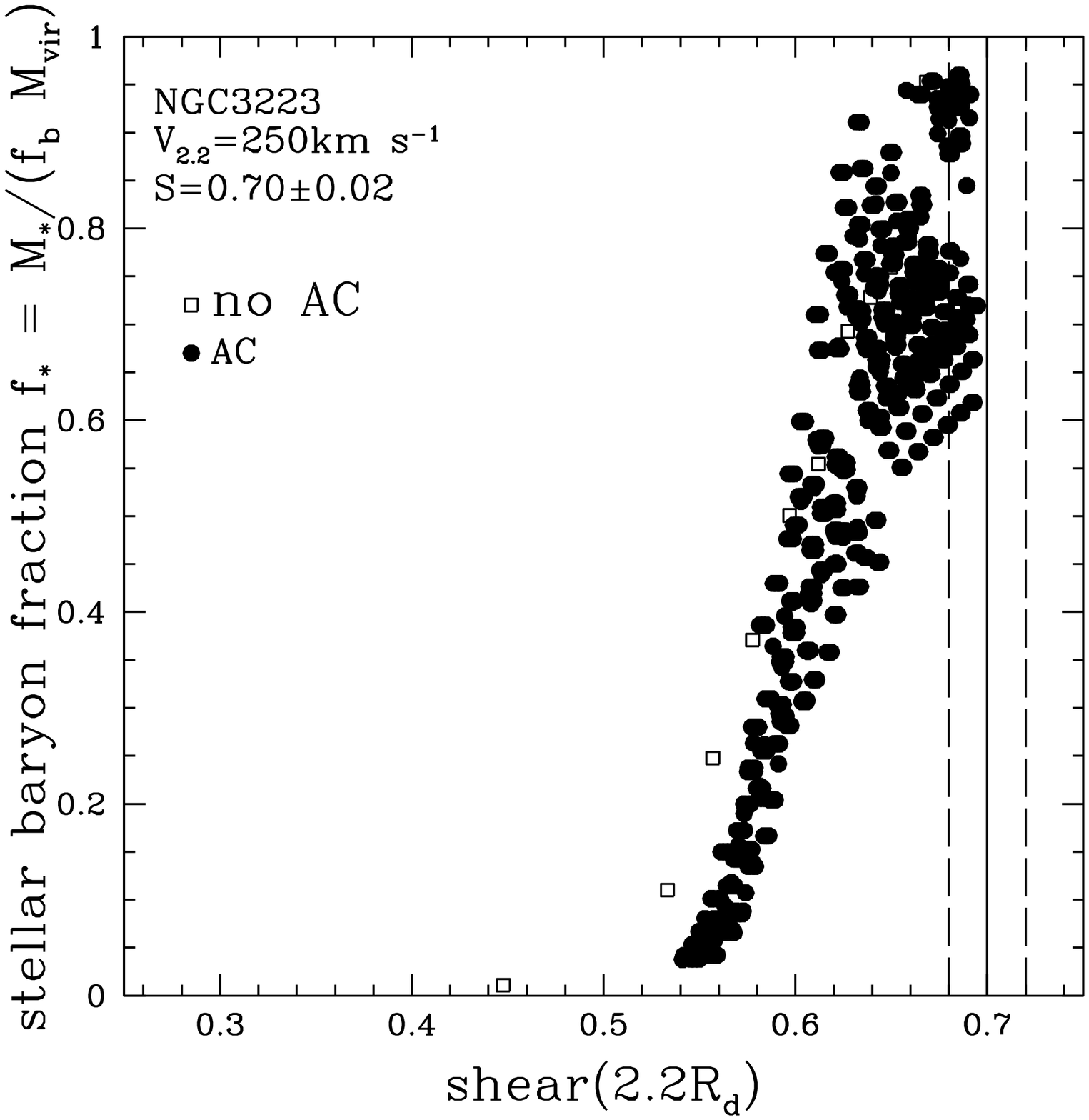}
\end{figure*}

\begin{figure*}
\includegraphics[width=5.4cm]{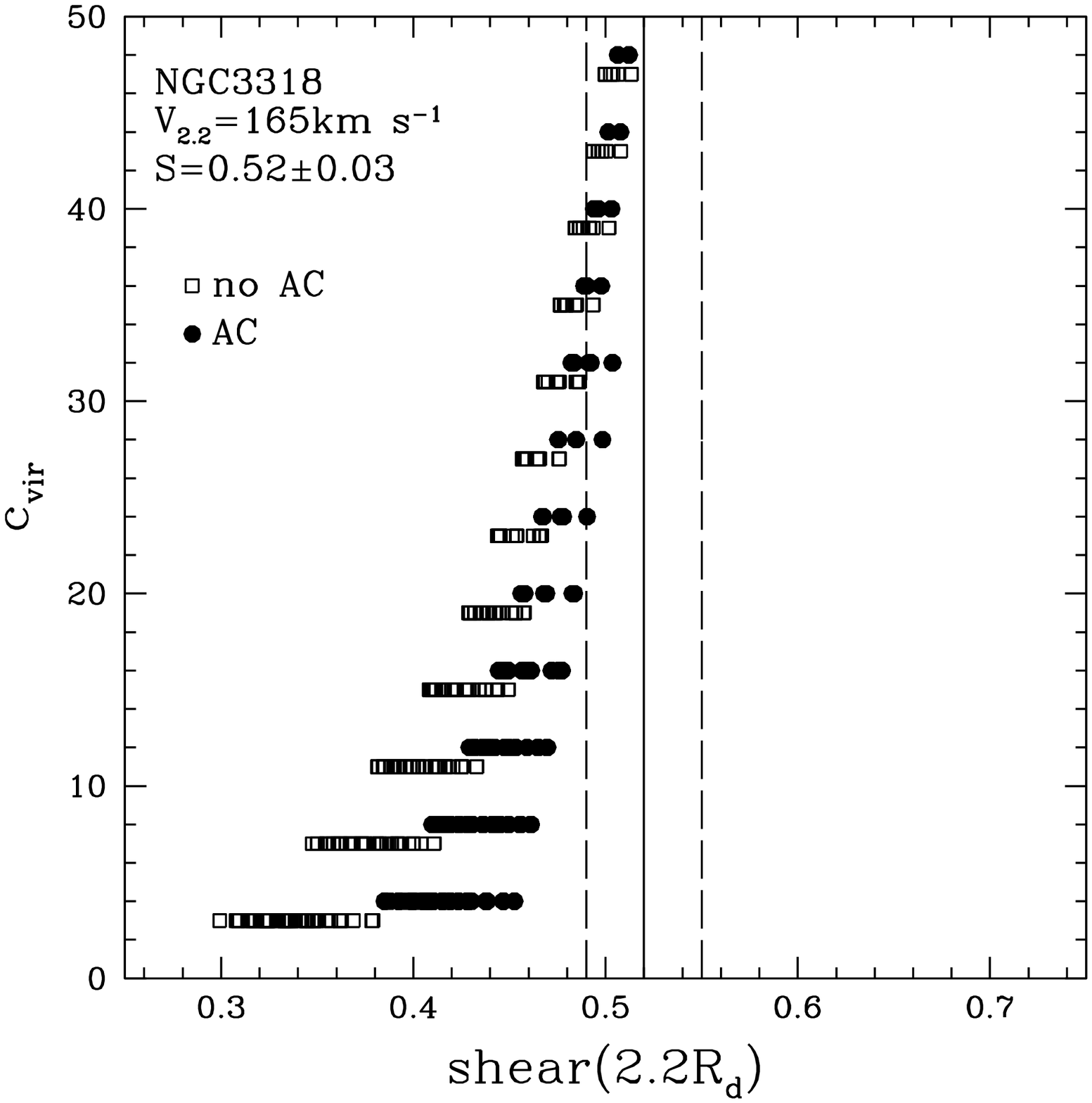}
\includegraphics[width=5.4cm]{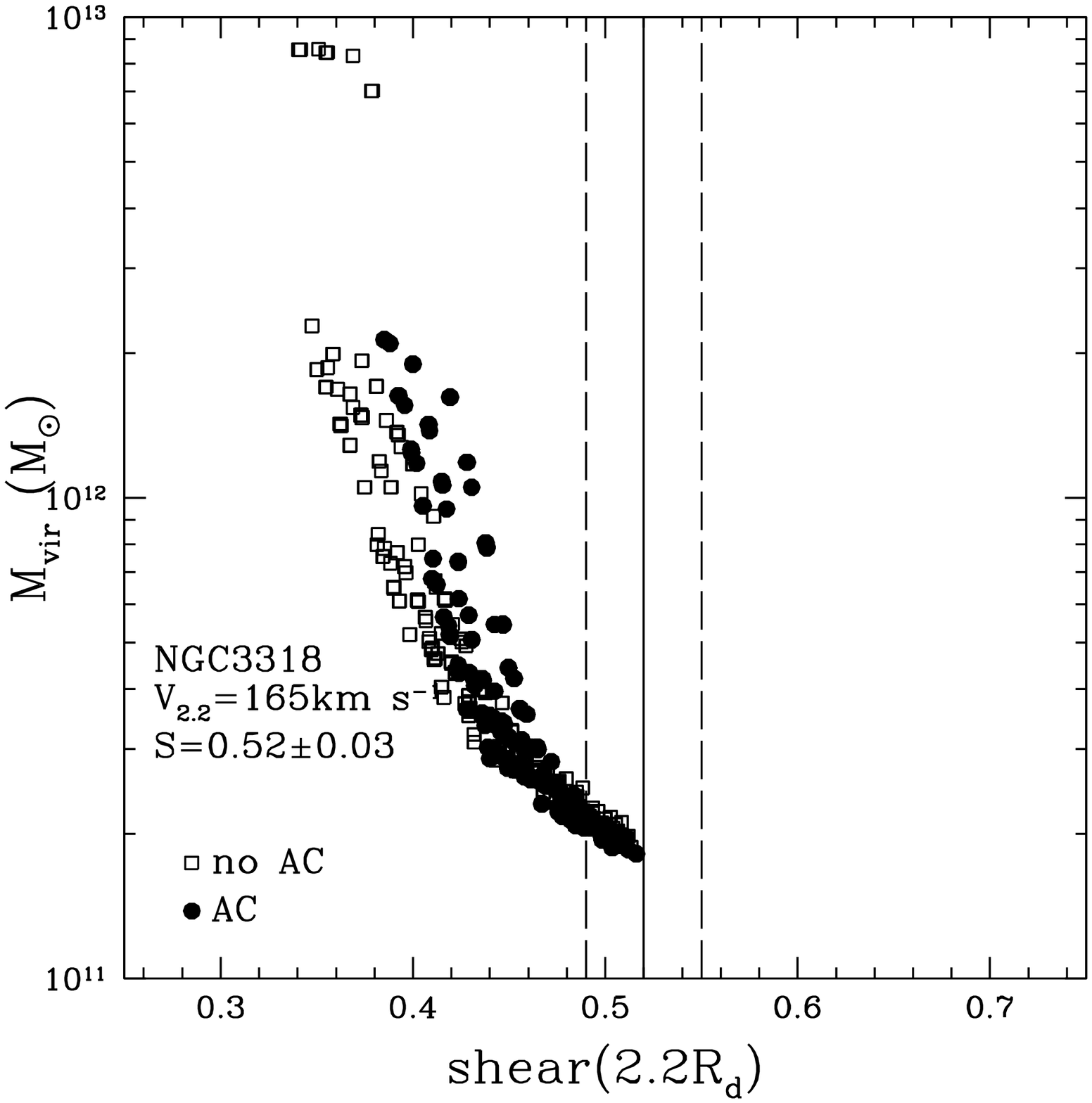}
\includegraphics[width=5.4cm]{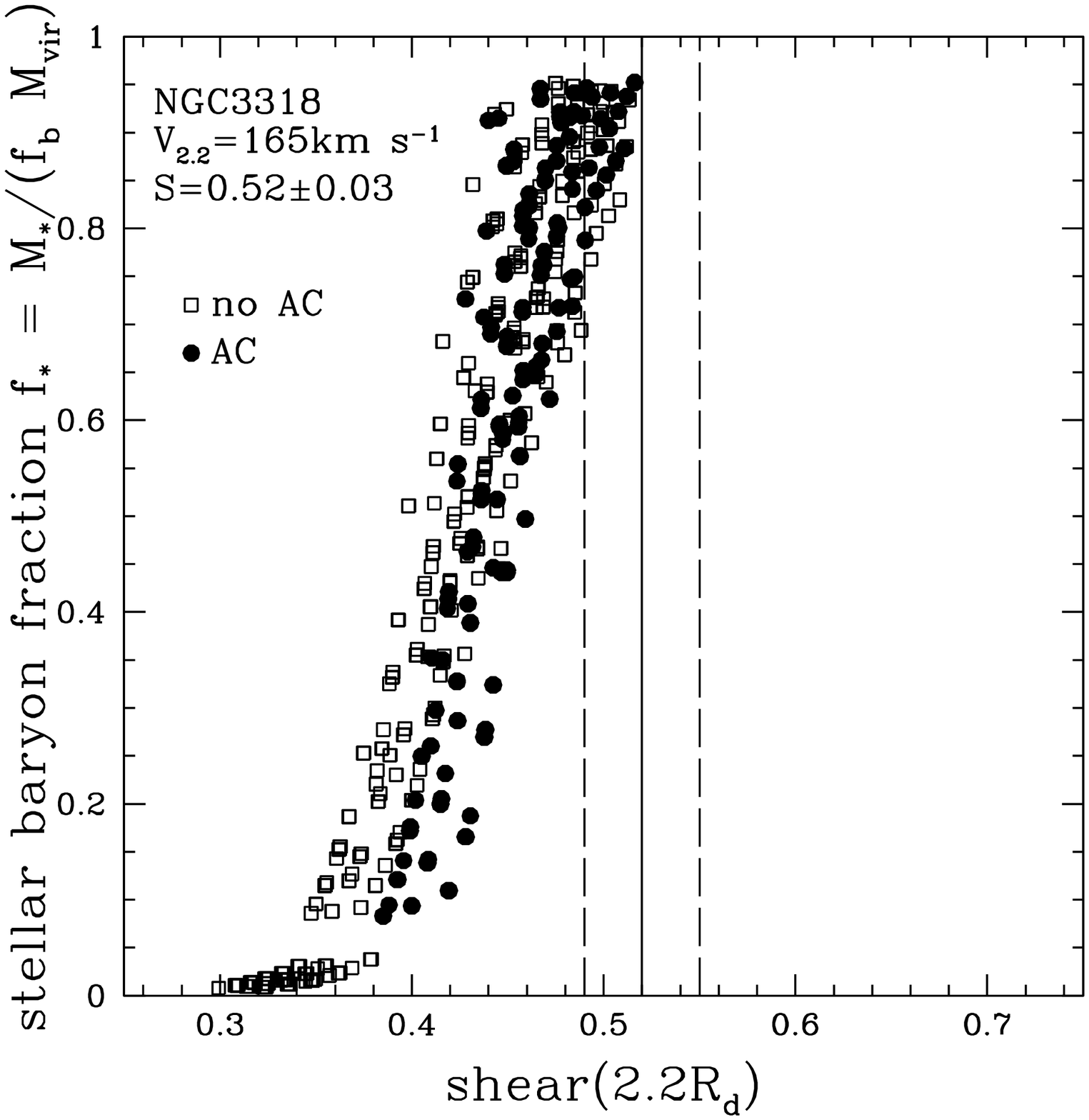}
\caption{Model results for 13 galaxies. {\em Left}: NFW concentration vs.\ shear; {\em center}: virial mass vs.\ shear; {\em right} stellar baryon fraction vs.\ shear.  The open squares represent the model without an adiabatically contracted dark matter halo (non-AC), and the filled circles represent the model with adiabatic contraction (AC).  The vertical lines represent the measured shear (solid line) with the 1$\sigma$ error added and subtracted ({\em dashed lines}).}
\end{figure*}

\begin{deluxetable*}{lccccccccccc}
\label{table2}
\tabletypesize{\scriptsize}
\tablecolumns{6}
\tablewidth{0pc}
\tablecaption{Galaxy model results}
\tablehead{
\colhead{Galaxy} & \colhead{$c_{\rm vir}$} & \colhead{$M_{\rm vir}$}               & \colhead{$f_*$}     & \colhead{$c_{\rm tot}$}  & \colhead{$c_{\rm DM}$}  \\
\colhead{Name}   & \colhead{}             & \colhead{($\times 10^{12}$ M$_{\odot}$)} & \colhead{}          & \colhead{}              & \colhead{}          \\
}
\startdata 
ESO 009-G010  & 6--42  & 0.05--1.05 & 0.10--0.95 & 0.02--0.30 & 0.02--0.18 \\
ESO 582-G012  & 18--48 & 0.05--0.25 & 0.50--0.90 & 0.18--0.31 & 0.13--0.23 \\
IC 4808       & 42--46 & 0.32--0.39 & 0.90--1.00 & 0.32--0.34 & 0.16--0.18 \\
NGC 150       & $>$48  & 0.14--0.29 & 0.38--0.47 & 0.34--0.42 & 0.26--0.28 \\
NGC 578       & 40--48 & 0.87--0.95 & 0.90--0.95 & 0.32--0.33 & 0.23--0.24 \\
NGC 908       & 34--48 & 0.21--0.23 & 0.90--0.95 & 0.29--0.30 & 0.16--0.20 \\
NGC 1292      & 25--48 & 0.22--1.53 & 0.05--0.30 & 0.06--0.24 & 0.04--0.21 \\
NGC 1300      & 30--36 & 0.21--0.23 & 0.80--0.93 & 0.19--0.21 & 0.14--0.16 \\
NGC 1353      & 6--19  & 1.31--21.0 & 0.01--0.12 & 0.01--0.08 & 0.01--0.05 \\
NGC 1365      & 3--7   & 0.83--31.0 & 0.15--0.83 & 0.02--0.10 & 0.01--0.02 \\
NGC 1964      & 40--48 & 0.27--0.42 & 0.40--0.52 & 0.30--0.35 & 0.21--0.23 \\
NGC 3223      & 43--48 & 0.59--0.92 & 0.99--1.00 & 0.16--0.24 & 0.14--0.17 \\
NGC 3318      & 32--38 & 0.19--0.28 & 0.75--0.95 & 0.25--0.34 & 0.16--0.21 \\
\enddata
\tablecomments{
Model reults from the no-AC models.
}
\end{deluxetable*}

\begin{deluxetable}{lccccccccccc}
\label{table3}
\tabletypesize{\scriptsize}
\tablecolumns{3}
\tablewidth{0pc}
\tablecaption{Galaxy model results by direct rotation curve fitting}
\tablehead{
\colhead{Galaxy} & \colhead{$c_{\rm vir}$} & \colhead{$M_{\rm vir}$}               \\
\colhead{Name}   & \colhead{}             & \colhead{(M$_{\odot}$)} \\
}
\startdata 
ESO 009-G010  & 15.0  & $1.05\times10^{12}$ \\
ESO 582-G012  & 39.0  & $2.50\times10^{11}$ \\
IC 4808       & 47.0  & $3.87\times10^{11}$ \\
NGC 150       & 48.0  & $2.91\times10^{11}$ \\
NGC 578       & 47.0  & $9.50\times10^{11}$ \\
NGC 908       & 47.0  & $2.25\times10^{11}$ \\
NGC 1292      & 35.0  & $1.53\times10^{12}$ \\
NGC 1300      & 35.0  & $2.33\times10^{11}$ \\
NGC 1353      & 11.0  & $4.38\times10^{12}$ \\
NGC 1365      &  7.0  & $8.28\times10^{11}$ \\
NGC 1964      & 48.0  & $3.26\times10^{11}$ \\
NGC 3223      & 44.0  & $9.15\times10^{11}$ \\
NGC 3318      & 35.0  & $2.08\times10^{11}$ \\
\enddata
\end{deluxetable}

For the second class of model, adiabatic contraction (AC) is adopted 
(Blumenthal et al.\ 1986; Bullock et al.\ 2001a; Pizagno et al.\ 2005).
In this case the dark matter density profile initially follows an NFW
profile.  The baryons are then allowed to cool and settle into the halo
center, and this process is much longer than one typical orbital time.
This slow infall results in the halo density distrbution contracting
adiabatically, which gives rise to a more concentrated dark matter halo.
AC was originally discussed as a means of producing featureless rotation
curves for large disk galaxies (Rubin et al.\ 1985), but it has also
proven accurate in discribing spiral galaxy formation in N-body simulations
(e.g., Gnedin et al.\ 2004, 2010, and references therein).  Nevertheless,
the degree to which AC operates in real galaxies is uncertain (see Seigar
\& Berrier 2011 for a review).

For our AC model, we use the prescription described by Blumenthal et al.\
(1986).  Gnedin et al.\ (2004) suggest a slightly modiﬁed prescription, but the
differences between the two methods are small compared to the
differences between our AC model and our non-AC model. In principle, any 
observational probe that can distinguish between AC and non-AC-type scenarios 
provides an important constraint on
the nature of gas infall into galaxies (i.e., was it fast or was it slow?).

For each galaxy we iterate over the central and $\pm 1\sigma$ values
found in the bulge-disk decompositions for $R_d$ and $L_{\rm disk}$ and explore 
the five values of mass-to-light ratio discussed above,
$(M/L)=$ 0.7, 0.8, 0.9, 1.0, and 1.1,  assuming that $M/L$ 
remains constant with radius. For each choice of bulge-disk model
parameters and mass-to-light ratios, we allow the (initial) halo
NFW concentration parameter to vary over the range of viable
values, $c_{\rm vir} \geq 3$ (Bullock et al.\ 2001b). We then determine
the halo virial mass $M_{\rm vir}$ necessary to reproduce the value of $V_{2.2}$
for the galaxy and determine the implied fraction of the mass in
the system in the form of stars compared to that ``expected'' from the 
Universal baryon fraction, $f_{*}=M_{*}/(f_bM_{\rm vir})$. We make the
demand that $f_*$ lies within the range of plausible values 
$0.01f_b < f_* < f_b$.  Note that the $V_{2.2}$ is the rotation velocity (in
km s$^{-1}$) at 2.2 disk scalelengths.  We chose $V_{2.2}$ because it is
where the disk contribution to the rotation curve is a maximum.

For each chosen value of $c_{\rm vir}$ and adopted disk galaxy formation 
scenario, the chosen $V_{2.2}$ constraint deﬁnes a complete rotation curve 
and thus provides a shear at every radius. The three panels of Figure 3 show 
the results of this exercise for the 13 galaxies in our sample.
In each panel, open symbols are for the non-AC model, and the
filled symbols are for the AC model.  Each point represents
a distinct input combination of $R_d$, $L_{\rm disk}$ , and $M/L$.  The measured
shear rate illustrated by a solid vertical line, and the $\pm 1\sigma$ range
in the observe shear for each galaxy is shown by the two
vertical dashed lines in each panel.

Consider first the left panel of Figure 3.  Here we plot the dark matter halo
concentration parameter versus the shear measured at $2.2R_d$.  More 
concentrated halos generally produce higher shears, as expected. It can be 
seen that for a given NFW concentration, $c_{\rm vir}$, several values of 
shear are possible. This is due
to changes in the baryon contribution (i.e., the disk mass and
disk scale length) to the rotation curve. Whether an increase in
the baryon contribution causes the shear to increase or decrease
depends on the size of the disk (i.e., the disk scale length). The
same is true for all of our galaxies. In the AC model (filled circles), 
$c_{\rm vir}$ refers to the halo concentration before the halo is adiabatically
contracted. This is why the AC points tend to have higher shear
values at fixed $c_{\rm vir}$ compared to the non-AC case. Every point (or
$R_d$, $L_{\rm disk}$, $M/L$ input combination) has an associated dark halo 
virial mass $M_{\rm vir}$ and stellar baryon fraction $f_*$.  These values are
plotted versus shear rate in the middle and right panels in Figure 3.
Observe that with only the $V_{2.2}$ constraint imposed (all
points), a wide range of dark matter halo properties are allowed.
Once we constrain the models by forcing the predicted shear to
be consistent with the observed range, we prefer a much narrower range of
halo concentrations, virial masses, and stellar baryon fractions, although
favoring one type of model over another (i.e., our AC versus our non-AC model)
is unfortunately difficult in most cases, if not impossible.

Model results for the central mass concentration in each galaxy are shown in 
Figure 4. Here $c_{\rm tot}$ and $c_{\rm dm}$ correspond to
the fraction of the total and dark matter mass contained within 2.2
disk scalelengths of each galaxy.  The ranges of allowed values for each
of the five paramaters in Figures 3 and 4 are listed in Table 2 for all of the 
galaxies in our sample.

\section{Discussion}

The work presented in this paper clearly demonstrates that the shear rate 
adds an important
constraint on galaxy formation models compared to what can be
learned from standard Tully-Fisher constraints alone. 
In an earlier paper (Seigar et al.\ 2006) we showed some tantalizing results
for two galaxies, IC 2522 and ESO 582-G12 (one of which we have updated the
results for in this paper).  With a further 12 galaxies now analyzed using 
the shear-pitch angle relation, we have demonstrated that this technique
can be extremely useful for determining mass concentrations in galaxies,
especially for statistically large samples, where the statistics of the
sample outweighs any uncertainties in the derived parameters 
(e.g., $M_{\rm vir}$ and $c_{\rm vir}$) for any individual galaxy.

\begin{figure*}
\label{mod2}
\includegraphics[width=4.1cm]{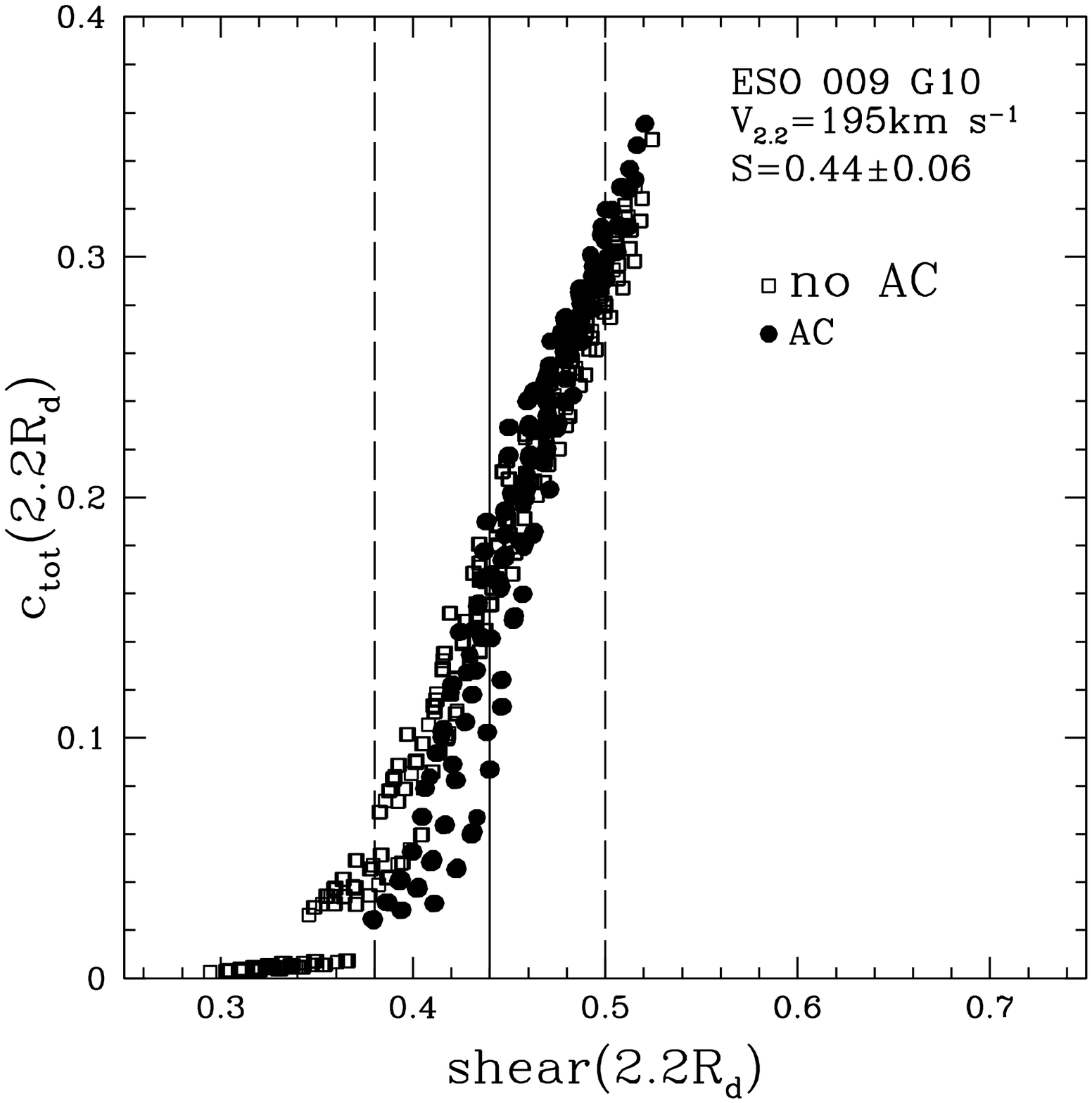}
\includegraphics[width=4.1cm]{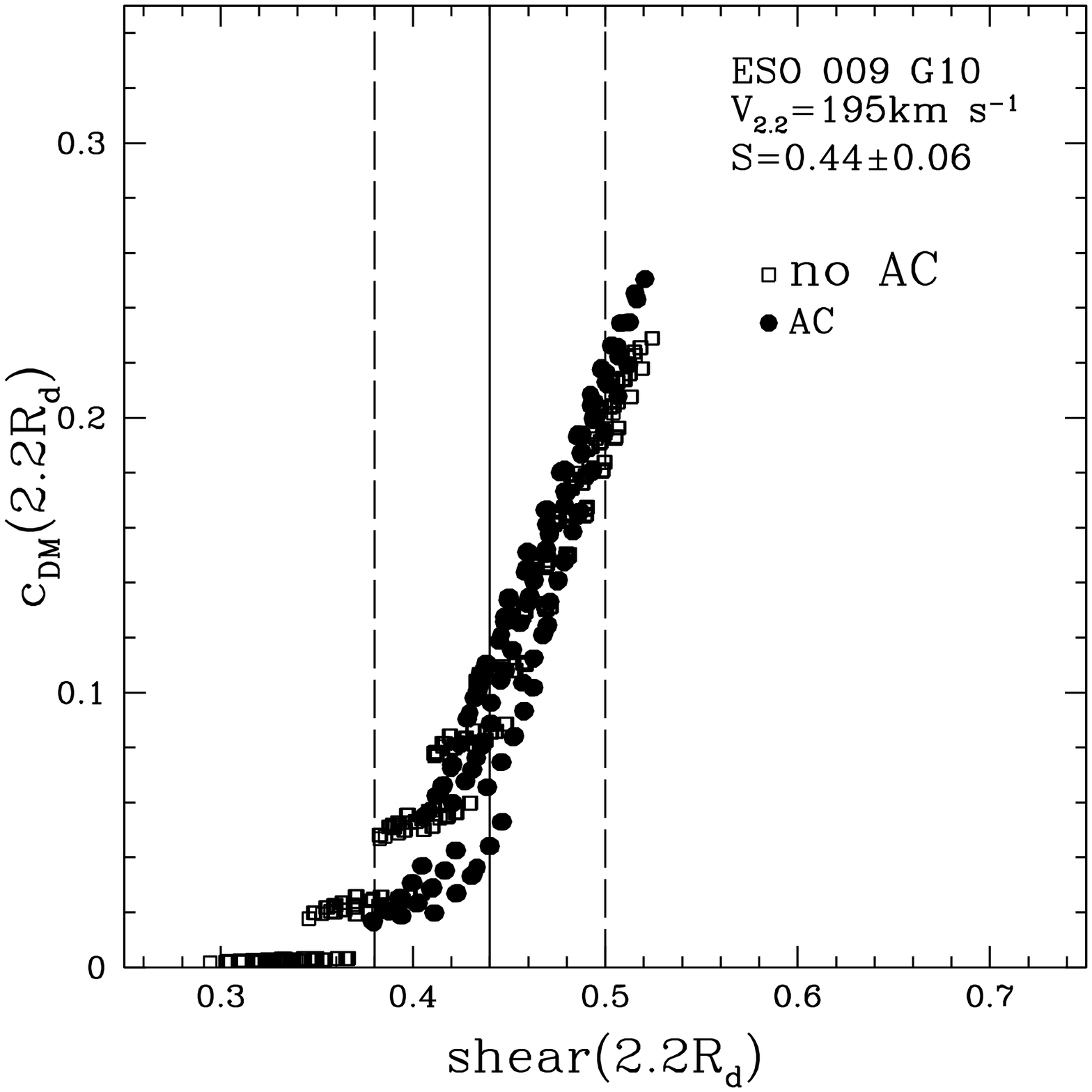}
\includegraphics[width=4.1cm]{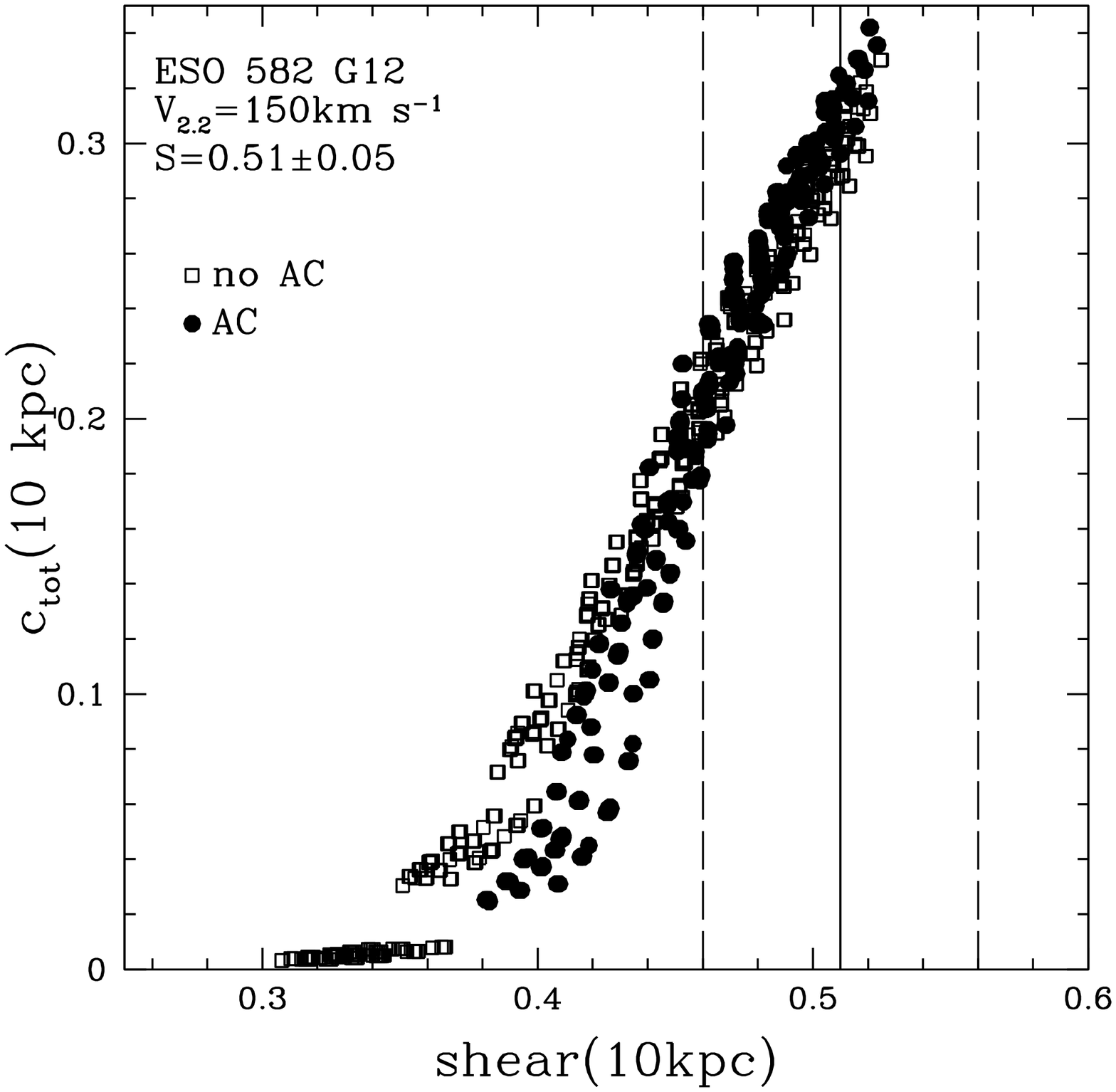}
\includegraphics[width=4.1cm]{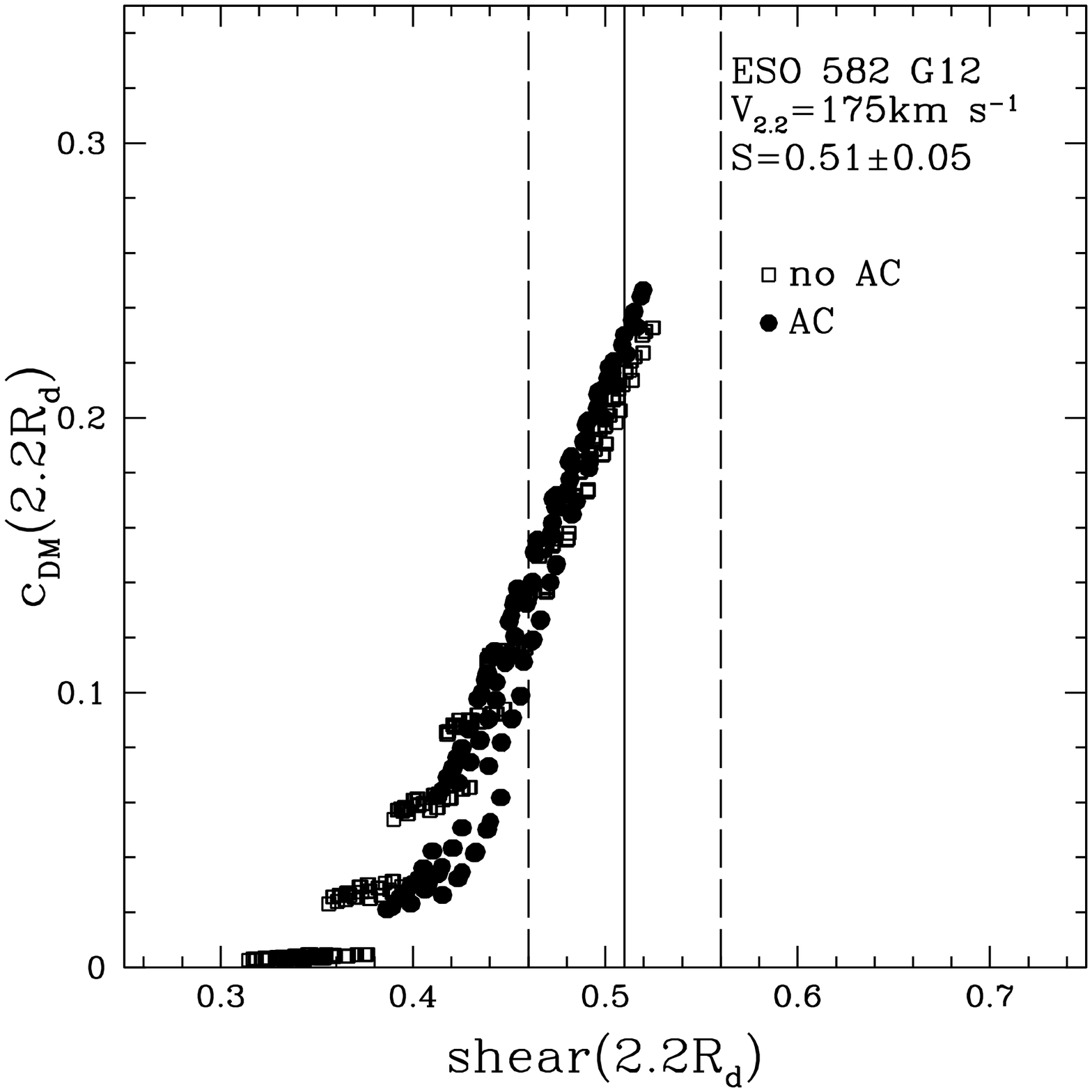}\\
\includegraphics[width=4.1cm]{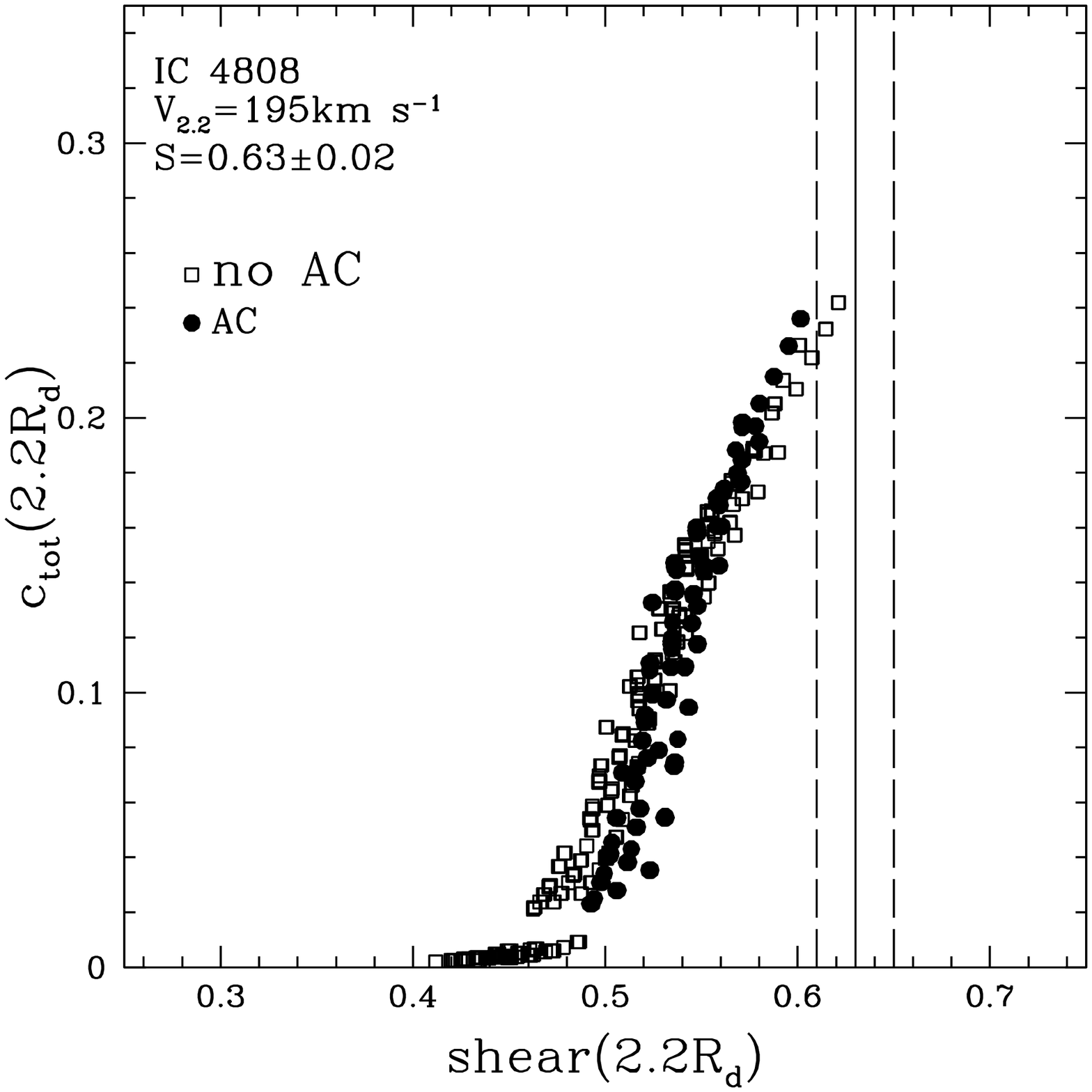}
\includegraphics[width=4.1cm]{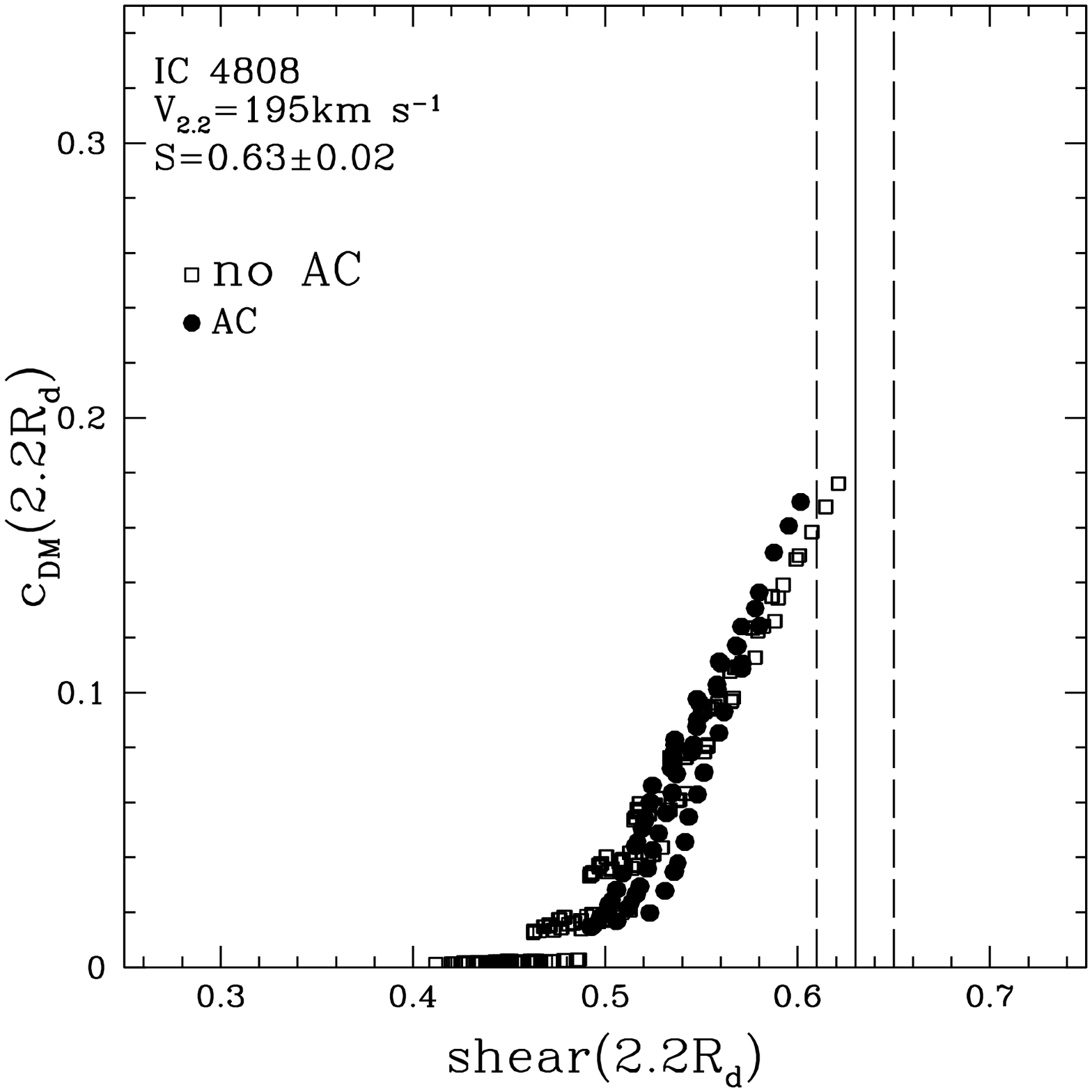}
\includegraphics[width=4.1cm]{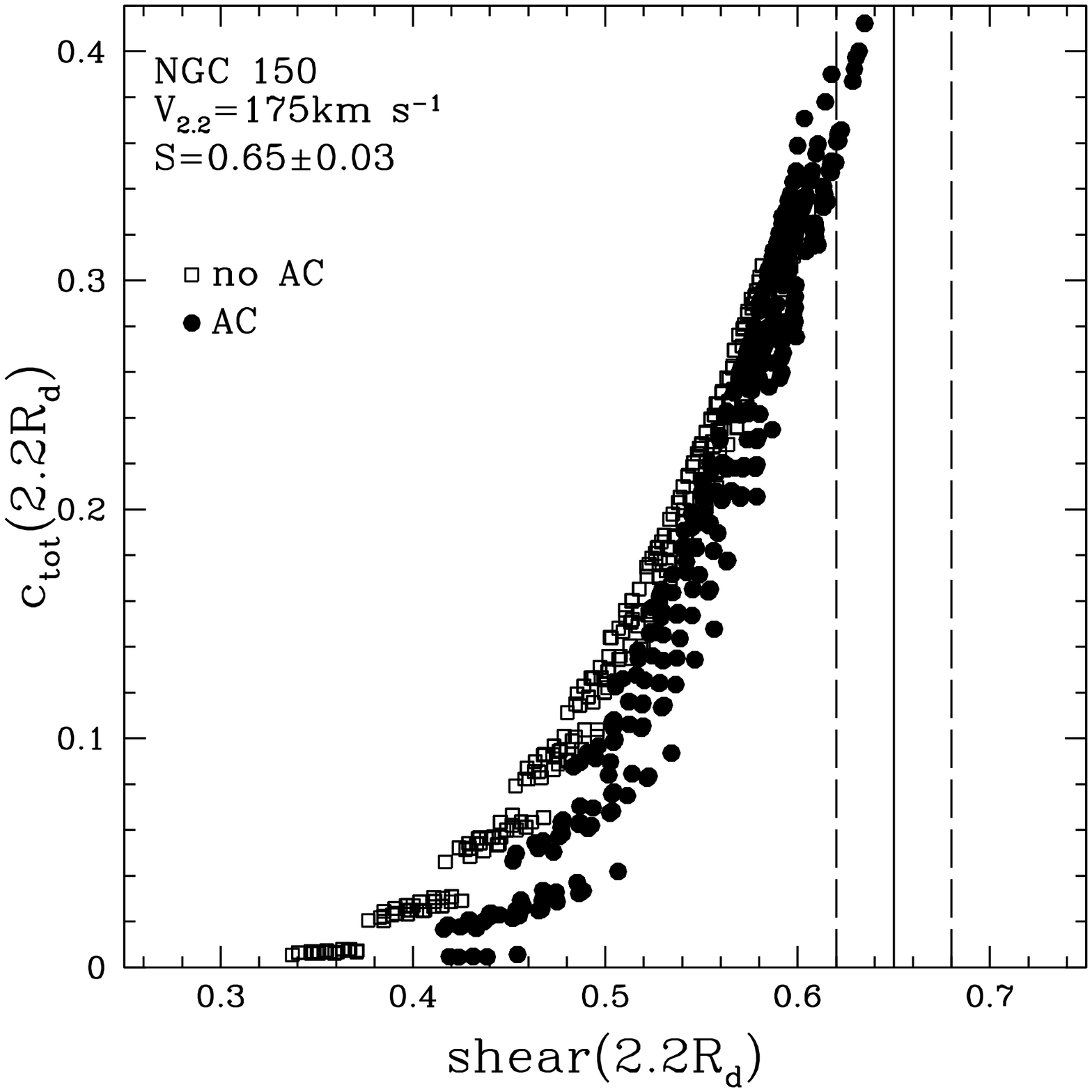}
\includegraphics[width=4.1cm]{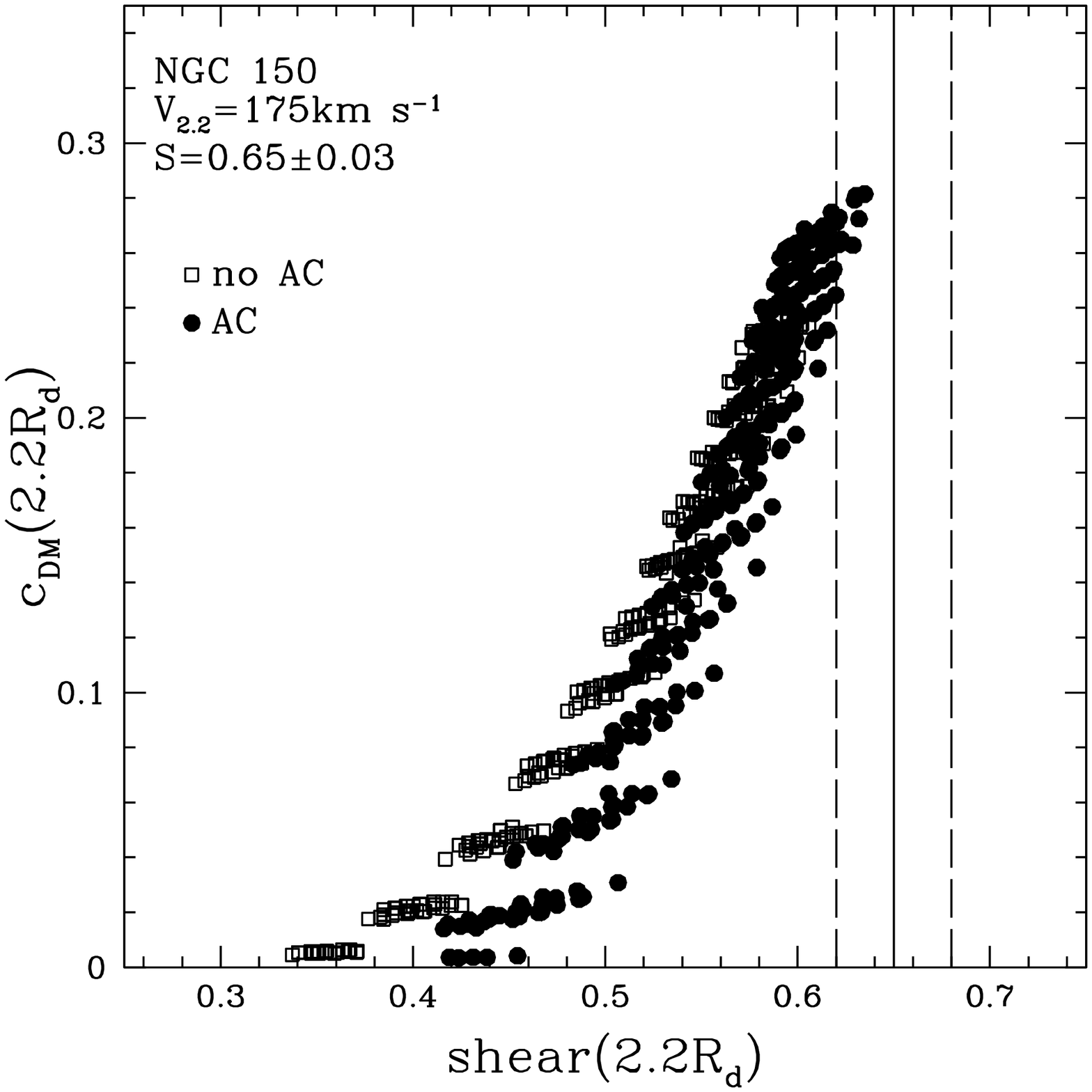}\\
\includegraphics[width=4.1cm]{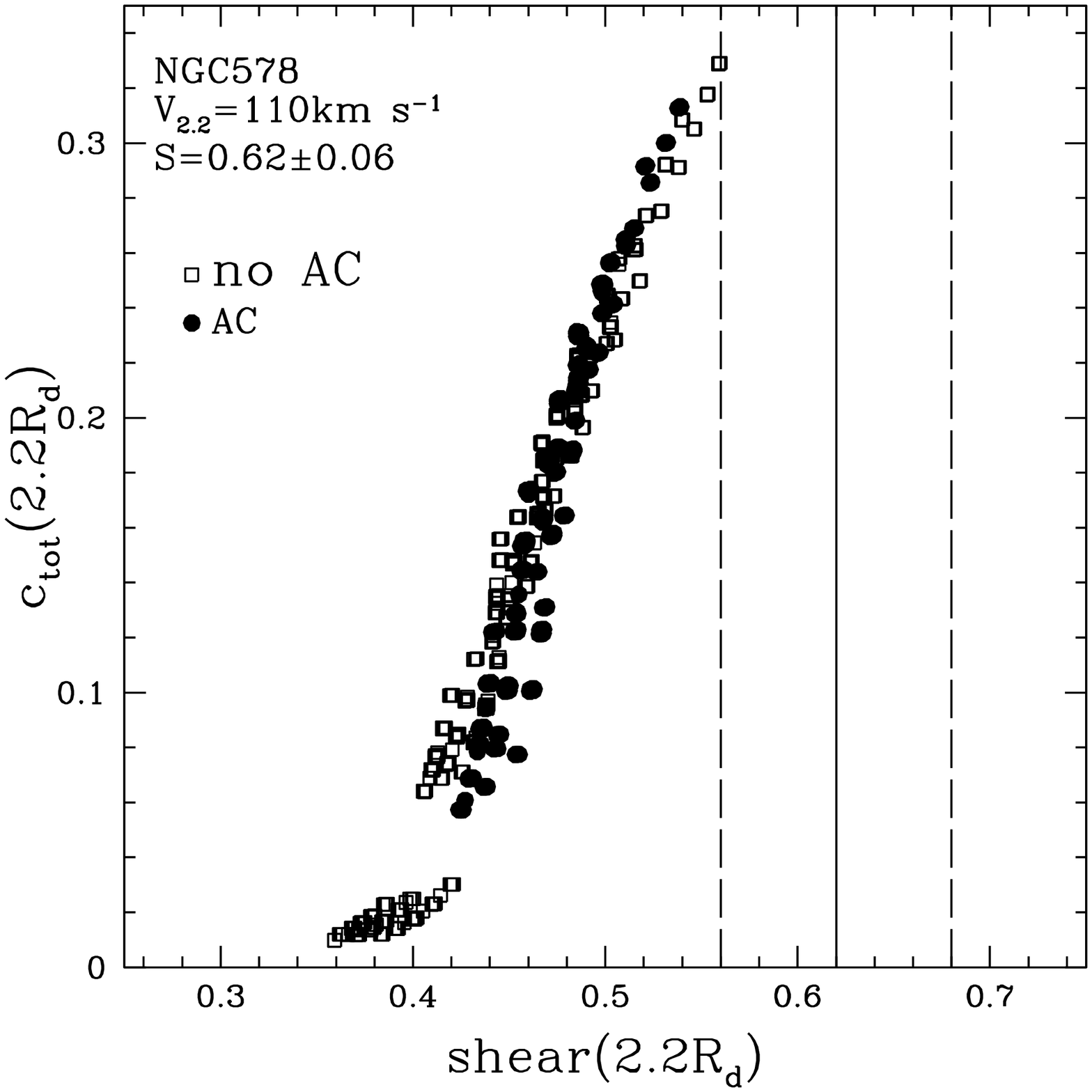}
\includegraphics[width=4.1cm]{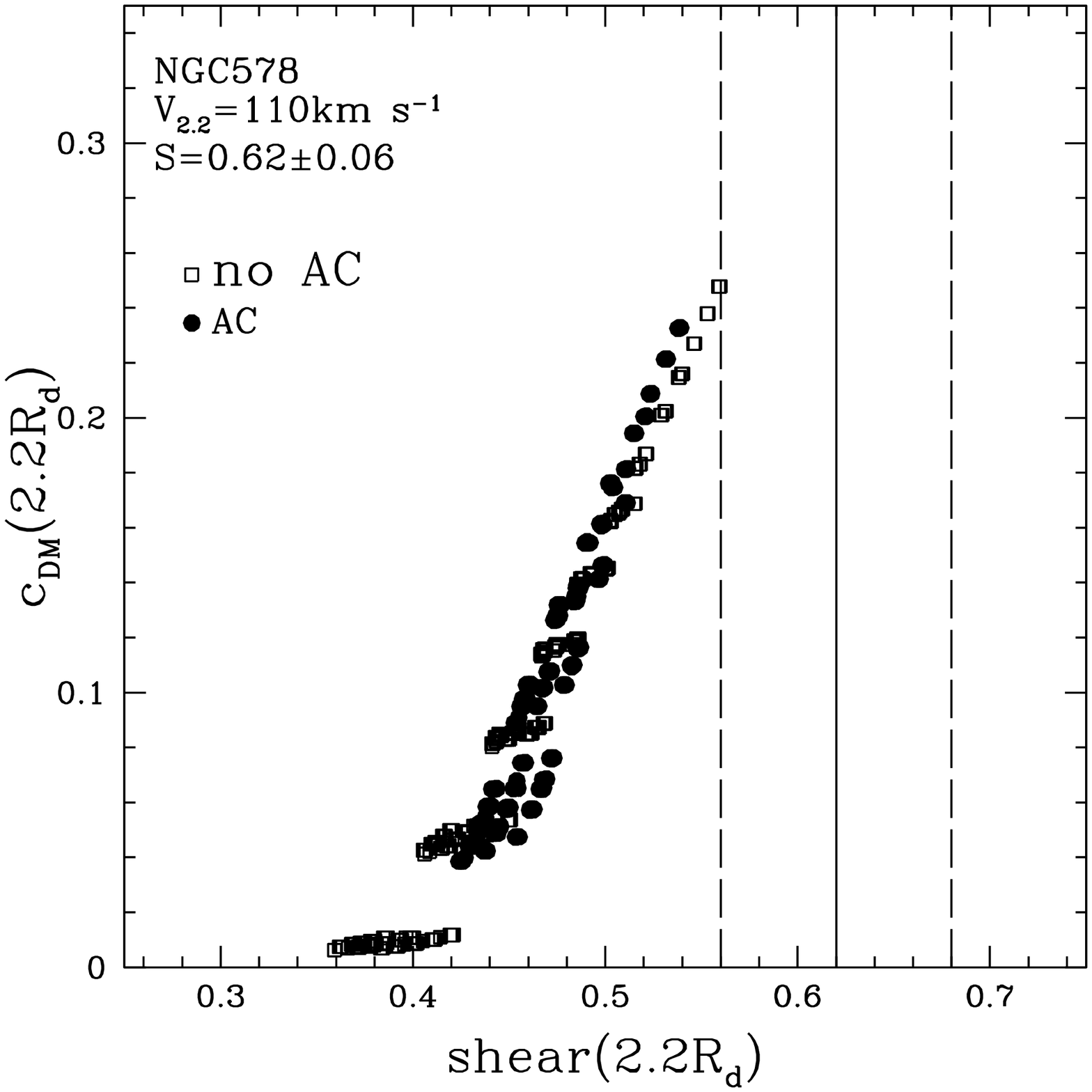}
\includegraphics[width=4.1cm]{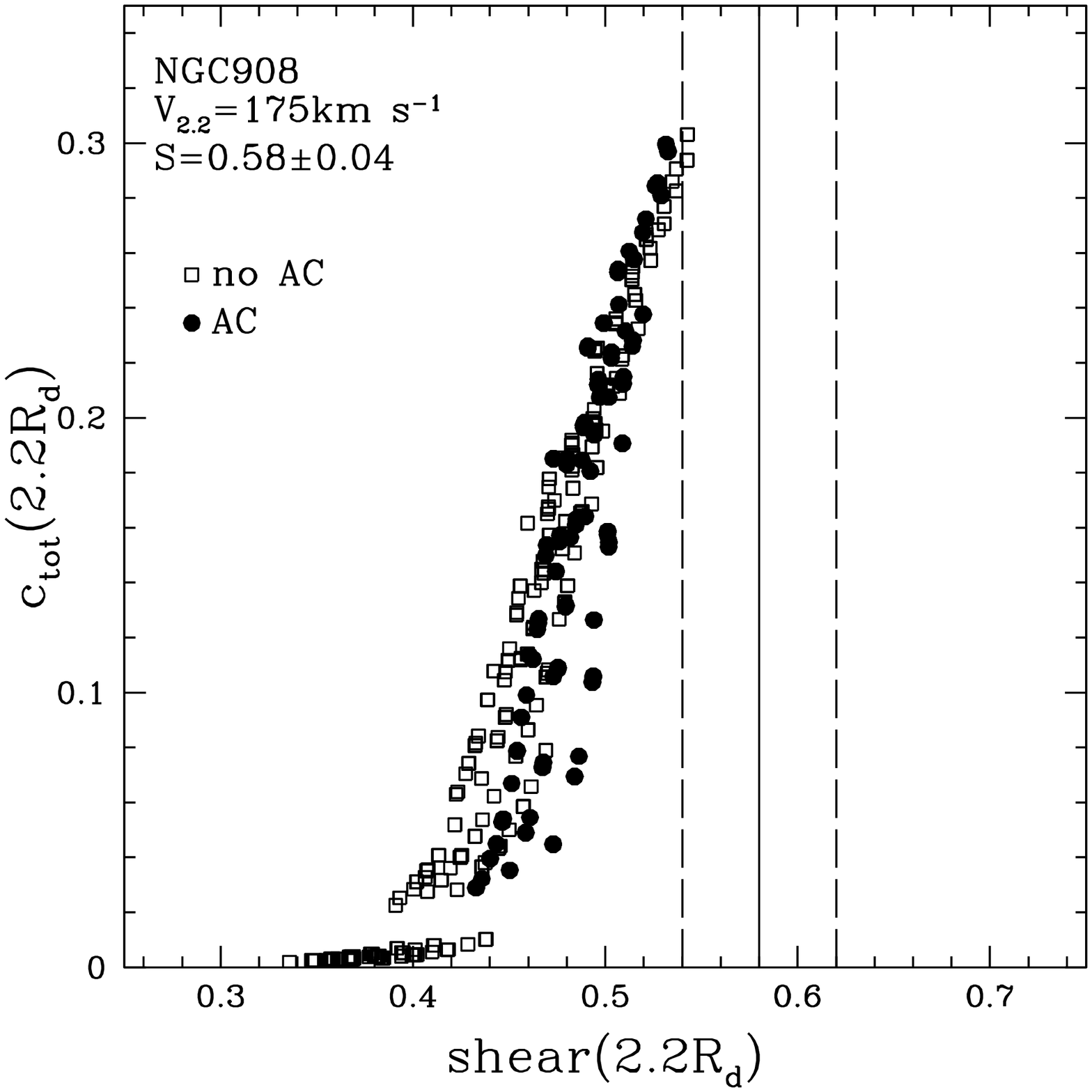}
\includegraphics[width=4.1cm]{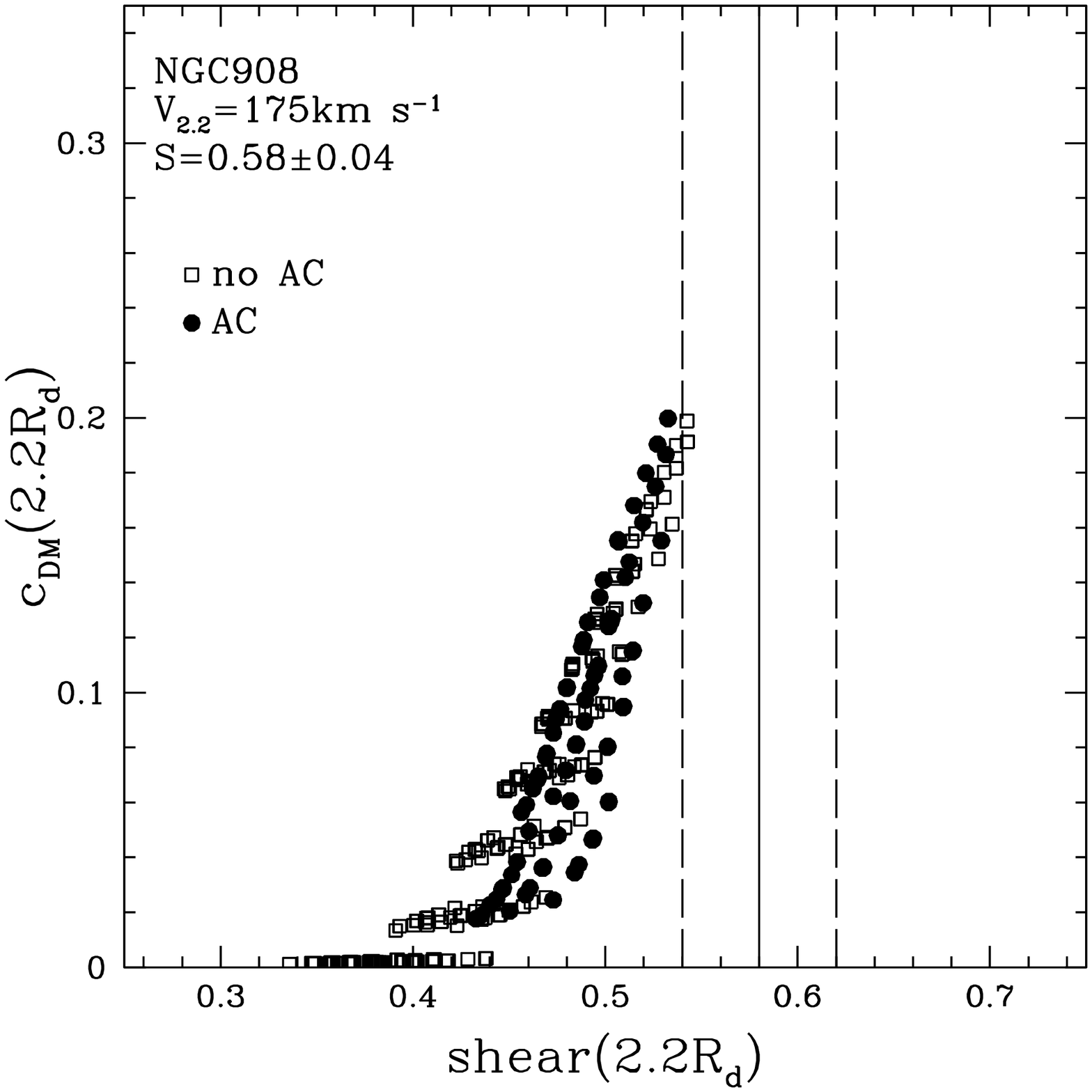}\\
\includegraphics[width=4.1cm]{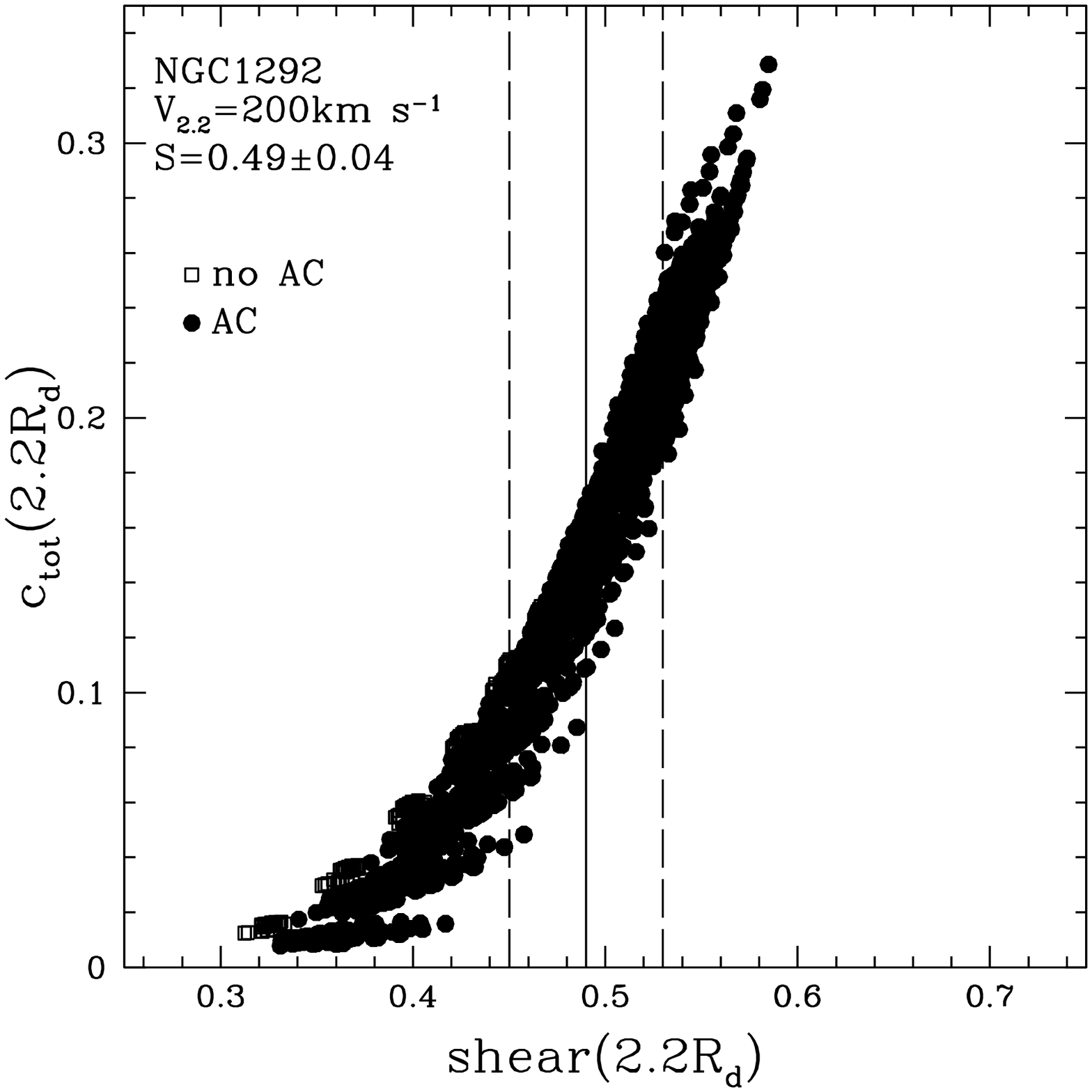}
\includegraphics[width=4.1cm]{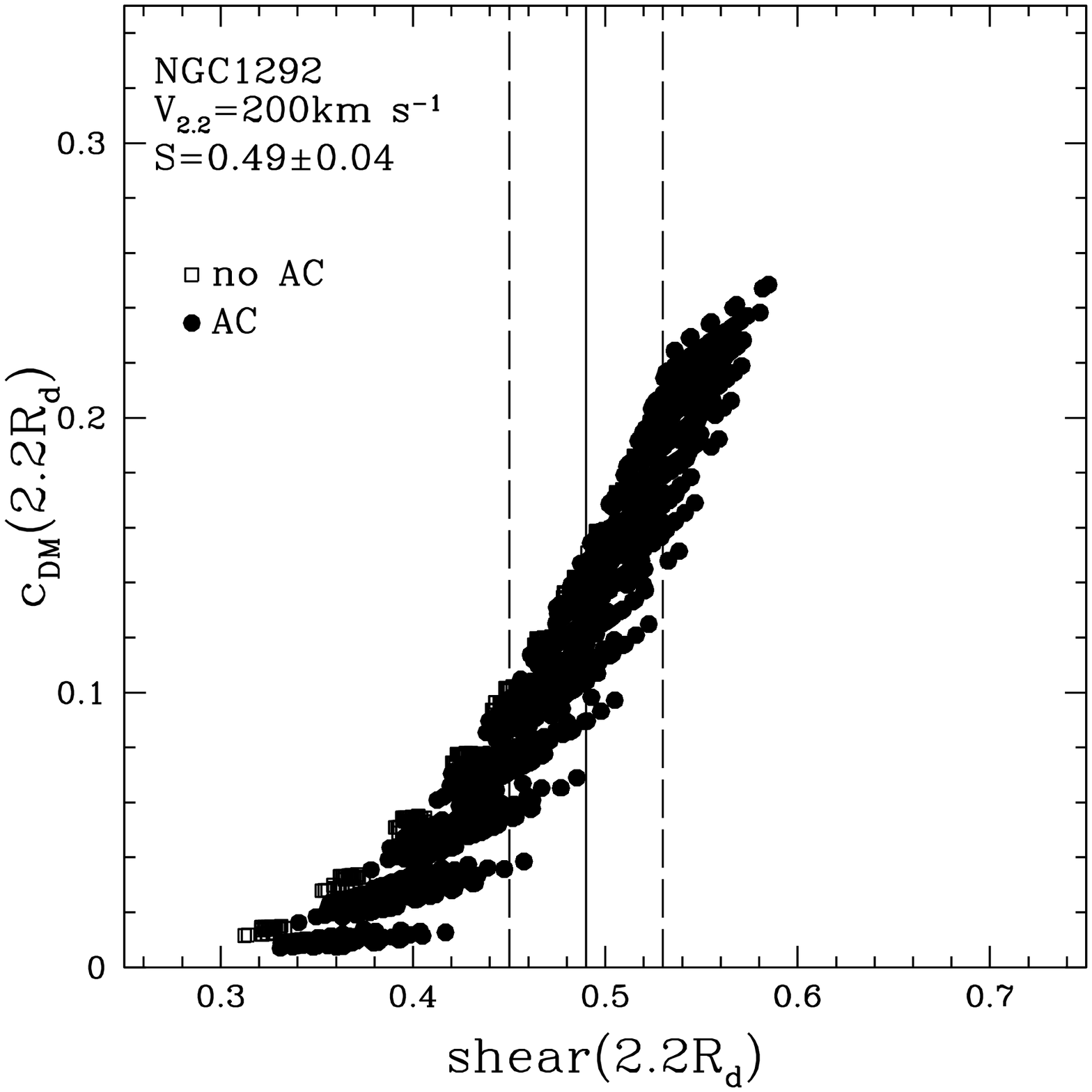}
\includegraphics[width=4.1cm]{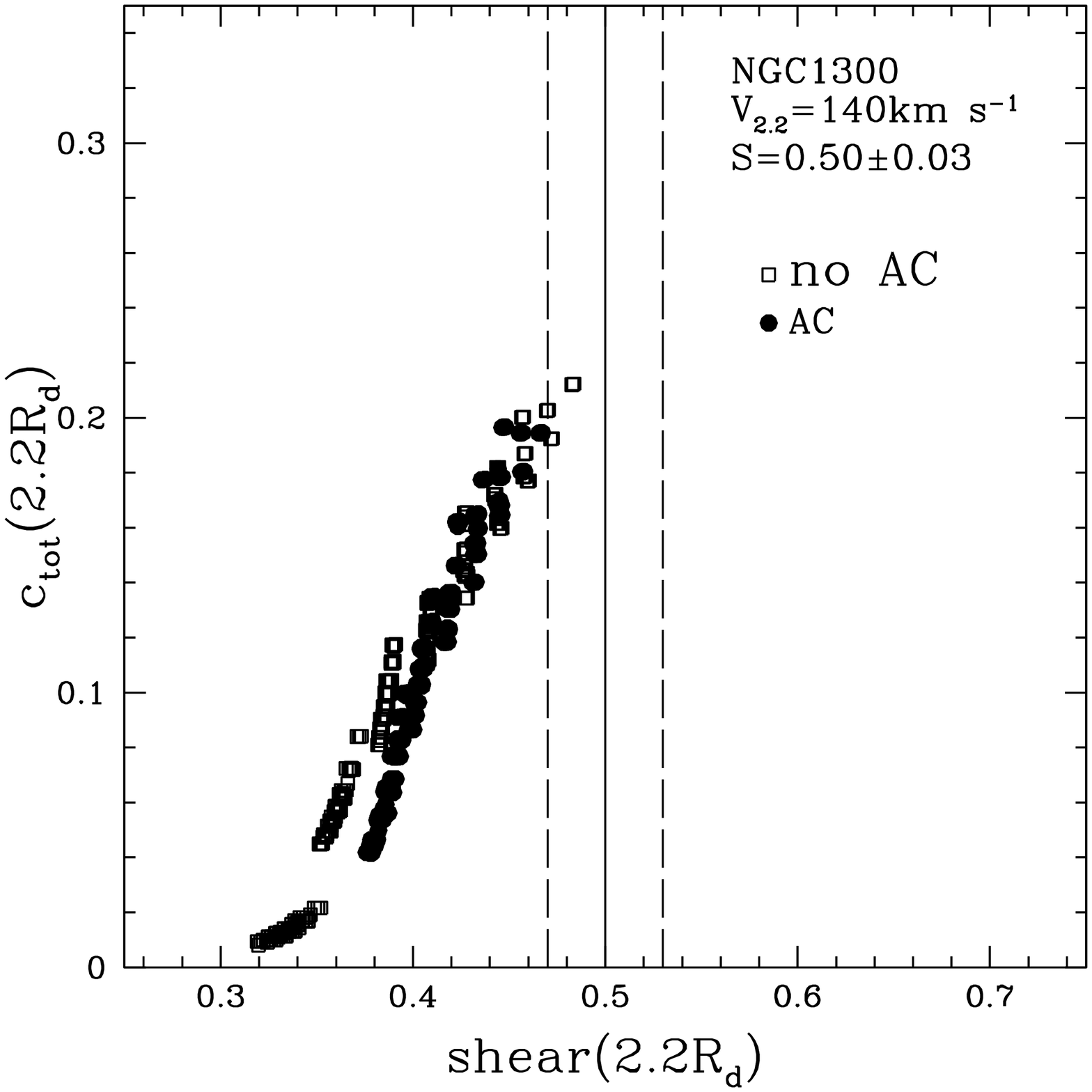}
\includegraphics[width=4.1cm]{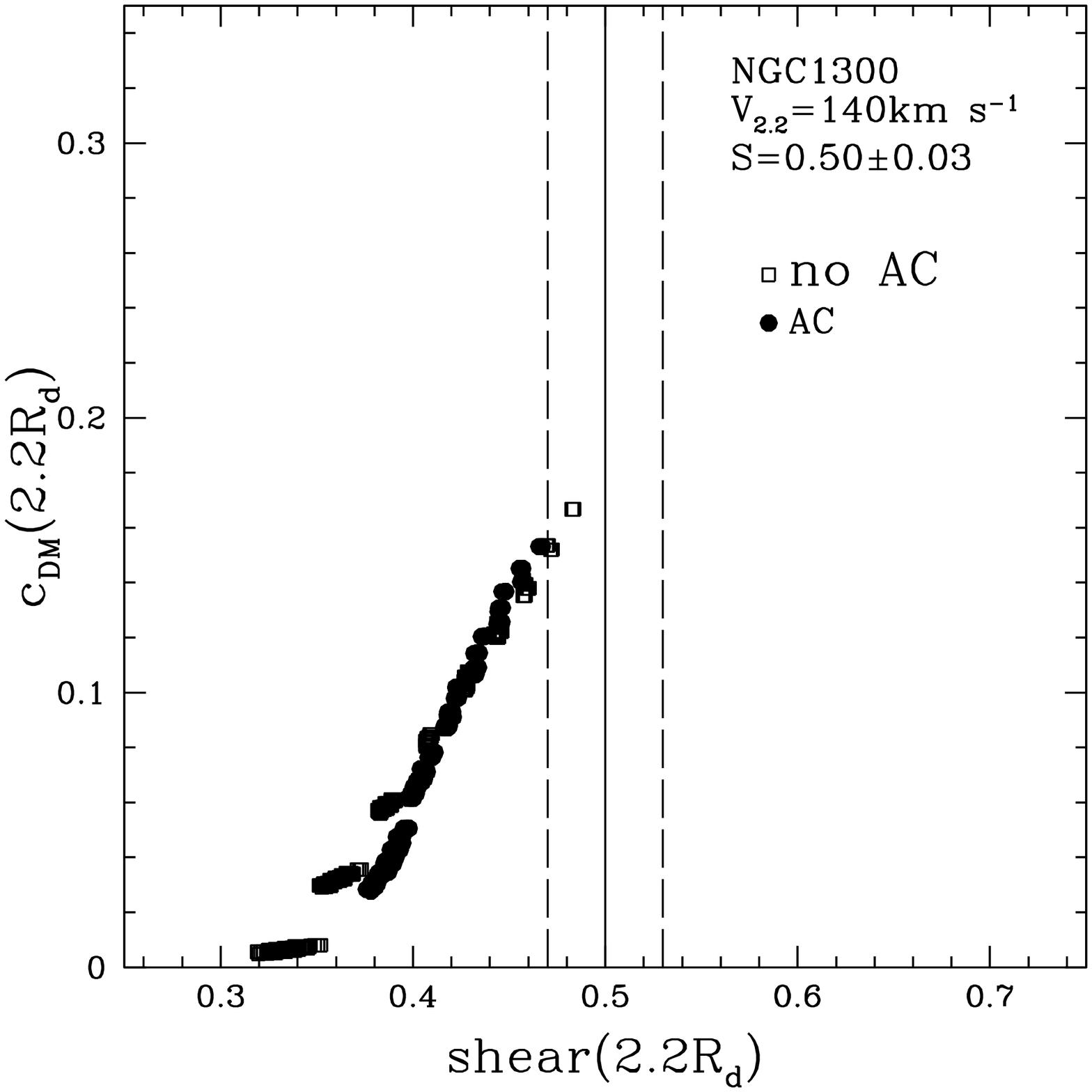}\\
\includegraphics[width=4.1cm]{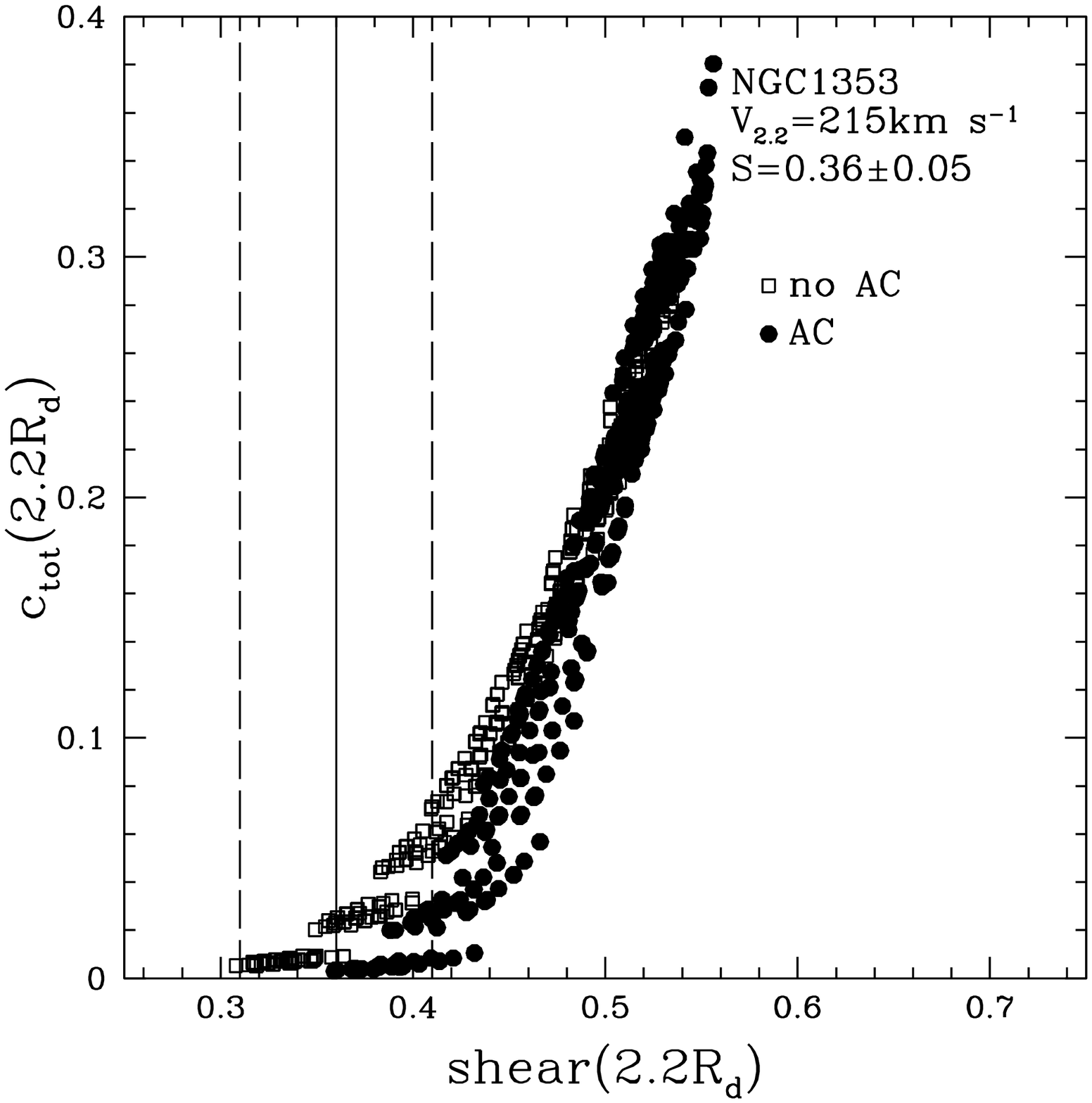}
\includegraphics[width=4.1cm]{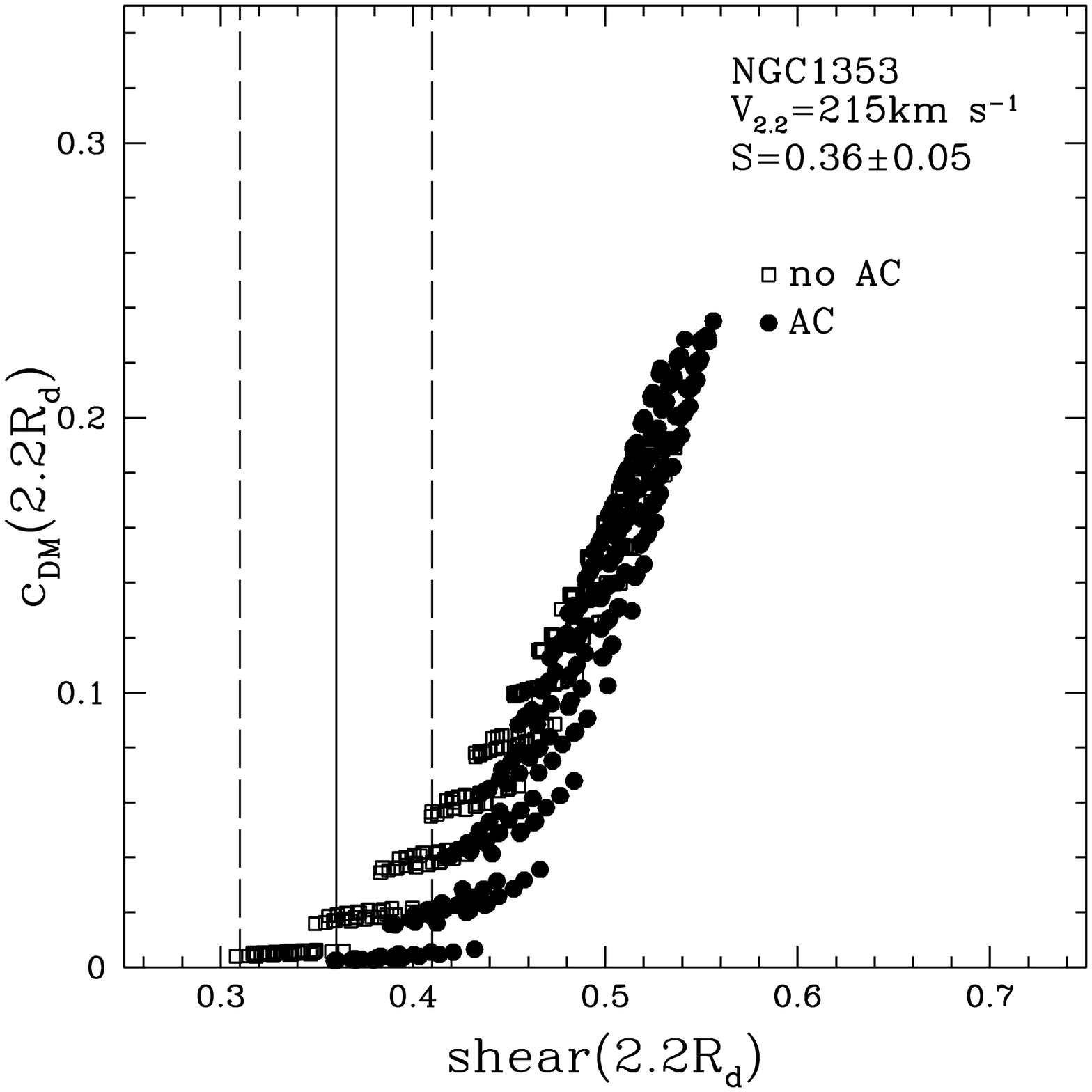}
\includegraphics[width=4.1cm]{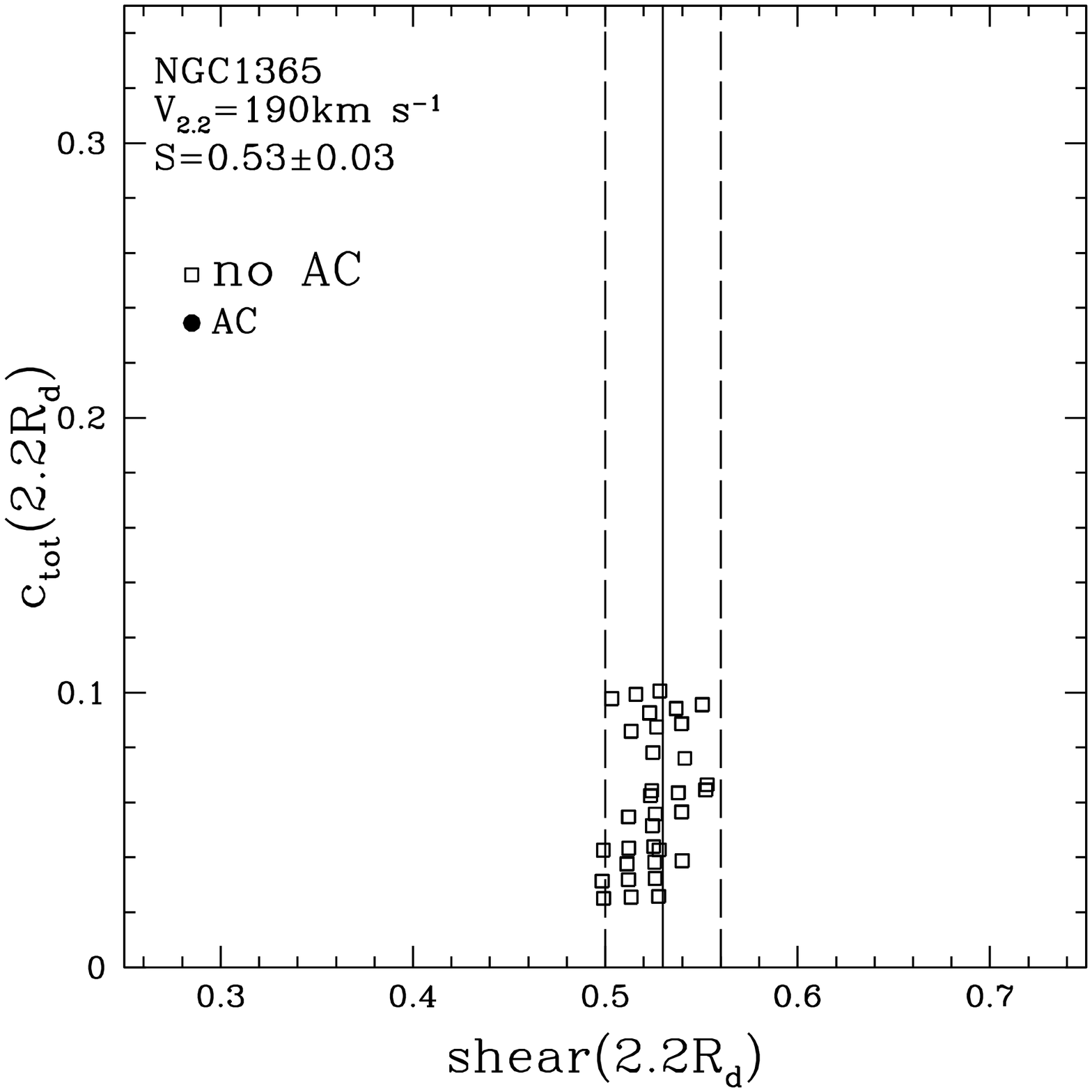}
\includegraphics[width=4.1cm]{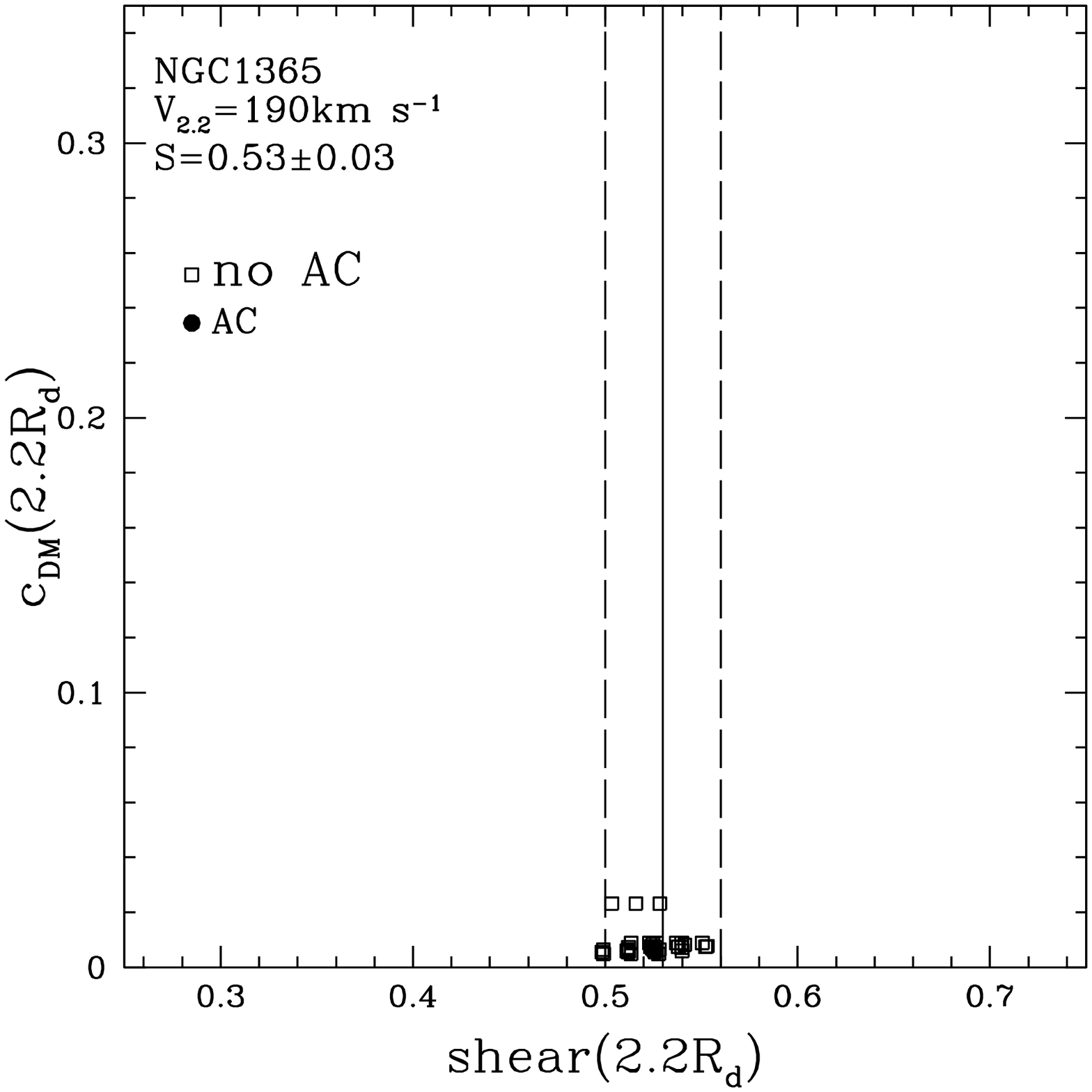}\\
\includegraphics[width=4.1cm]{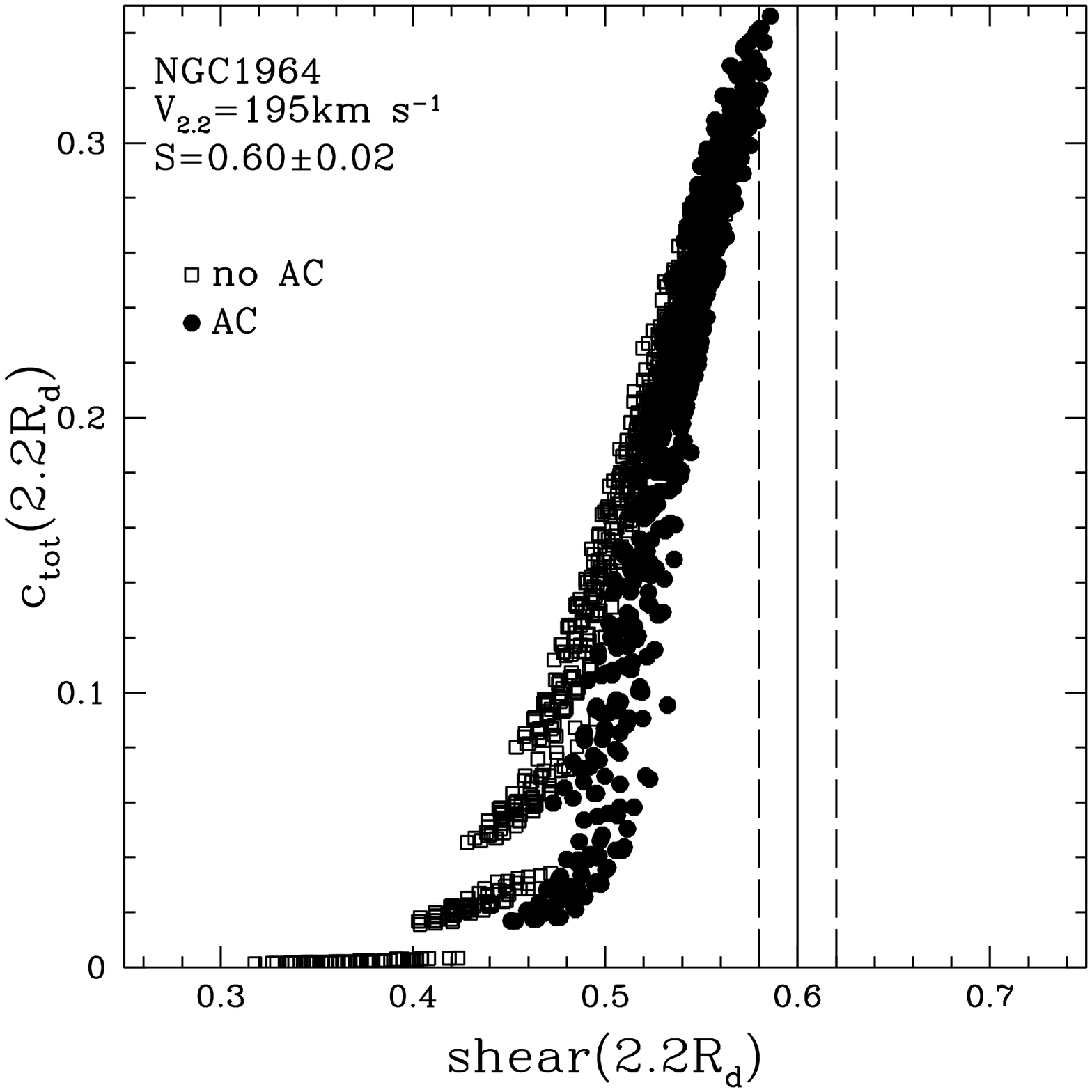}
\includegraphics[width=4.1cm]{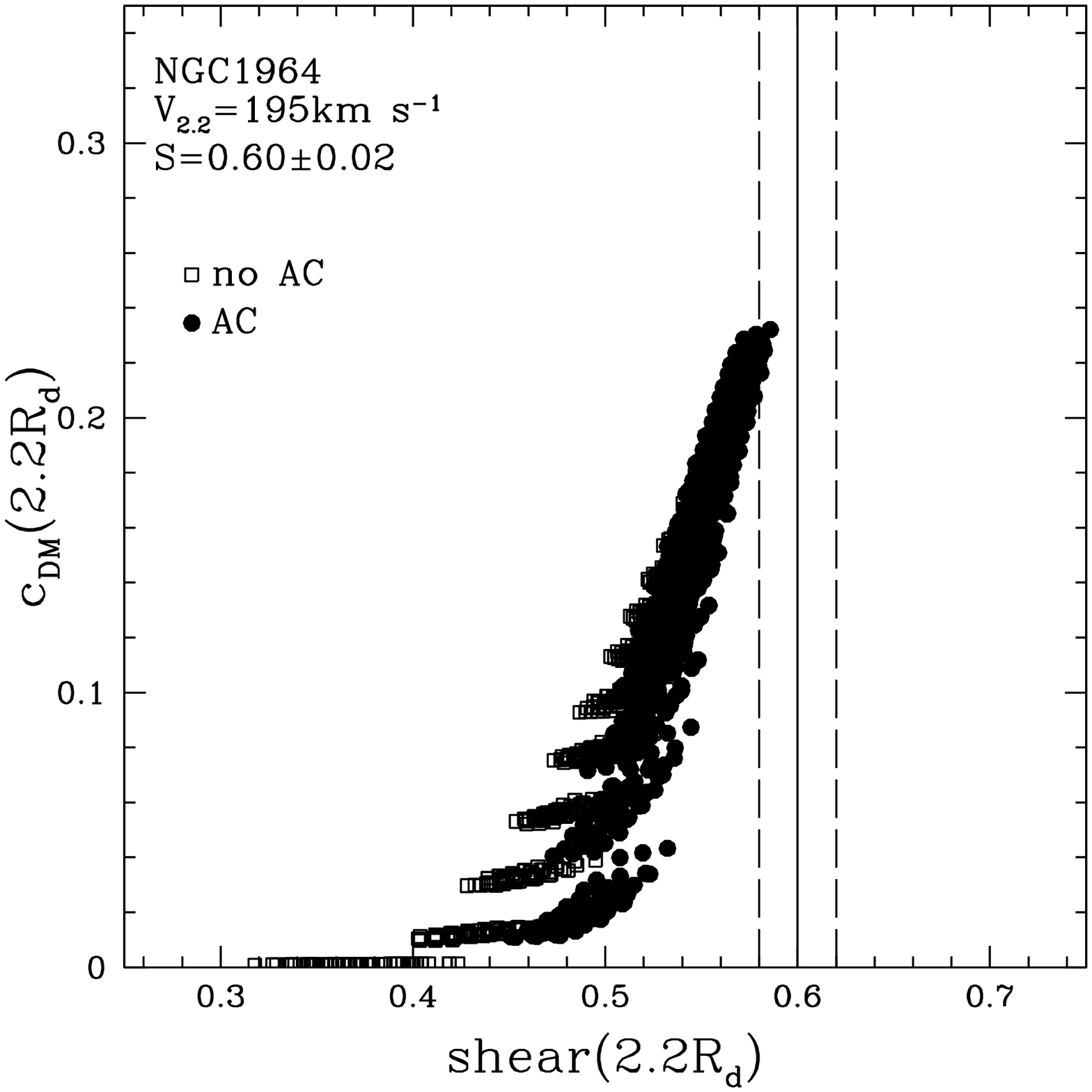}
\includegraphics[width=4.1cm]{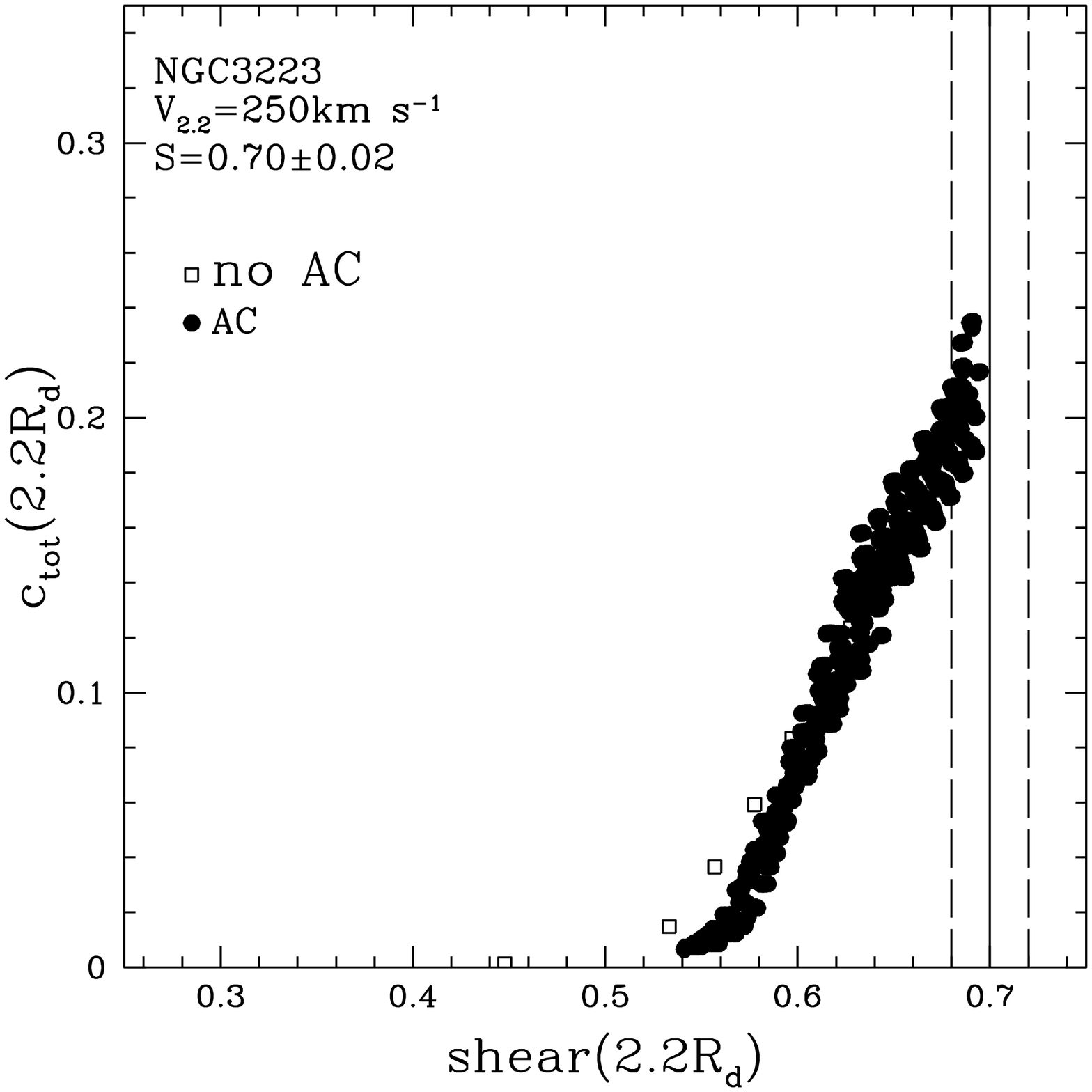}
\includegraphics[width=4.1cm]{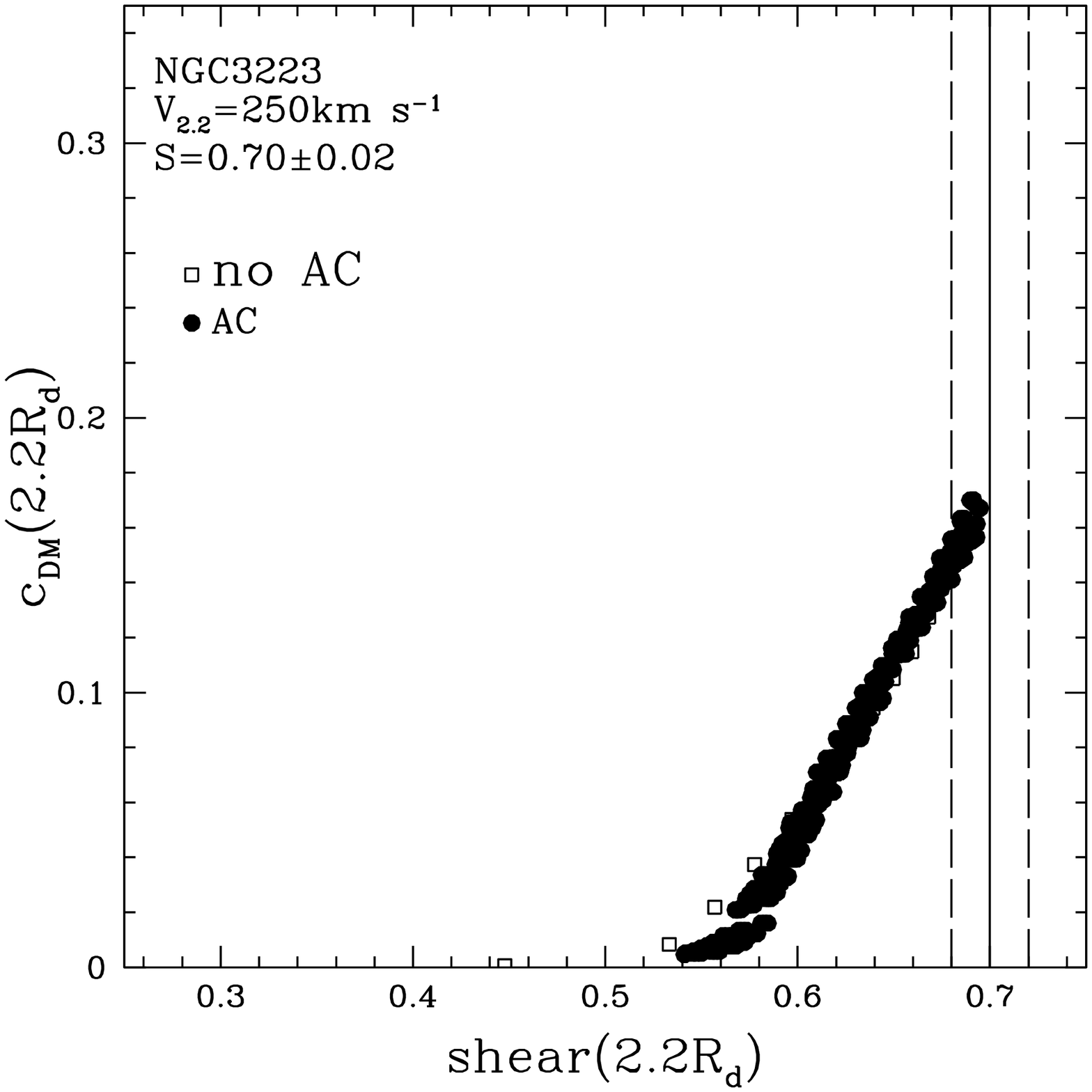}
\end{figure*}

\begin{figure*}[t]
\includegraphics[width=4.1cm]{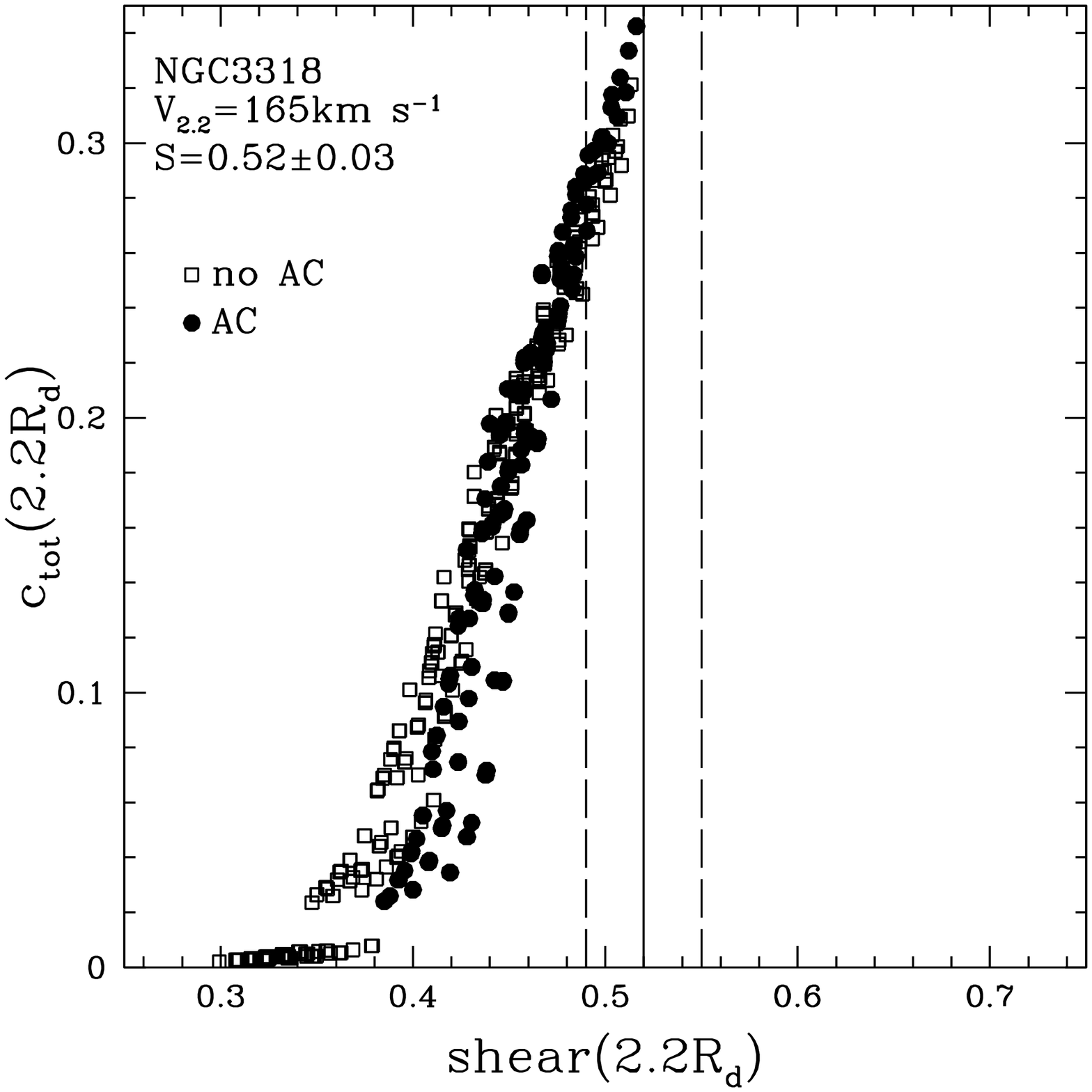}
\includegraphics[width=4.1cm]{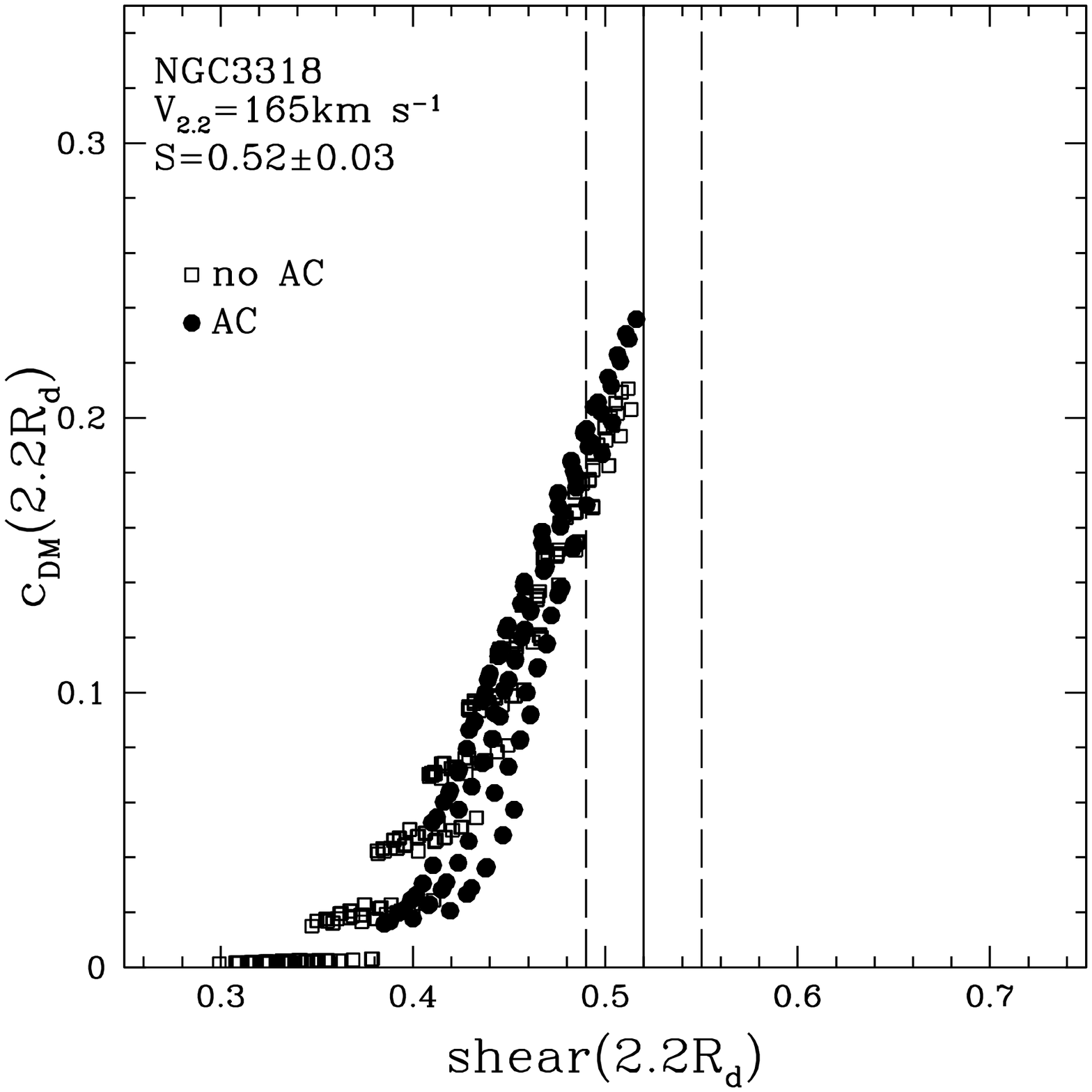}
\caption{Model results for 13 galaxies. {\em Left}: Central mass concentration (mass fraction within 2.2$R_d$) vs.\ shear; {\em right}: dark matter concentration (dark matter mass fraction within 2.2$R_d$) vs.\ shear. The open squares represent the model without adiabatic contraction of the stellar halo, and filled circles represent the model with adiabatic contraction. The vertical lines present the measured shear ({\em solid line}) with the 1$\sigma$ error added and subtracted ({\em dashed lines}).}
\end{figure*}

\begin{figure*}
\label{rcurv}
\includegraphics[width=4.1cm]{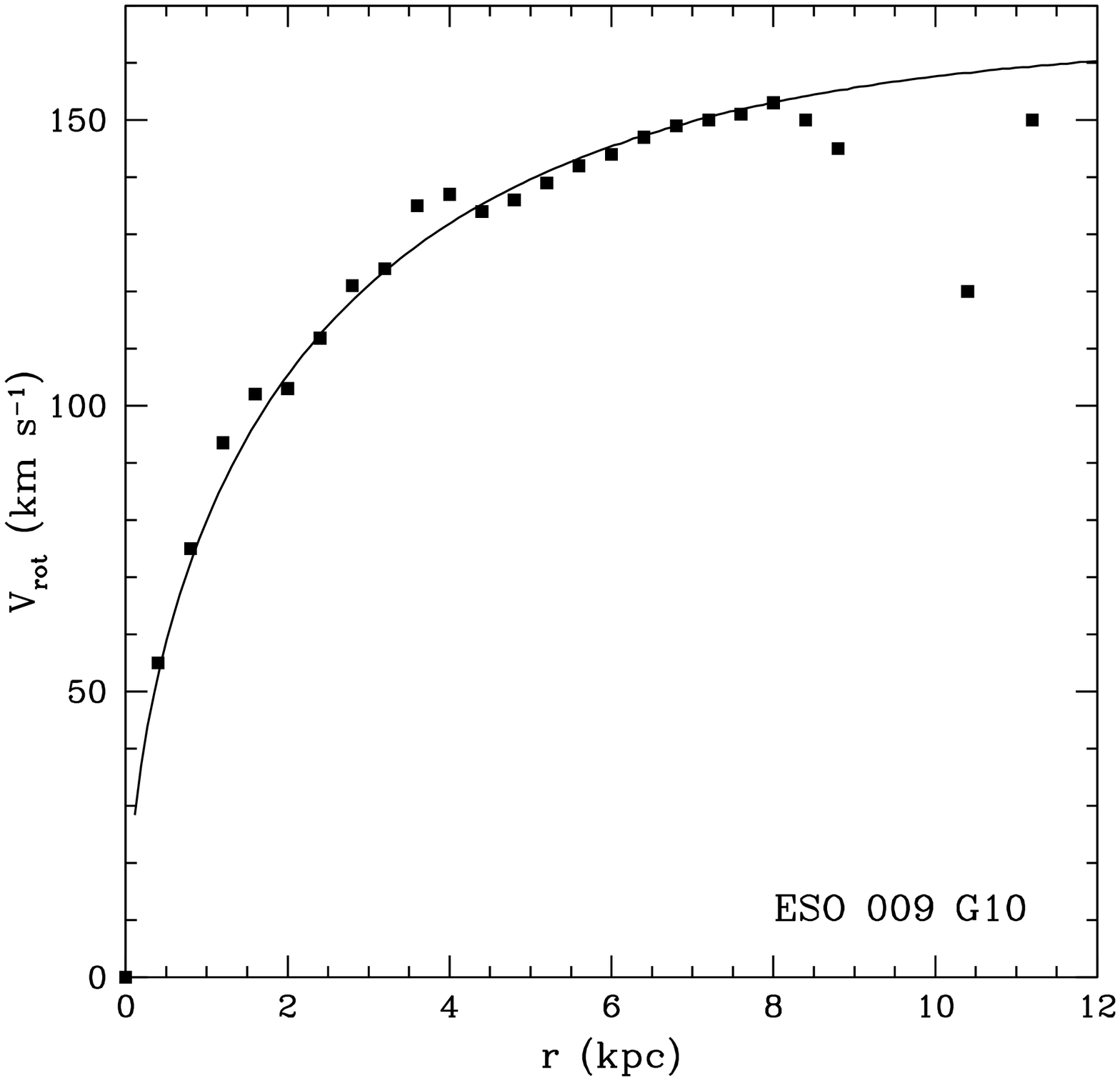}
\includegraphics[width=4.1cm]{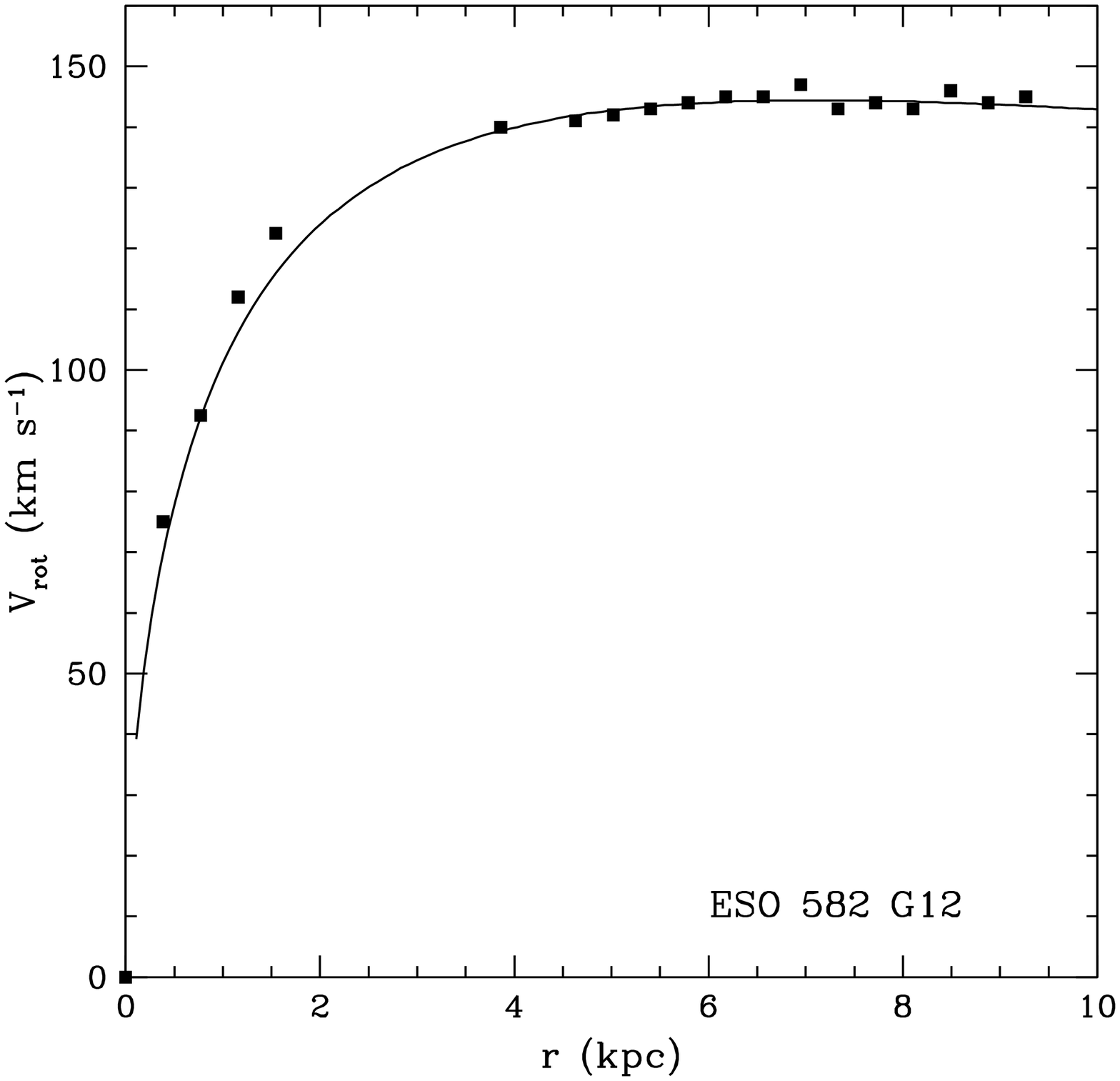}
\includegraphics[width=4.1cm]{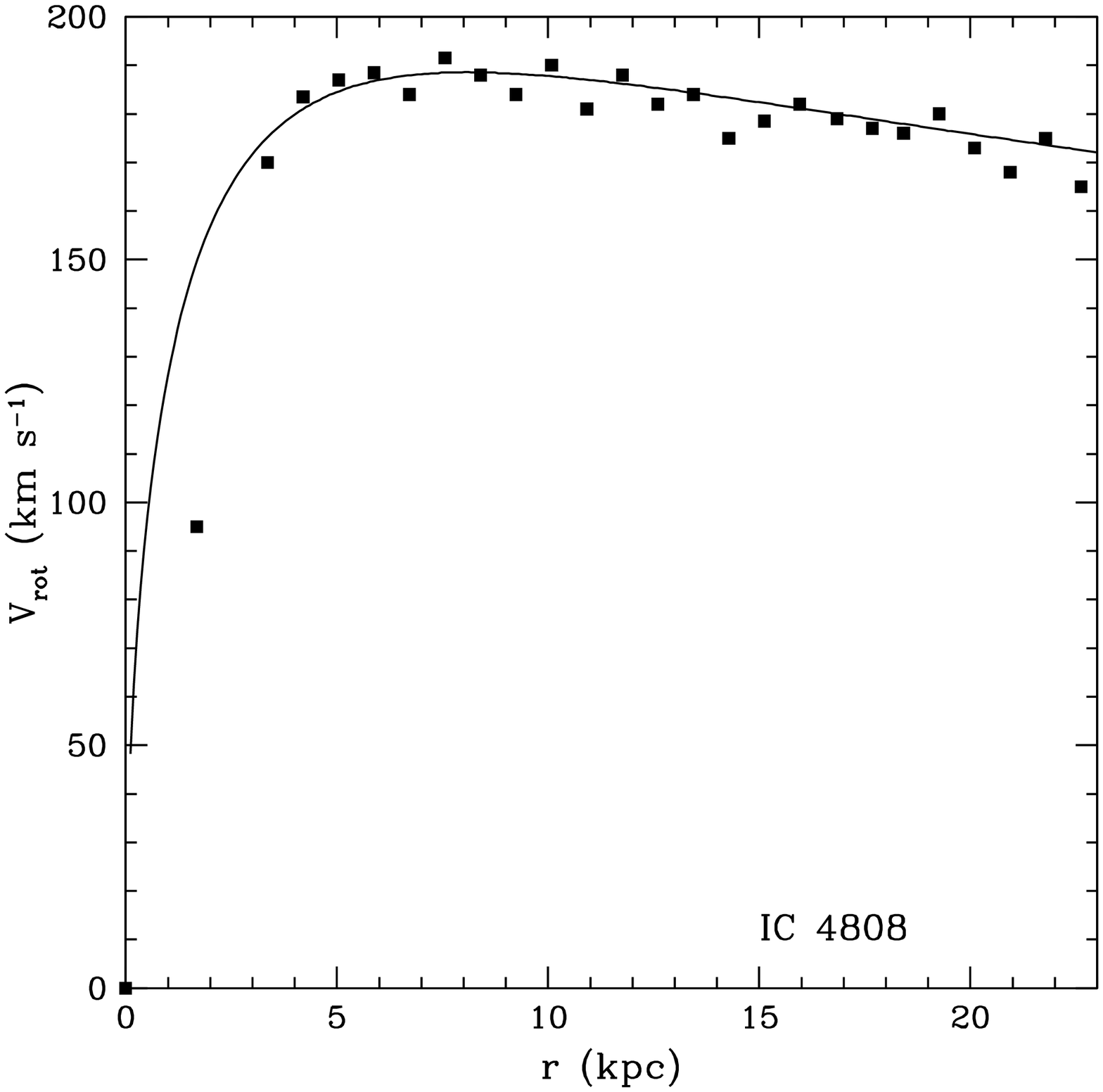}
\includegraphics[width=4.1cm]{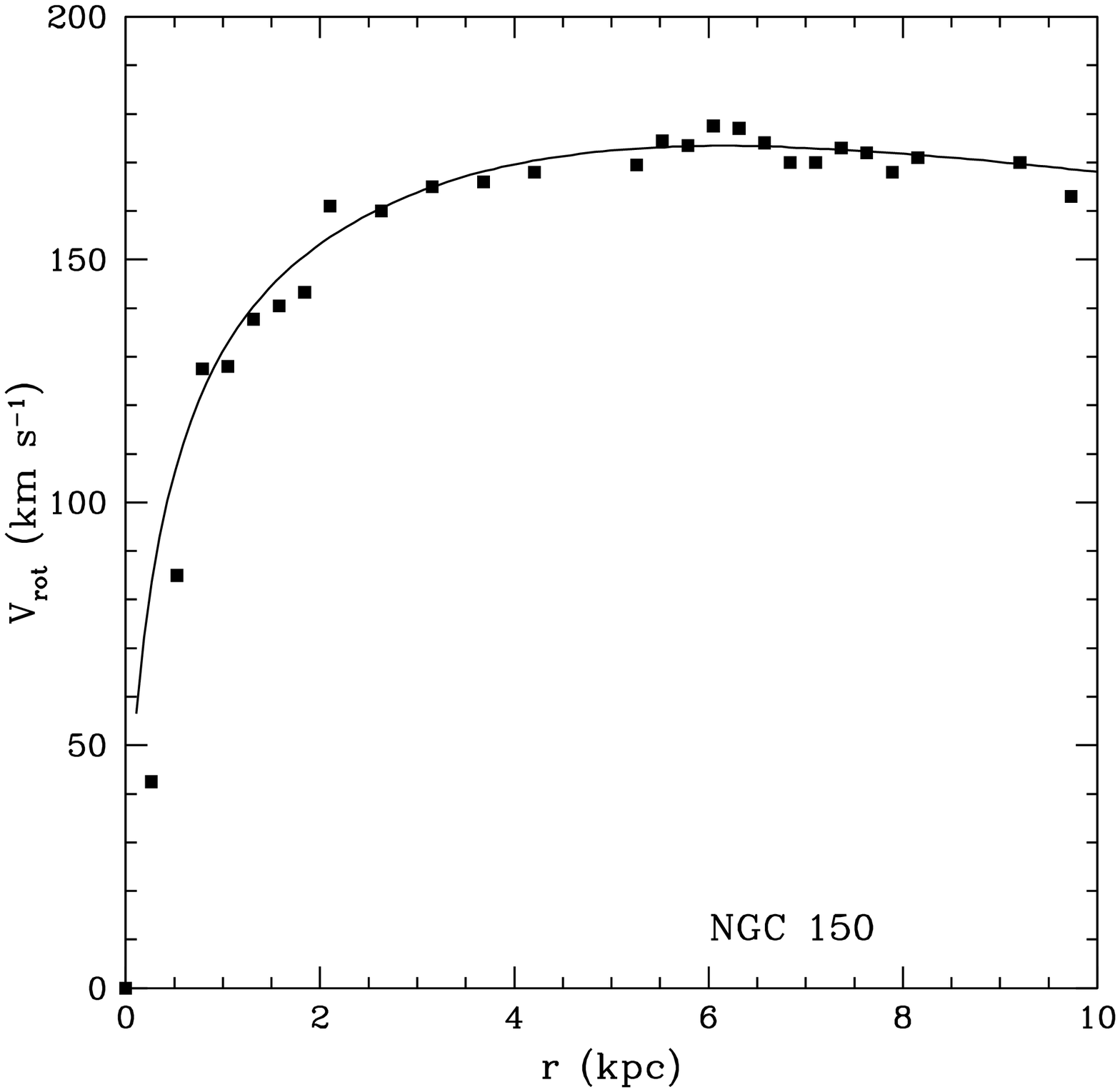}\\
\includegraphics[width=4.1cm]{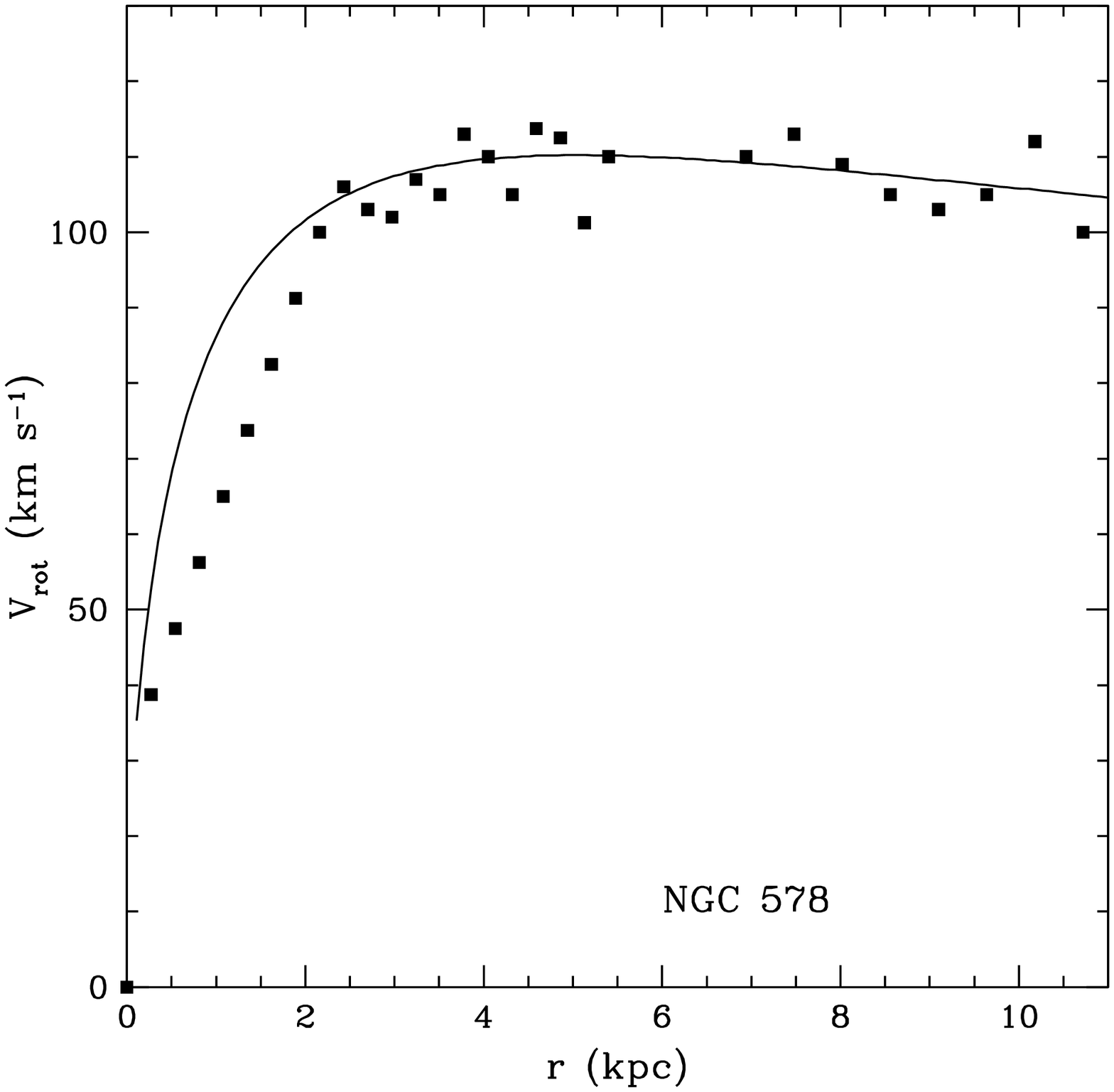}
\includegraphics[width=4.1cm]{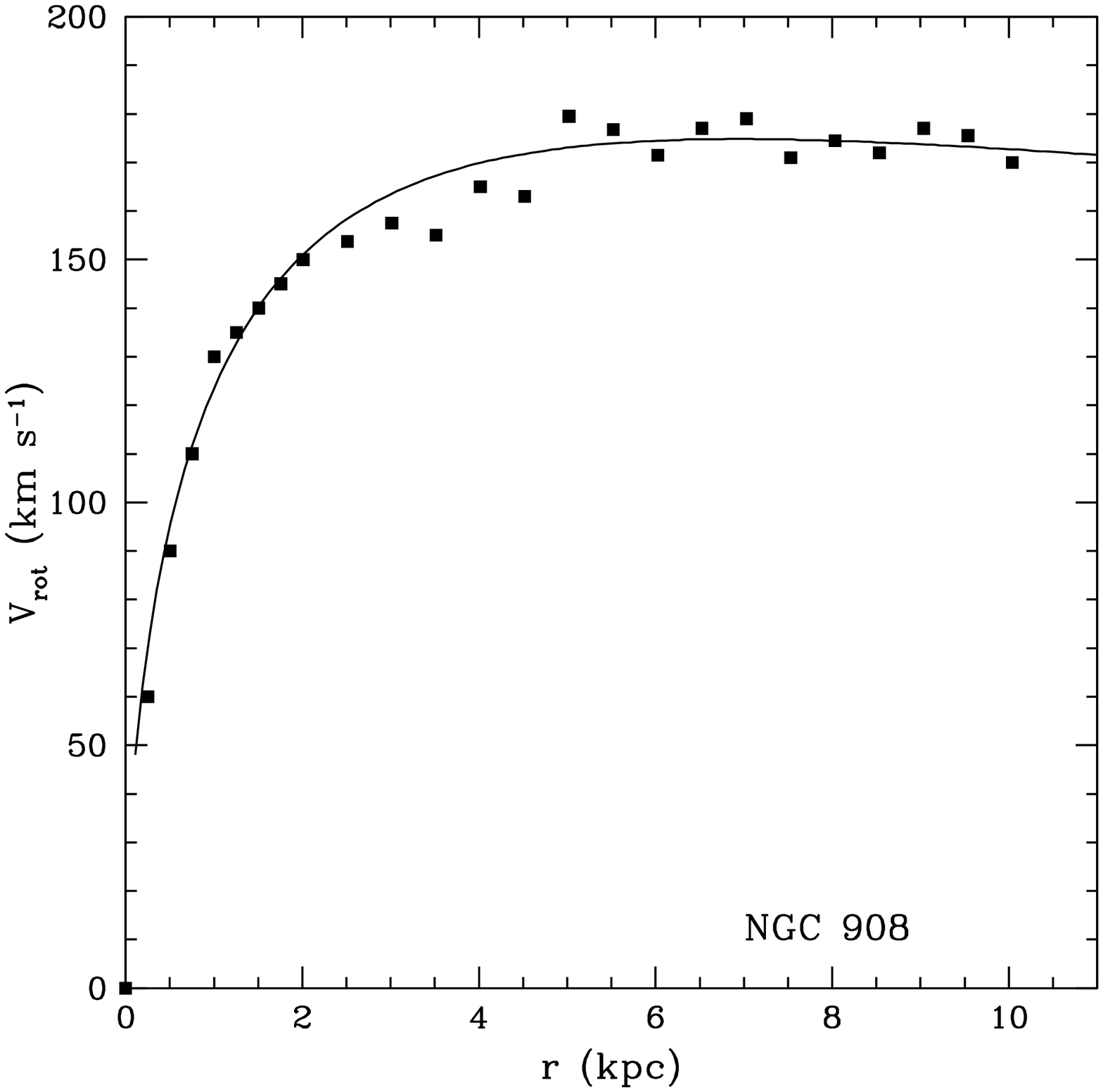}
\includegraphics[width=4.1cm]{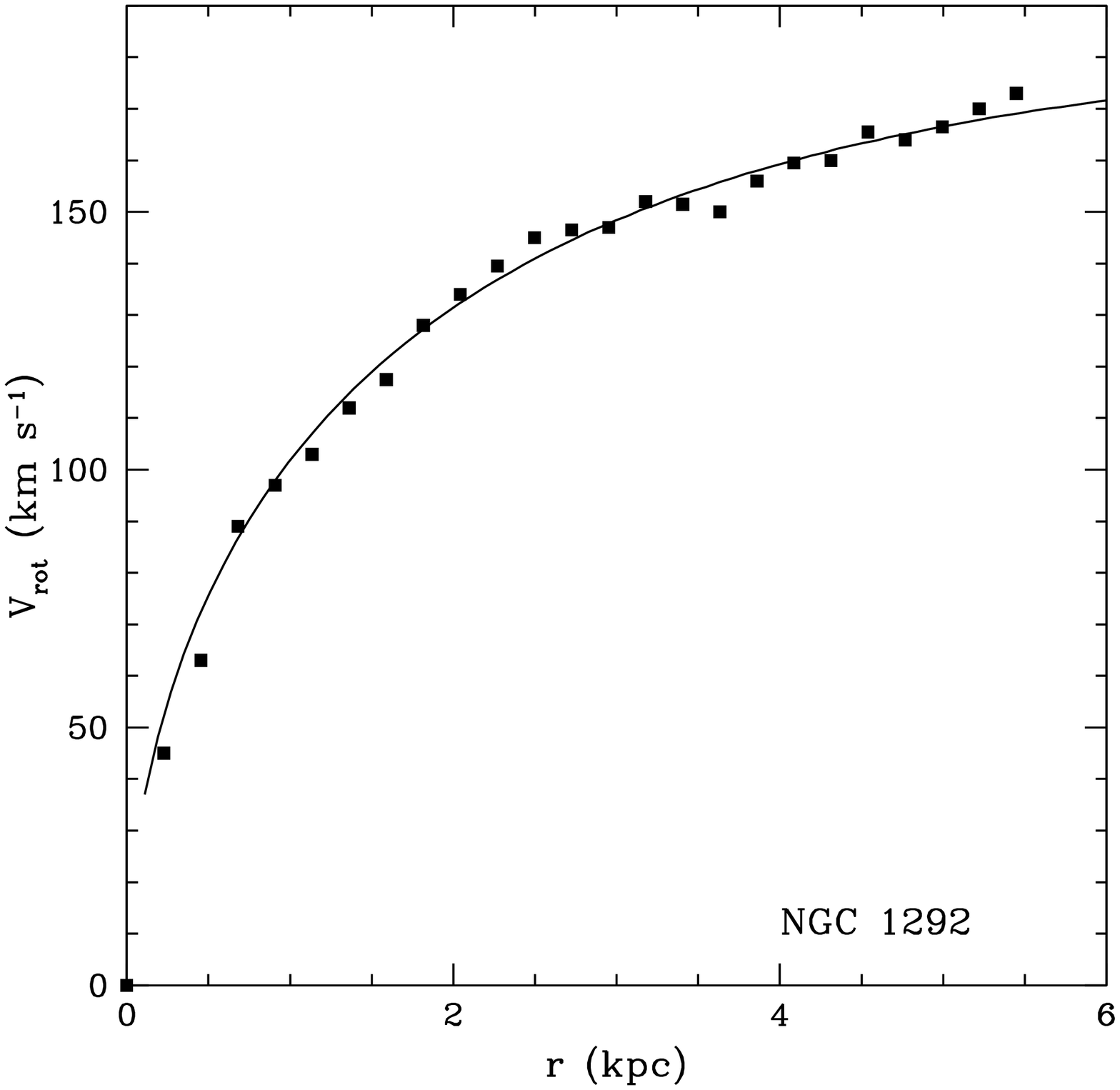}
\includegraphics[width=4.1cm]{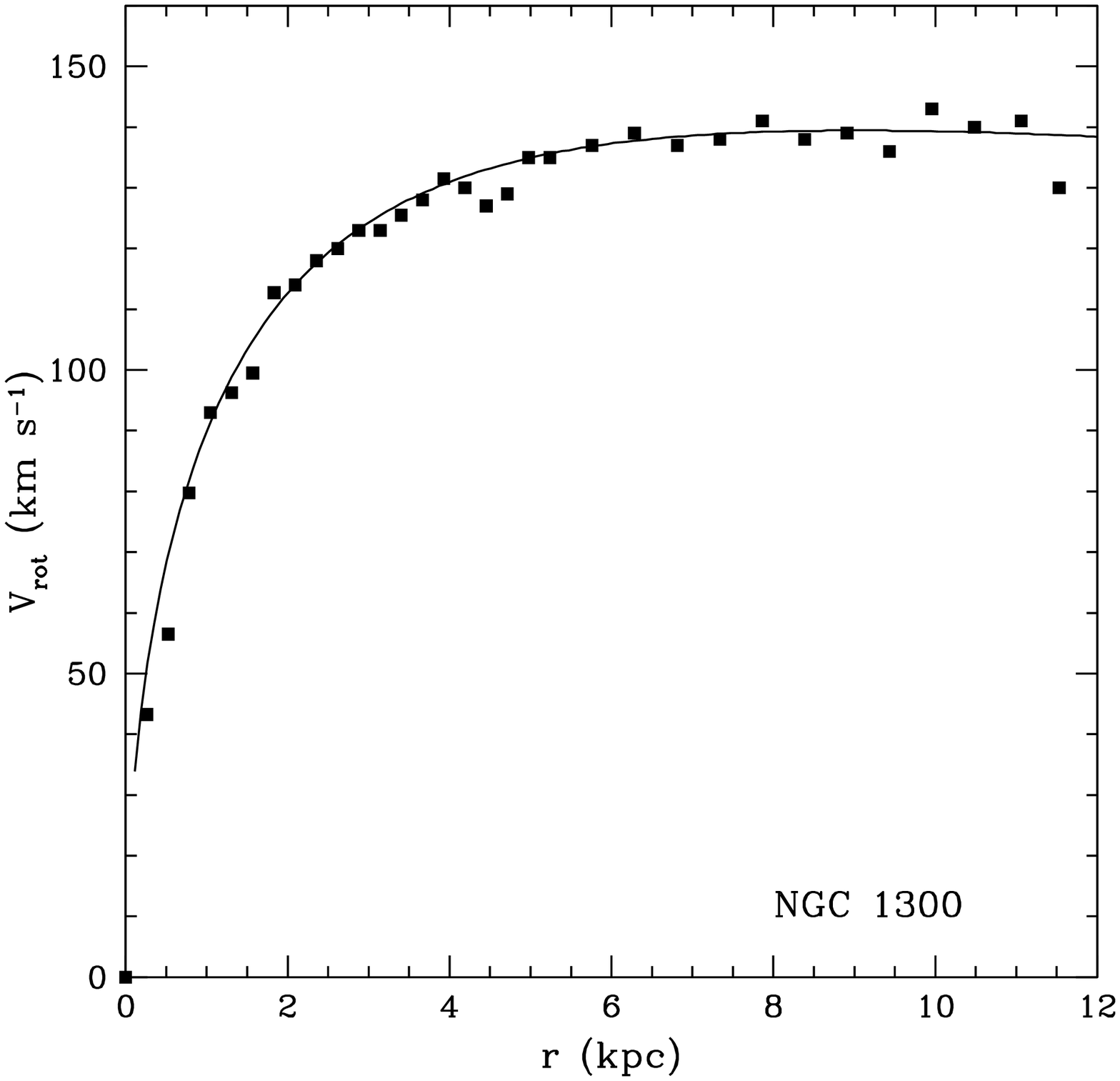}\\
\includegraphics[width=4.1cm]{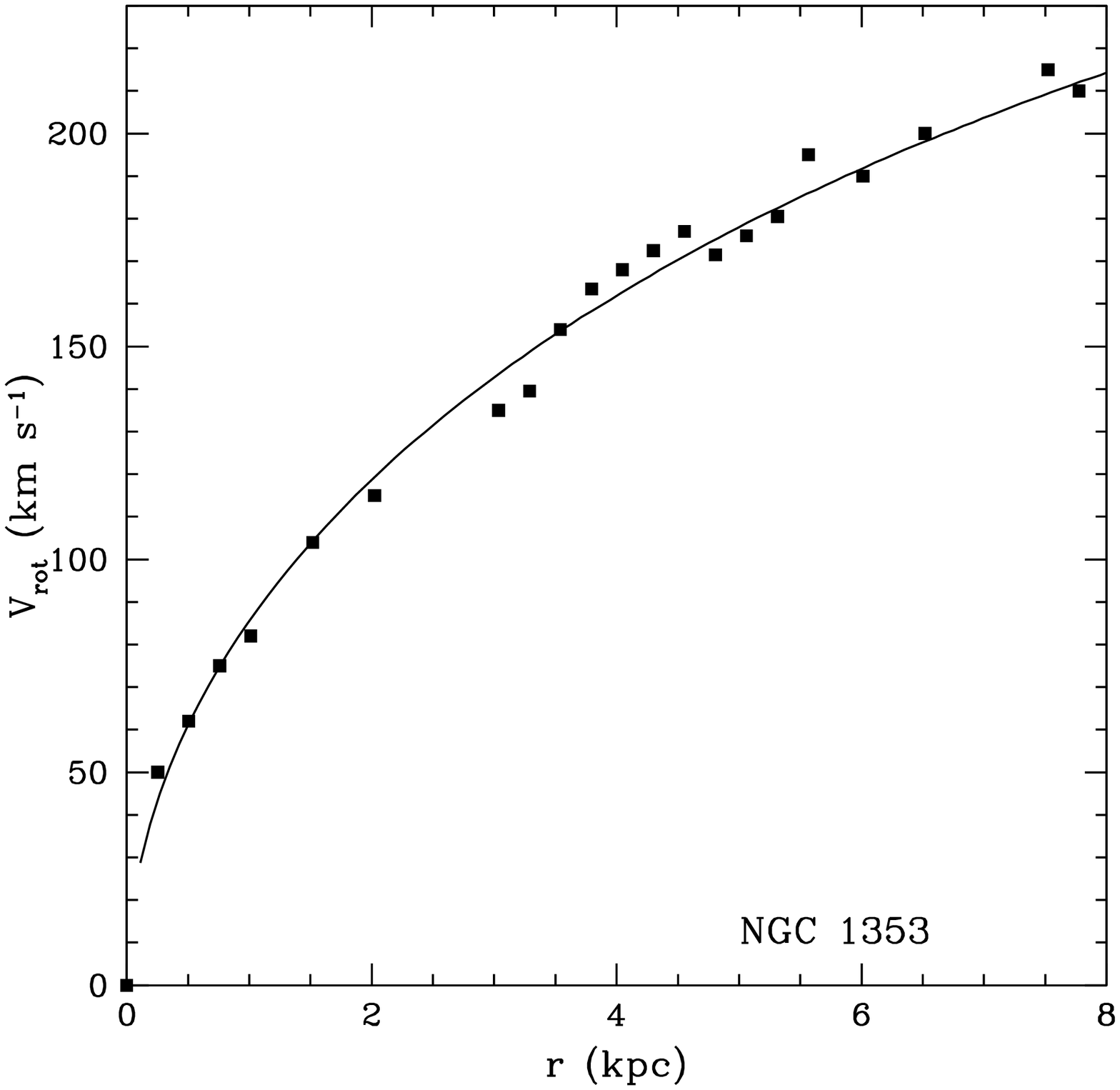}
\includegraphics[width=4.1cm]{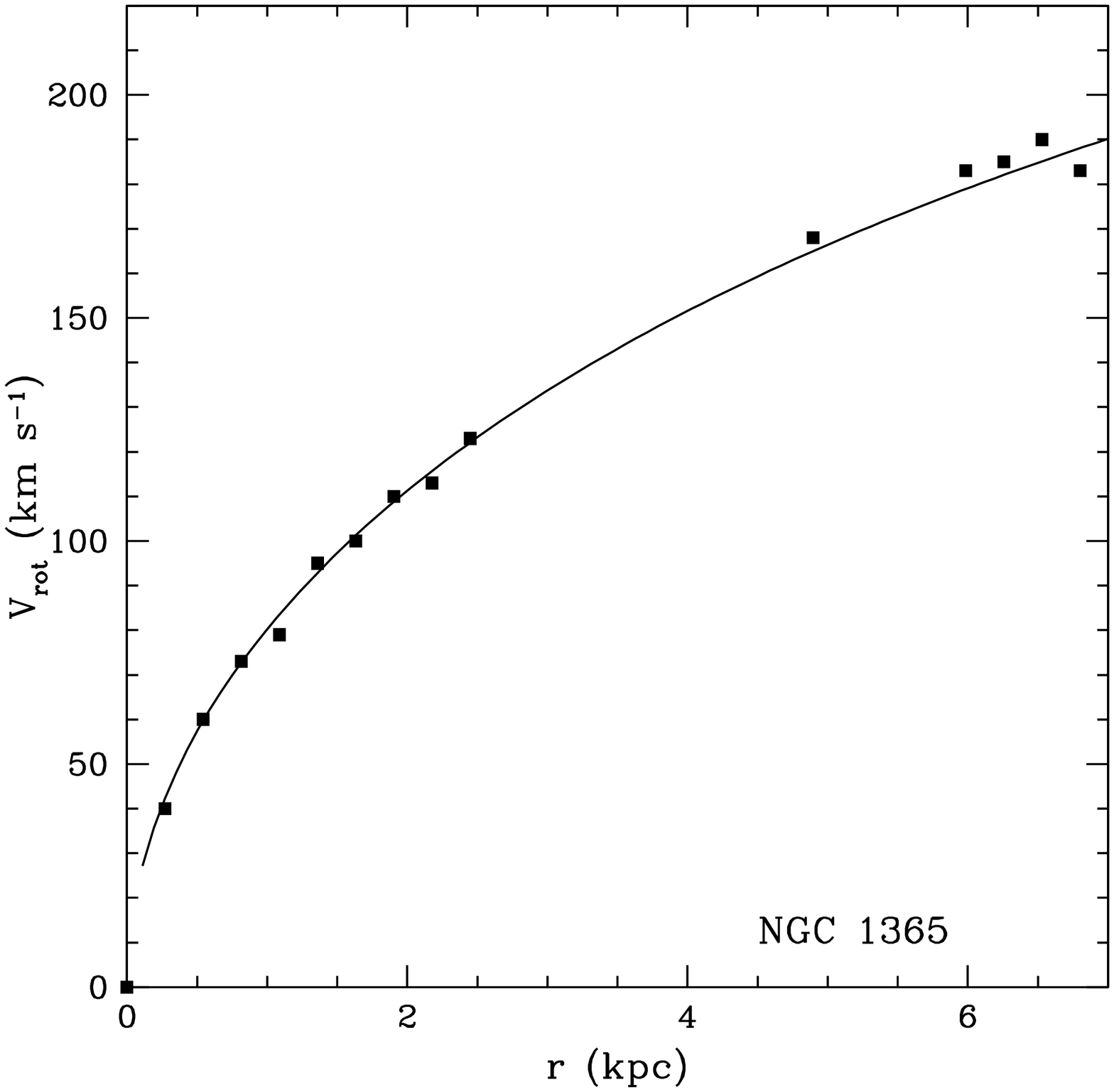}
\includegraphics[width=4.1cm]{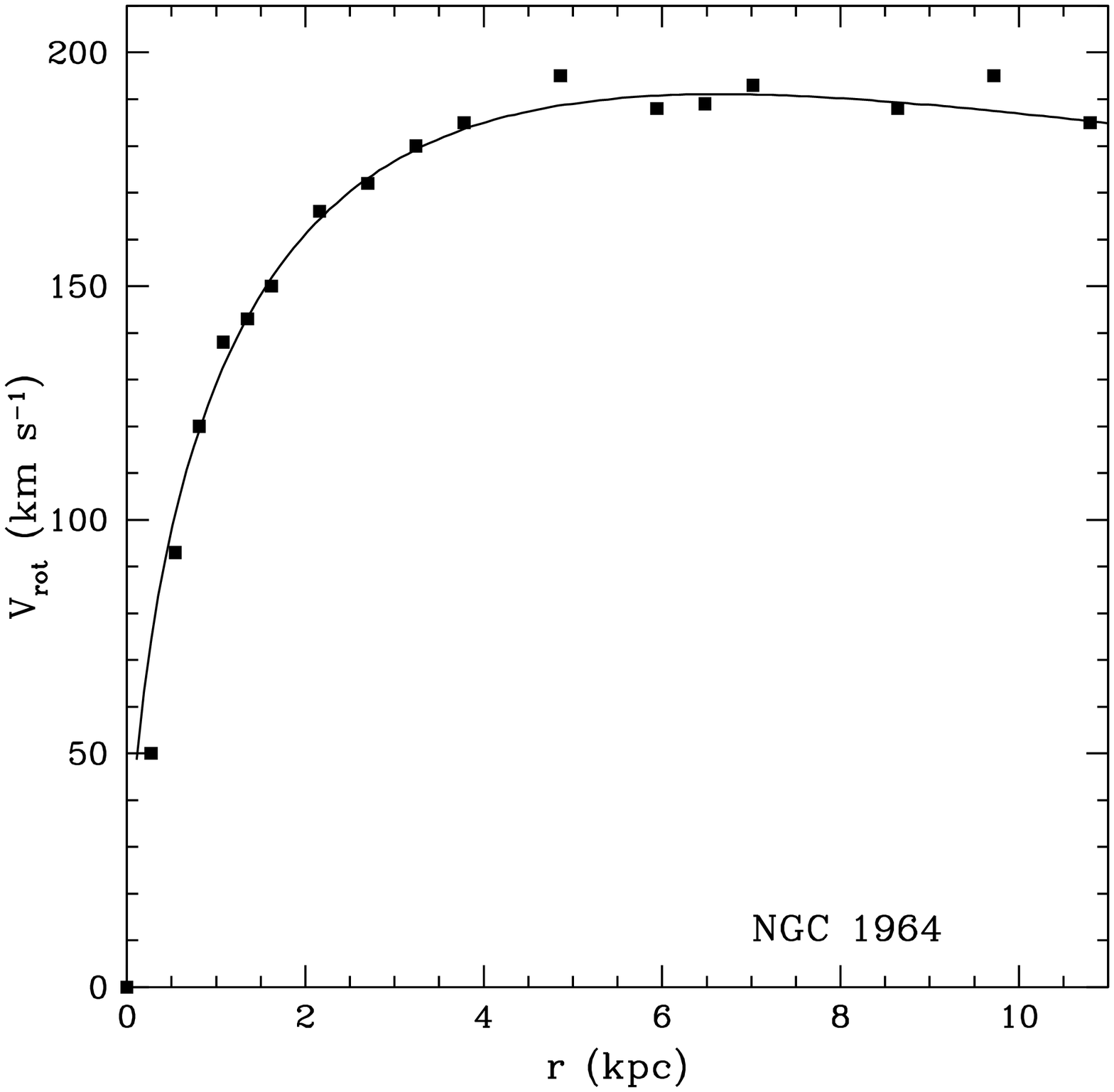}
\includegraphics[width=4.1cm]{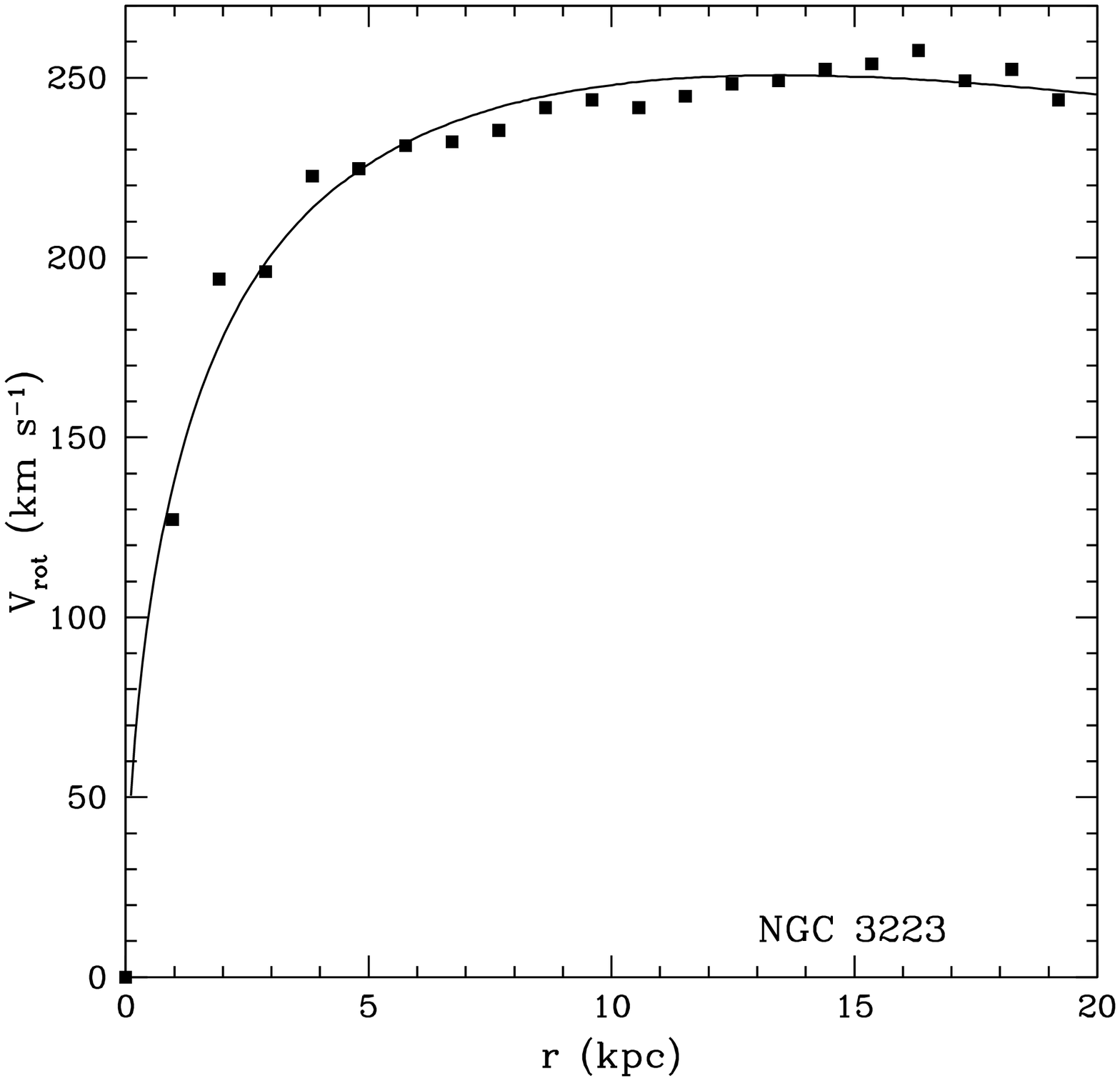}\\
\includegraphics[width=4.1cm]{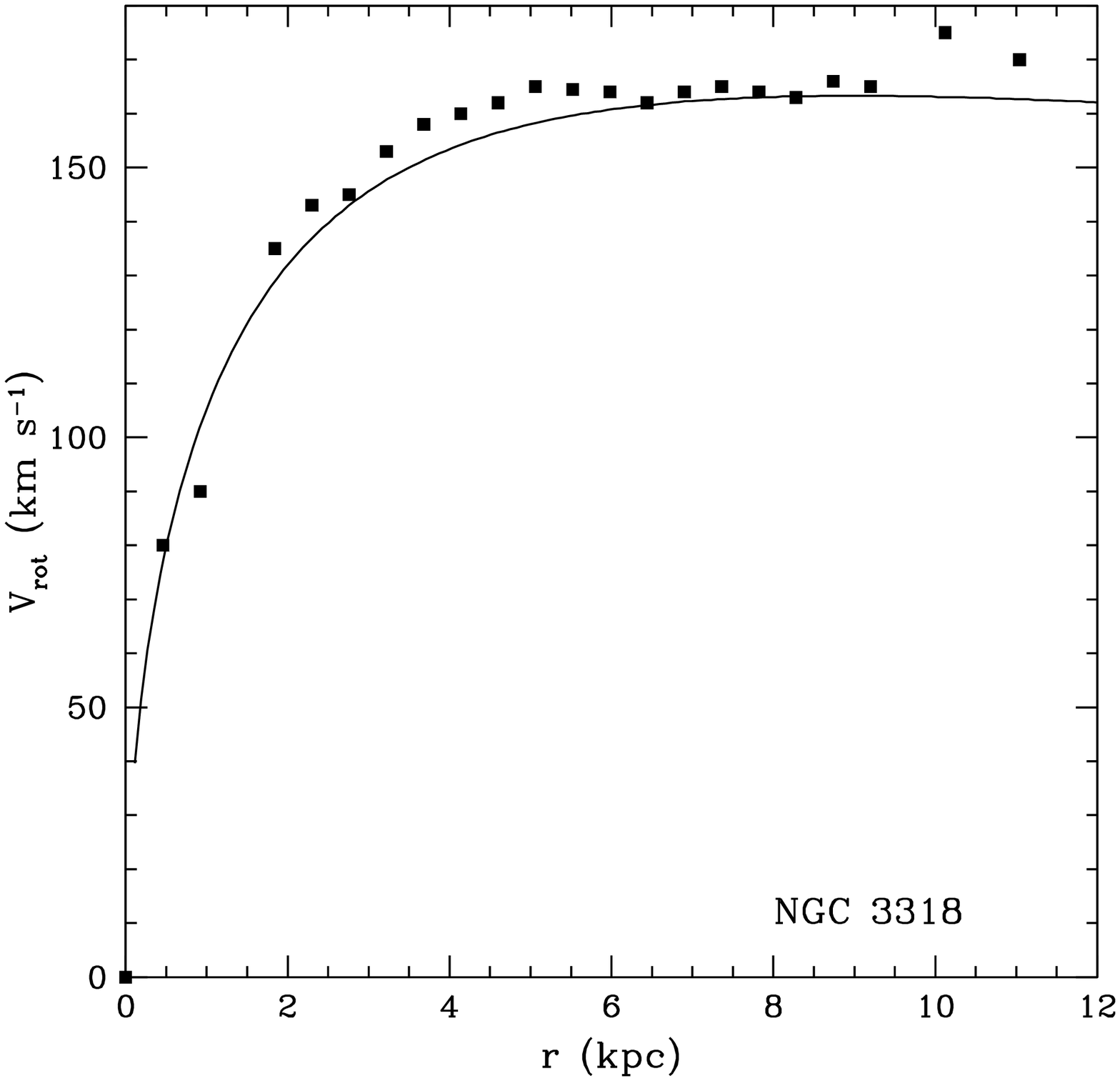}
\caption{Observed rotation curves with the overlaid model rotation curves for 13 galaxies. The errors on the data points are typically $<10$\% (Persic \& Salucci 1995). The solid lines represent the rotation curve that best matches the observed shear at 2.2$R_d$.}
\end{figure*}

Before we proceed to perform such an analysis (which will actually be presented
in a forthcoming paper), we first proceed with a further test of the modeling
presented here.  Since the sample of galaxies we present here have observed
H$\alpha$ rotation curves, 
an important test of our modeling is how well the modeled
rotation curves fit the data.  In this case, we take the no-AC models that
describe the shear most accurately (i.e., those with a combination of
$R_d$, $L_{\rm disk}$, and $M/L$ that accurately predict the shear as determined
from Figures 3 and 4), and find which model best fits the observed H$\alpha$ 
rotation curves from Persic \& Salucci (1995).  We have chosen to work with
only our no-AC models here as we could not rule out one type of model over the
other (i.e., AC model versus no-AC model), and for consistency we chose
the pure NFW model here (i.e., the no-AC model).  Furthermore, several papers
(e.g., Kassin et al.\ 2008a, b) have shown that AC rarely operates, and that
rotation curves can be described without the need for AC.  Nevertheless some
examples (e.g., M31) have been shown to require AC (e.g., Klypin et al.\ 2002;
Seigar et al.\ 2008a).  The virial mass and NFW concentrations resulting from
this are listed in Table 3.

The results of our model fits to the observed data are presented in
Figure 5, which shows the Persic \& Salucci (1995) H$\alpha$ rotation
overlaid with our modeled
rotation curves. In all cases the rotation curve is modeled well
in the outer parts, where we are measuring the shear.  Furthermore, in all but
one case (NGC 578), the inner parts of the rotation curve are modeled extremely
well.

Given the further constraints placed on the NFW concentration parameter
($c_{\rm vir}$) by finding the model that best fits the observed rotation curve
(figure 5), we can now compare this parameter with other properties of the
galaxy.  Recent results have suggested that there may be a link between
the mass of supermassive black holes (SMBHs) in galaxy nuclei and the dark
halo concentration parameter.  For example, Booth \& Schaye (2009) explored
a potential connection between SMBH mass and dark matter halos using a series
of simulations.  Treuthardt et al.\ (2012) developed this further using a
combination of imaging data and simulated galaxies, but could not come to 
any strong conclusions due to a lack of dynamical information.  Furthermore,
observations of AGN (active galactic nuclei) in galaxies with little or no
bulge, suggest that the dark matter halo concentration may play a role in the
determination of SMBH mass (Satyapal et al.\ 2007, 2008).  These arguments
have been used to explain the recent discovery of a correlation between
SMBH mass and spiral arm pitch angle in disk galaxies (Seigar et al.\ 2008b;
Berrier et al.\ 2012).  If the fundamental factor in determining SMBH mass
is dark matter halo concentration (e.g., Ferrarese 2002; Satyapal et al.\ 2007,
2008), then all of the observed relationships naturally follow.  Nevertheless
it is important to point out that Kormendy \& Bender (2011) have provided
arguments that suggest that the dark matter concentration does not have any
direct correlation with the properties of the SMBH.  However, their paper did
not take into account the theoretical work of Booth \& Schaye (2009), which
suggests that the SMBH growth is regulated by the dark matter halo 
concentration.  Shortly after the publication of Kormendy \& Bender (2011)
further observational constraints provided evidence that the SMBH mass is
correlared with properties of the dark matter halo (Volonteri et al.\ 2011).

\begin{figure}
\label{pitchc}
\includegraphics[width=7.6cm]{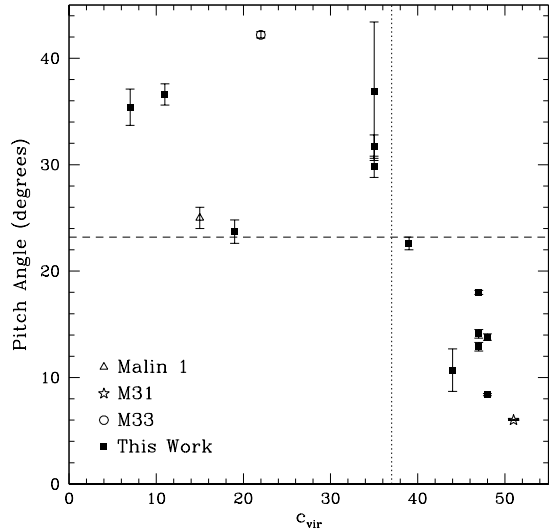}
\caption{Spiral arm pitch angle vs.\ NFW concentration.  The open triangle represents Malin 1 (Seigar 2008), the star represents M31 (Seigar et al. 2008), the open circle represents M33 (Seigar 2011) and the filled squares represent the current sample.}
\end{figure}

Given the apparent controversy over the link between SMBH mass and dark matter
concentration, we realized that the parameters we had derived could be used
to shed some light on this question.  Using the concentration parameters
derived by fitting our models directly to the observed rotation curves, we 
tested the correlation between NFW concentration ($c_{\rm vir}$) and the 
spiral arm pitch angles of these galaxies, using the fact that pitch angle can
be used as a proxy for SMBH mass (Seigar et al.\ 2008b; Berrier et al.\ 2012).
The result can be seen in Figure 6.

Note that the NFW concentrations here are derived from our no-AC models, as
described in the direct rotation curve fitting above in section 4.  Figure 6
also includes data for M31 (Seigar et al.\ 2008), M33 (Seigar 2011), 
and Malin 1 (Seigar 2008).  Although these authors report that an AC model
works best for Malin 1 and M31, for consistency, here we use the concentration
parameter for the no-AC model that they used.  In the case of M33, the authors
only report a pure NFW concentration.

Figure 6 does not show a strong correlation between spiral arm pitch angle
(or SMBH mass) and the NFW concentration parameter.  However, there does
appear to be a trend whereby galaxies with very open spirals (i.e., with a 
pitch angle, $P>23^{\circ}$) seem to have lower concentrations than those with
tightly wound spirals (with $P<23^{\circ}$).  There is a word of caution 
however.  The cutoff for the concentration parameter corresponding to this 
pitch angle is $c_{\rm vir} \simeq 37$.  For the eight galaxies with 
$P<23^{\circ}$, the concentration is in the range $37 \leq c_{\rm vir} \leq 51$.
These are extremely high concentrations.  This range corresponds to a
$\sim 3\sigma - \sim 4.4\sigma$ outliers from the expected concentration
distribution for large spiral galaxies (e.g., Bullock et al.\ 2001a).
Indeed, high concentrations such as these have been used as an argument against
a pure NFW dark matter density profile, and instead it has been suggested
that in these cases, an adiabatically contracted profile should be
adopted (see for example the case of M31 in Seigar et al.\ 2008a).  Indeed,
if we chose to adopt our AC models in these cases (i.e., for tight spirals
or high concentrations), Figure 3 shows that there is a tendency for the
NFW concentration to decrease.  One should note that this is the initial
NFW concentration, but it is in better agreement with the expected distribution
of $c_{\rm vir}$ as given by Bullock et al.\ (2001a), although the halo
does become more concentrated due to AC.

\section{Summary and concluding remarks}

In this paper we have examined {\em Spitzer IRAC} 3.6-$\mu$m images of 13
galaxies that have H$\alpha$ rotation curves in the literature.  Using these
galaxies we have shown that:
\begin{itemize}
\item The relation between spiral arm pitch angle and rotation curve shear still exists, although with a slightly larger scatter than had been presented in Seigar et al.\ (2006).
\item Using shear (or pitch angle as a proxy for shear) as well as a maximum rotation velocity (or galaxy luminosity as a proxy) we can constrain various parameters related to the dark matter halo (e.g., virial mass, NFW concentration, etc.) and parameters relating to the total mass concentration of these galaxies.
\item There appears to be a weak trend between spiral arm pitch angle and NFW concentration parameter, whereby galaxies with tightly wound arms (i.e., $P<23^{\circ}$) have particularly high concentrations.  For these galaxies we suggest that AC may play a role, which would help to bring their dark matter concentrations down into a range that is more consistent with dark matter halo simulations for spiral galaxies.
\end{itemize}

It should be noted that the role of the baryon-dominated central spheroidal component (i.e., the bulge) has remain untouched in this paper.  However, we have shown that we can derive useful dynamical information from imaging data alone 
(we can get a maximum rotation velocity from the galaxy luminosity, and we 
can get the slope of the rotation curve --- or shear --- from a spiral arm
pitch angle) and  in future papers we intend to apply this method to a much larger sample of 
galaxies.  Some forthcoming projects we have in mind are as follows:
\begin{itemize}
\item In a recent paper by Davis et al.\ (2014), we defined a volume-limited
sample of Southern Hemisphere 
spiral galaxies with a limiting absolute B band magnitude of 
$M_B=-19.528$ and a redshift limit of $z=0.0068$.  We now intend to measure
the 3.6-$\mu$m spiral arm pitch angles of these galaxies, and determine
their mass concentrations using the methods described here.  We will also investigate whether chosing one exponential scalelength, $R_d$, rather than $2.2R_d$ for the radius at which shear is measured, significantly affects the best-fitting dark matter halo concentrations ($c_{\rm vir}$) and virial masses ($M_{\rm vir}$).  This may give us some handle on the affect of the bulge component, and therefore tell us something about whether the pitch angles and SMBH masses depend on $c_{\rm vir}$ or a combination of $c_{\rm vir}$ and the bulge mass.
 This work will be
published in Paper 3 of this series.
\item A fourth (and final) paper in this series will measure spiral arm 
pitch angles of disk galaxies in the GOODS-S and -N fields.  This will result
in a determination of mass concentrations in galaxies as a function of 
lookback time.
\end{itemize}

\acknowledgments

This research has made use of the NASA/IPAC Extragalactic Database (NED) 
which is operated by the Jet Propulsion Laboratory, California Institute 
of Technology, under contract with the National Aeronautics and Space 
Administration.
The authors wish to thank Paolo Salucci for supplying the rotation curve
data from which shear and the $V_{2.2}$ velocities were derived for some
of the galaxies in this paper.
We also thank the referee whose comments improved the content of this
paper.

\end{document}